\documentclass[iop,revtex4]{emulateapj}
\usepackage{apjfonts}
\usepackage{amssymb,appendix,multirow,morefloats,amsmath}
\usepackage[figuresright]{rotating}
\usepackage[flushleft]{threeparttable}
\usepackage{subfigure}
\usepackage{epstopdf}
\usepackage{graphicx}

\begin{document}

\def\lsim{\mathrel{\lower .85ex\hbox{\rlap{$\sim$}\raise
.95ex\hbox{$<$} }}}
\def\gsim{\mathrel{\lower .80ex\hbox{\rlap{$\sim$}\raise
1.0ex\hbox{$>$} }}}

\title{The Norma Arm Region \textit{Chandra} Survey Catalog: X-ray Populations in the Spiral Arms}

\shorttitle{The NARCS Catalog: X-ray Populations in the Spiral Arms}

\shortauthors{Fornasini et al.}
\slugcomment{{\sc Accepted to ApJ: } September 4, 2014}

\author{Francesca M. Fornasini\altaffilmark{1,2}, John A. Tomsick\altaffilmark{2}, Arash Bodaghee\altaffilmark{2,3}, Roman A. Krivonos\altaffilmark{2,4}, Hongjun An\altaffilmark{5}, Farid Rahoui\altaffilmark{6,7}, Eric V. Gotthelf\altaffilmark{8}, Franz E. Bauer\altaffilmark{9,10,11}, and Daniel Stern\altaffilmark{12}}

\altaffiltext{1}{Astronomy Department, University of California, 601 Campbell Hall, Berkeley, CA 94720, USA (e-mail: f.fornasini@berkeley.edu)}
\altaffiltext{2}{Space Sciences Laboratory, 7 Gauss Way, 
University of California, Berkeley, CA 94720, USA}
\altaffiltext{3}{Georgia College and State University, CBX 082, Milledgeville, GA 31061, USA}
\altaffiltext{4}{Space Research Institute, Russian Academy of Sciences, Profsoyuznaya 84/32, 117997 Moscow, Russia}
\altaffiltext{5}{Department of Physics, McGill University, Rutherford Physics Building, 3600 University Street, Montreal, QC H3A 2T8, Canada}
\altaffiltext{6}{European Southern Observatory, Karl Schwarzschild-Strasse 2, 85748 Garching bei M\"{u}nchen, Germany}
\altaffiltext{7}{Department of Astronomy, Harvard University, 60 Garden Street, Cambridge, MA 02138, USA}
\altaffiltext{8}{Columbia Astrophysics Laboratory, Columbia University, 550 West 120th Street, New York, NY 10027, USA}
\altaffiltext{9}{Instituto de Astrof\'{\i}sica, Facultad de F\'{i}sica, Pontificia Universidad Cat\'{o}lica de Chile, 306, Santiago 22, Chile} 
\altaffiltext{10}{Millennium Institute of Astrophysics, Vicu\~{n}a Mackenna 4860, 7820436 Macul, Santiago, Chile} 
\altaffiltext{11}{Space Science Institute, 4750 Walnut Street, Suite 205, Boulder, Colorado 80301}
\altaffiltext{12}{Jet Propulsion Laboratory, California Institute of Technology, 4800 Oak Grove Drive, MS 169-506, Pasadena, CA 91109}

\begin{abstract}

We present a catalog of 1415 X-ray sources identified in the Norma arm region \textit{Chandra} survey (NARCS), which covers a 2$^{\circ} \times 0\fdg8$ region in the direction of the Norma spiral arm to a depth of $\approx$20 ks.  Of these sources, 1130 are point-like sources detected with $\geq3\sigma$ confidence in at least one of three energy bands (0.5-10, 0.5-2, and 2-10 keV), five have extended emission, and the remainder are detected at low significance.  Since most sources have too few counts to permit individual classification, they are divided into five spectral groups defined by their quantile properties.  We analyze stacked spectra of X-ray sources within each group, in conjunction with their fluxes, variability, and infrared counterparts, to identify the dominant populations in our survey.  We find that $\sim$50\% of our sources are foreground sources located within 1-2 kpc, which is consistent with expectations from previous surveys.  Approximately 20\% of sources are likely located in the proximity of the Scutum-Crux and near Norma arm, while 30\% are more distant, in the proximity of the far Norma arm or beyond.  We argue that a mixture of magnetic and nonmagnetic CVs dominates the Scutum-Crux and near Norma arms, while intermediate polars (IPs) and high-mass stars (isolated or in binaries) dominate the far Norma arm.  We also present the cumulative number count distribution for sources in our survey that are detected in the hard energy band.  A population of very hard sources in the vicinity of the far Norma arm and active galactic nuclei dominate the hard X-ray emission down to $f_X\approx10^{-14}$ erg cm$^{-2}$ s$^{-1}$, but the distribution curve flattens at fainter fluxes.  We find good agreement between the observed distribution and predictions based on other surveys.

\end{abstract}


\hspace{0.1in}\textit{Online-only material} : catalog

\section{Introduction}
\label{sec:intro}
X-ray observations of the Galactic stellar population provide an important probe of several stages of stellar evolution. The brightest stellar X-ray sources are associated with compact stellar remants.  Neutron stars (NS), black holes (BH), and white dwarfs (WD) that are accreting matter from a binary companion are bright X-ray emitters.  Isolated neutron stars are also bright X-ray sources when they are young and hot, or if they accelerate particles in strong magnetic fields ($10^{12} \lesssim B \lesssim 10^{14}$ G).  We can learn about earlier stages of stellar evolution from X-ray observations as well.  Massive OB and Wolf-Rayet stars can produce X-rays through shocks in their stellar winds, and are sometimes more luminous than X-ray sources associated with compact stellar remnants.  Low-mass main sequence stars can produce low levels of X-ray emission in their magnetic coronae, and young stellar objects can produce X-rays due to their strong magnetic fields.  

Over the past decade, studies of X-ray source populations in several Galactic regions have been carried out using observations from the \textit{Chandra X-ray Observatory}.  These surveys have targeted the Galactic center \citep{wang02,muno09}, the Galactic bulge \citep{hong09}, the Orion region \citep{grosso05}, the Carina arm \citep{townsley11}, and a ``typical" region of the Galactic plane without point sources brighter than $2\times10^{-13}$ ergs cm$^{-2}$ s$^{-1}$ \citep{ebisawa05}.  Although some young, X-ray emitting massive stars have been discovered in the Galactic center \citep{mauerhan10}, the Galactic center and bulge are dominated by old X-ray stellar populations.  In contrast, the Orion region is a well-known star-forming region, and it has been argued that the Carina region is also a very young star-forming region since there is no evidence of a supernova explosion having occurred there yet \citep{smith07}.  

\begin{figure*}[t]
\makebox[\textwidth]{ %
\centering
\includegraphics[width=1.1\textwidth]{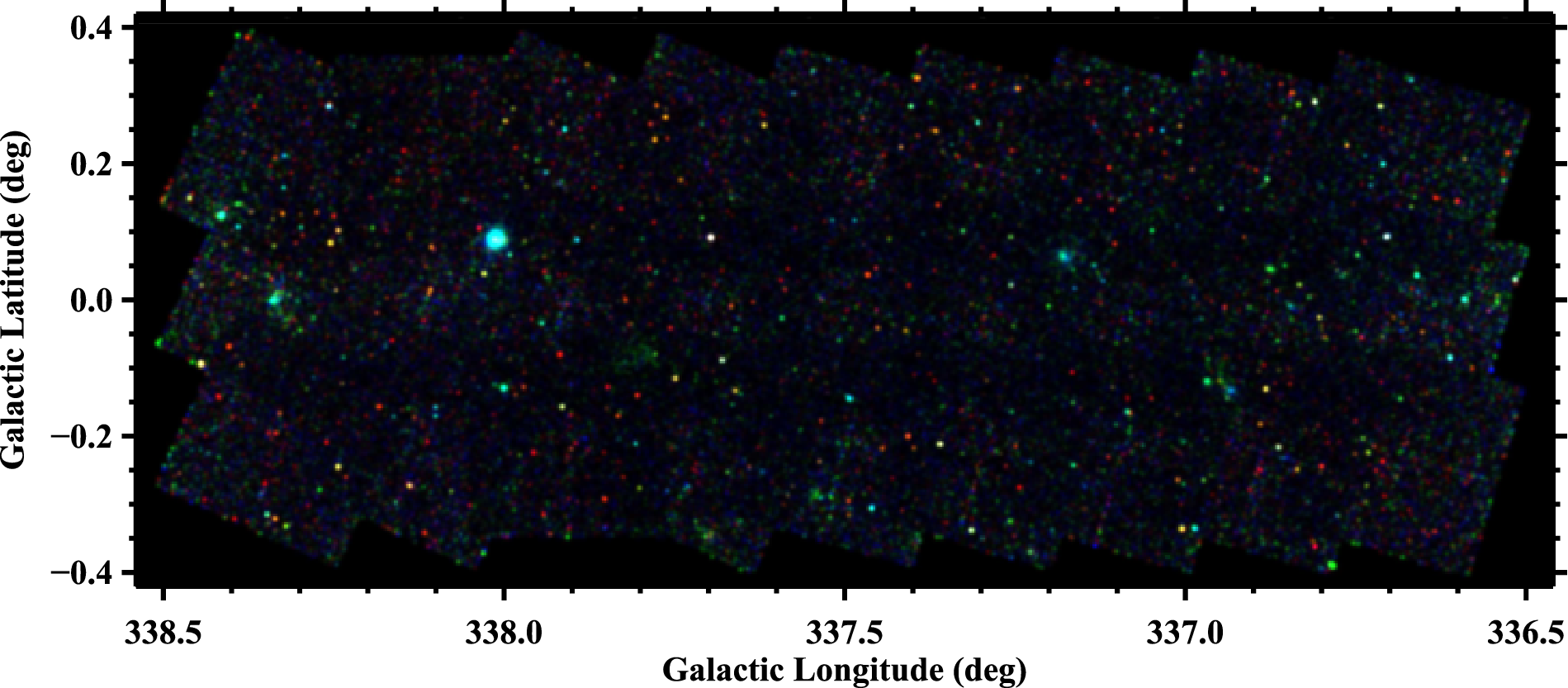}}
\caption{Three-color image of the Norma survey area.  Red is 0.5-2 keV, green is 2-4.5 keV, and blue is 4.5-10 keV.  Each energy band was smoothed using \texttt{aconvolve}.  Some artifacts are present at the chip edges. (A color version of this figure is available in the online journal.)}
\label{fig:mosaic}
\end{figure*}

We have conducted a \textit{Chandra} survey of the Norma spiral arm, which complements the aforementioned surveys since Norma's stellar population is likely more evolved than that of the Orion and Carina regions, but less evolved than that of the Galactic center and bulge.  The line-of-sight tangent to the Norma spiral arm contains the highest number of massive star-forming complexes in the Galaxy \citep{russeil03}, demonstrating there is a significant population of young stars in this arm.  Further evidence that this is a region of recent star formation is the presence of warm dust emission in \textit{Spitzer} 8 $\mu$m images and several HII regions mapped by 1420 MHz continuum emission from the Southern Galactic Plane Survey \citep{georgelin96}.  However, the supernova remnants discovered in this region \citep{green04,funk07,combi05} suggest that the Norma stellar population is older than the populations in the Orion and Carina regions observed with \textit{Chandra}.  Thus, the relative fractions of X-ray populations in the Norma arm may differ from those in other regions of the Galactic plane; in particular, the fraction of X-ray sources associated with evolved massive stars may be enhanced in Norma. 

The Norma Arm Region Chandra Survey (NARCS) consists of a 1.3 deg$^2$ region in the direction of the Norma spiral arm observed to a depth of 20 ks.  We present a catalog of all sources detected in NARCS (\S \ref{sec:obs}).  We divide the sources into groups based on their quantile properties (\S \ref{sec:specanalysis}) and analyze the photometric and spectral properties of each group to attempt to identify the dominant X-ray populations in this survey (\S \ref{sec:discussion}).  We calculate the NARCS number-flux distribution, correcting for the Eddington bias, the incompleteness of our detection procedure, and the variations in sensitivity across the surveyed area (\S \ref{sec:methodology}).  We compare the resulting number-flux distribution to predictions based on previous surveys of Galactic X-ray sources (\S \ref{sec:logNlogS}). 

\begin{table*}
\centering
\footnotesize
\caption{Observations of the Norma Region}
\begin{threeparttable}
\begin{tabular}{cccccc} \hline \hline
ObsID & R.A. (deg) & Dec. (deg) & Roll (deg) & Start Time (UT) & Exposure (ks)\\ 
(1) & (2) & (3) & (4) & (5) & (6) \\
\hline
12507&250.373201&-46.662951&342.217237&2011 Jun 6 10:15:53&18.8\\
12508&250.155011&-46.530604&342.217232&2011 Jun 6 15:57:13&18.5\\
12509&249.937805&-46.397816&342.217176&2011 Jun 6 21:22:23&19.4\\
12510&250.180190&-46.812896&342.217230&2011 Jun 9 12:29:02&19.9\\
12511&249.961646&-46.681456&333.217149&2011 Jun 17 11:15:19&19.3\\
12512&249.743370&-46.550407&317.716418&2011 Jun 27 04:52:55&20.5\\
12513&249.984947&-46.965904&317.716539&2011 Jun 27 11:00:21&20.2\\
12514&249.767582&-46.829470&342.217317&2011 Jun 10 16:07:39&19.8\\
12515&249.550110&-46.695978&342.217265&2011 Jun 10 22:04:48&19.5\\
12516&249.790838&-47.111874&342.217386&2011 Jun 11 03:46:38&19.5\\
12517&249.572205&-46.978413&342.217342&2011 Jun 11 09:28:28&19.5\\
12518&249.354673&-46.844540&342.217308&2011 Jun 11 15:10:18&19.5\\
12519&249.594334&-47.262081&333.217286&2011 Jun 13 04:25:13&19.3\\
12520&249.375577&-47.128273&333.217242&2011 Jun 13 10:13:08&19.0\\
12521&249.157932&-46.994022&333.217206&2011 Jun 13 15:46:38&19.0\\
12522&249.396933&-47.410725&333.217339&2011 Jun 13 21:20:08&19.0\\
12523&249.178061&-47.276529&333.217293&2011 Jun 14 02:53:38&19.0\\
12524&248.960334&-47.141940&333.217275&2011 Jun 14 08:27:08&19.5\\
12525&249.198427&-47.559064&333.217397&2011 Jun 14 14:08:58&19.5\\
12526&248.979417&-47.424468&333.217332&2011 Jun 14 19:50:48&19.0\\
12527&248.761625&-47.289491&333.217351&2011 Jun 15 19:36:46&19.3\\
12528&248.998831&-47.707016&333.217482&2011 Jun 16 01:24:35&19.0\\
12529&248.779750&-47.572056&333.217441&2011 Jun 16 06:58:05&19.0\\
12530&248.561776&-47.436667&333.217382&2011 Jun 16 12:31:35&19.3\\
12531&248.798050&-47.854617&333.217515&2011 Jun 16 18:09:14&19.5\\
12532&248.578823&-47.719259&333.217450&2011 Jun 16 23:51:04&19.5\\
12533&248.360823&-47.583518&333.217451&2011 Jun 17 05:32:54&19.5\\
 \hline
\end{tabular}
\begin{tablenotes}

\item (1) Observation ID number.

\item (2) Right ascension (J2000.0) of observation pointing. 

\item (3) Declination (J2000.0) of observation pointing.

\item (4) Roll angle of observation.

\item (5) Start time of observation.

\item (6) Exposure time of observation. 
\end{tablenotes}
\end{threeparttable}
\label{tab:obs}
\end{table*}

\section{Observations and Source Catalog}
\label{sec:obs}

We performed \textit{Chandra} ACIS-I observations in faint mode of a 2$^{\circ} \times 0\fdg8$ region of the Norma spiral arm in June 2011.  The primary goal of this survey was to discover faint High-Mass X-ray Binaries (HMXBs) that may have been missed in previous surveys performed with instruments with less sensitivity and angular resolution than \textit{Chandra}.  The Norma spiral arm, likely due to its evolutionary state, was chosen as the target for this search because it hosts the largest number of known HMXBs of any line-of-sight through the Galaxy \citep{bodaghee12c}.  Even though HMXBs are more common in the Norma arm than elsewhere, they are rare sources compared to other X-ray populations; thus, our ongoing efforts to identify new HMXBs will be discussed in future papers while here we will focus on studying the dominant classes of X-ray sources in this survey. 

Our field was subdivided into 27 pointings; Table \ref{tab:obs} reports their coordinates and exposure times and Figure \ref{fig:mosaic} is a mosaic image of the survey.  Our observing strategy was to cover a wide area with relatively uniform flux sensitivity and good spatial resolution; therefore, we chose field centers spaced by 12$^{\prime}$, which provided roughly 70 arcmin$^2$ of overlap on the outskirts of adjacent observations such that the additional exposure time in these overlapping regions partly made up for the worsening point-spread function (PSF) at large off-axis angles.  There are only four archival \textit{Chandra} imaging observations of $>$20 ks which fall within the area of our survey region.  We chose not to incorporate these data into our survey because they do not add much value compared to the modifications we would have to make to our analysis strategy, and because they might bias our study of faint X-ray populations in the Norma region since they only provide deeper coverage of a fourth of the surveyed area which may not be representative of the region as a whole.    

The ACIS-I consists of four 1024 $\times$ 1024 pixel CCDs, covering a 17$^{\prime}$ $\times$ 17$^{\prime}$ field of view \citep{garmire03}.  The on-axis spatial resolution of the ACIS-I is fully sampled by the $0\farcs492 \times 0\farcs492$ CCD pixel but it increases greatly off-axis.  The PSF increases in size and becomes more elliptical at large off-axis angles, such that at an off-axis angle of 10$^{\prime}$, the PSF has ellipticity $\approx$ 0.3 and semi-major axis $\approx$ 15$^{\prime}$ for an ECF of 90\% for 4.5 keV photons \citep{allen04}.  The CCDs are sensitive to incident photons with energies in the 0.3-10.0 keV range, and have a resolution of about 50-300 eV.  The time resolution of the CCDs, which is determined by the read-out time, is 3.2 s.  

We analyzed the data using standard tools from the CIAO package, version 4.4.\footnotemark\footnotetext{Available at http://cxc.harvard.edu/ciao4.4/}  We used \texttt{chandra\_repro} to reprocess the level 1 event lists provided by the \textit{Chandra} X-Ray Center (CXC).  This tool calls on \texttt{acis\_process\_events} to clean the cosmic ray background for very faint mode observations and also applies the sub-pixel event repositioning algorithm EDSER.  Background flares accounted for $<$1\% of the exposure time and were all relatively weak so, for simplicity, they were not removed.  

\subsection{Source Detection and Localization}
\label{sec:detection}
We searched for X-ray sources in each observation separately using the wavelet detection algorithm \texttt{wavdetect}.  For each observation, we generated images in three energy bands and four spatial resolutions.  The three energy bands were the full 0.5-10 keV band (FB), the soft 0.5-2 keV band (SB), and the hard 2-10 keV band (HB); these three bands were chosen to make our source search sensitive to spectrally different sources.  For each energy band, we made one image with the full resolution (0.5$^{\prime\prime}$), one binned by a factor 2 (1$^{\prime\prime}$), one binned by a factor of 4 (2$^{\prime\prime}$), and another binned by a factor of 8 (4$^{\prime\prime}$).  Exposure maps for each of the three energy bands and four spatial resolutions were also generated applying the spectral weights for a power-law model with $\Gamma = 2.0$ and $N_{\mathrm{H}} = 5.0 \times 10^{22}$ cm$^{-2}$, a column density appropriate for sources in the far Norma arm at $10-12$ kpc distances;\footnotemark\footnotetext{See \S \ref{sec:specanalysis} for details on the calculation of $N_{\mathrm{H}}$ to a given distance.} the \texttt{wavdetect} exposure threshold was set to 0.1 to minimize spurious detections at detector locations with low exposure times.  PSF maps of the 39.3\% enclosed-count fraction (ECF)\footnotemark\footnotetext{As recommended by \citet{freeman}} at 4.5 keV (for the full and hard bands) and at 1.5 keV (for the soft band), computed using \texttt{mkpsfmap}, were also supplied to \texttt{wavdetect}.  We used wavelet scales that increased by a factor of $\sqrt{2}$ from 1 to 16; this range of wavelet scales and four different spatial resolutions were chosen because the ACIS-I PSF varies significantly with offset from the aim point.  We selected the Mexican Hat wavelet, and a sensitivity threshold equal to one over the number of pixels in the image, which corresponds to the chance of detecting approximately one spurious source per image assuming a spatially uniform background.  Therefore, we expect \texttt{wavdetect} to identify 324 spurious sources (27 observations $\times$ 3 energy bands $\times$ 4 resolutions); we find 315 sources are detected at $<3\sigma$, which is in rough agreement with the expected number of spurious sources, so the sample of sources detected at $\geq3\sigma$ in at least one energy band is probably mostly free of spurious sources.  

We combined the source lists from the four images with different spatial resolutions for each observation and energy band.  Sources were identified as matches if the separation between them was smaller than the 50\% encircled energy contour for the PSF of 4.5 keV photons at the detector position of the source plus the uncertainty in position due to the pixel size in the lower resolution image (e.g. when comparing sources in the full resolution image and the image binned by a factor of 2, this uncertainty is 2 pixels$\times \sqrt{2} \times 0\farcs492$/pixel=$1\farcs39$).  When duplicate sources were identified, we only retained the position of the source detected in the highest resolution image.  

Then we made background maps for each of the observations and energy bands by removing the counts of sources detected in that particular observation and energy band and filling these regions in with a number of counts determined from the average local background.  The source regions we removed were defined as circles centered on the source position, with a radius ($r_{96\%}$) equal to the 96\% enclosed count fraction (ECF) PSF for 4.5 keV photons (for the full and high energy bands) or for 1.5 keV (for the low energy band).  In addition, we manually defined regions for the extended sources present in observations ObsID 12508, 12516, 12523, 12525, 12526, and 12528.  The background regions were defined as annuli with an inner radius equal to the radius of the punched-out region, and an outer radius twice as large; if an annulus overlapped a punched-out region, that overlapping segment was removed from the annulus.  The number of background counts to be randomly distributed within the punched-out region was calculated by multiplying the counts in the background region by the ratio of the region areas and the ratio of the region mean effective areas, as determined from the exposure maps.  After the punched-out regions had been filled in with the appropriate number of background counts, the background maps were smoothed using \textit{csmooth} with Gaussian kernels of sizes ranging from 20 to 50 pixels.  For the observations containing extended sources, first, a smoothed background map was made with both the extended and points sources removed.  This map was passed to \texttt{csmooth} as a background map, and a smoothed background map was made with only the point sources removed.

We then employed \texttt{wavdetect} again to search for X-ray sources in each observation, but this time we used the smoothed background maps we made instead of defaulting to the background maps automatically generated by \texttt{wavdetect}.  We found that when the background maps we made were used, a larger fraction of sources was detected in higher resolution images than with the automatically-generated maps.  As before, for each observation, we combined the source lists from the four images with different spatial resolutions.  Then we combined the source lists from the three energy bands.  When two sources were identified as a match, only one source entry was retained; preference was given to sources detected in the full energy band and then the soft band, because the PSF size is smaller at low energies, allowing better source localization. 

\begin{table}[b]
\centering
\footnotesize
\caption{Refined Astrometry}
\begin{threeparttable}
\begin{tabular}{cccccc} \hline \hline
ObsID & R.A. (deg) & Dec. (deg) & Roll (deg) & Unc. & \# Counterparts\\ 
(1) & (2) & (3) & (4) & (5) & (6) \\
\hline
12507&250.373197&-46.666299&342.207886&$0\farcs38$&14\\
12508&250.155064&-46.530768&342.256256&$0\farcs16$&6\\
12509&249.937691&-46.397888&342.258575&$0\farcs42$&26\\
12510&250.180484&-46.812861&342.220398&$0\farcs32$&21\\
12511&249.961539&-46.681483&333.150848&$0\farcs39$&7\\
12512&249.743528&-46.550463&317.725342&$0\farcs38$&14\\
12513&249.985150&-46.965931&317.737030&$0\farcs35$&24\\
12514&249.767578&-46.829508&342.200439&$0\farcs39$&9\\
12515&249.550029&-46.696055&342.115234&$0\farcs35$&22\\
12516&249.790920&-47.111803&342.203583&$0\farcs35$&10\\
12517&249.572277&-46.978334&342.136200&$0\farcs54$&5\\
12518&249.354557&-46.844589&342.191071&$0\farcs35$&23\\
12519&249.594337&-47.262048&333.187683&$0\farcs36$&19\\
12520&249.375668&-47.128341&333.142365&$0\farcs35$&3\\
12521&249.157996&-46.994096&333.227539&$0\farcs32$&24\\
12522&249.396838&-47.410790&333.249512&$0\farcs34$&18\\
12523&249.178005&-47.276545&333.162964&$0\farcs31$&6\\
12524&248.960241&-47.141987&333.209045&$0\farcs34$&24\\
12525&249.198264&-47.559033&333.258545&$0\farcs29$&15\\
12526&248.979187&-47.424375&333.234863&$0\farcs28$&5\\
12527&248.761554&-47.289507&333.188904&$0\farcs37$&19\\
12528&248.998873&-47.707003&333.211639&$0\farcs39$&19\\
12529&248.779969&-47.572088&333.210205&$0\farcs30$&9\\
12530&248.561714&-47.436723&333.186920&$0\farcs39$&15\\
12531&248.797988&-47.854470&333.221191&$0\farcs31$&17\\
12532&248.578674&-47.719259&333.231171&$0\farcs34$&9\\
12533&248.360765&-47.583555&333.165527&$0\farcs32$&18\\
 \hline
\end{tabular}
\begin{tablenotes}

\item (1) Observation ID number.

\item (2) Right ascension (J2000.0) after astrometric correction. 

\item (3) Declination (J2000.0) after astrometric correction.

\item (4) Roll angle after astrometric correction.

\item (5) Average systematic uncertainty between IR and X-ray positions after astrometric refinement.

\item (6) Number of VVV counterparts used to refine astrometry.
\end{tablenotes}
\end{threeparttable}
\label{tab:reproject}
\end{table} 

\begin{sidewaystable}
\centering
\footnotesize
\caption{Catalog of Point and Extended Sources: Detection and Localization}
\begin{threeparttable}
\begin{tabular}{ccccccccccccc} \hline \hline
No. & Source & ObsID & R.A. & Dec. & Unc. & Offset & Sig. & Sig. & Sig. & Radius & PSF & Flags\\ 
& (CXOU J) & (125**) & (deg) & (deg) & (arcsec) & (arcmin) & FB & SB & HB & (arcsec) & (arcsec) & \\
(1) & (2) & (3) & (4) & (5) & (6) & (7) & (8) & (9) & (10) & (11) & (12) & (13) \\
\hline
1&163228.2-473755&33&248.117829&-47.632173&4.03&10.3&3.6&2.3&2.5&13.8&13.8&...\\
2&163241.5-474039&33&248.172944&-47.677522&1.79&9.5&9.0&10.4&2.5&11.9&11.9&...\\
3&163244.6-474133&33&248.186065&-47.692513&3.63&9.6&2.4&0.0&2.8&12.2&12.2&...\\
4&163248.7-473017&33&248.203151&-47.504857&1.41&7.9&8.4&1.2&8.5&9.2&9.2&...\\
5&163251.0-474135&33&248.212798&-47.693198&3.16&8.9&3.0&5.0&0.0&10.7&10.7&...\\
6&163253.0-474201&33&248.221111&-47.700286&2.26&9.0&5.2&1.8&4.8&10.9&10.9&...\\
7&163259.0-473819&33&248.246176&-47.638806&1.13&5.7&7.5&7.2&4.1&5.2&5.2&...\\
8&163259.4-472804&33&248.247582&-47.467941&3.39&8.3&2.3&3.4&0.4&10.1&10.1&...\\
9&163303.2-472547&33&248.263337&-47.429724&13.98&10.0&0.0&0.8&0.0&14.0&14.0&...\\
10&163306.2-473239&33&248.276159&-47.544291&1.55&4.2&3.3&0.7&3.1&3.5&3.5&...\\
\hline
\end{tabular}
\begin{tablenotes}
\item \underline{Notes:} Table \ref{tab:catalog} is published in its entirety in the electronic edition of the Astrophysical Journal.  A portion is shown here for guidance regarding its form and content.  See Appendix \ref{app:firsttable} for detailed column descriptions.
\end{tablenotes}
\end{threeparttable}
\label{tab:catalog}
\end{sidewaystable}

\begin{sidewaystable}
\vspace{-2.0in}
\centering
\footnotesize
\caption{Catalog of Point and Extended Sources: Photometry}
\begin{threeparttable}
\begin{tabular}{ccccccccccccc} \hline \hline
No. & $C_{net}$ & $C_{net}$ & $C_{net}$ & $f_{\mathrm{ph}}$ FB & $f_{\mathrm{ph}}$ SB & $f_{\mathrm{ph}}$ HB & $E_{50}$ & $E_{25}$ & $E_{75}$ & $f_X$ FB & Phot. & Quantile \\ 
 & FB & SB & HB & (10$^{-6}$cm$^{-2}$s$^{-1}$) & (10$^{-6}$cm$^{-2}$s$^{-1}$) & (10$^{-6}$cm$^{-2}$s$^{-1}$) & (keV) & (keV) & (keV) & (10$^{-14}$erg cm$^{-2}$s$^{-1}$) & Flag & Group\\
(1) & (2-4) & (5-7) & (8-10) & (11-13) & (14-16) & (17-19) & (20-21) & (22-23) & (24-25) & (26-28) & (29) & (30) \\
\hline
1&16$^{+7}_{-6}$&6$^{+4}_{-3}$&10$^{+6}_{-5}$&3.76$^{+1.57}_{-1.32}$&0.76$^{+0.55}_{-0.40}$&2.47$^{+1.39}_{-1.13}$&2.9$\pm$1.8&1.6$\pm$0.8&5.6$\pm$2.4&1.75$^{+1.31}_{-1.25}$&- - -&C\\
2&47$^{+9}_{-8}$&37$^{+7}_{-6}$&10$^{+6}_{-5}$&10.55$^{+1.99}_{-1.76}$&4.68$^{+0.93}_{-0.80}$&2.28$^{+1.31}_{-1.07}$&1.4$\pm$0.1&1.0$\pm$0.1&1.8$\pm$0.5&2.35$^{+0.49}_{-0.44}$&- - -&A\\
3&9$^{+5}_{-4}$&3&9$^{+5}_{-4}$&3.48$^{+2.15}_{-1.71}$&0.69&3.88$^{+2.14}_{-1.69}$&5.4$\pm$0.9&4.5$\pm$1.3&6.1$\pm$0.9&3.03$^{+1.93}_{-1.56}$&- S -&D\\
4&36$^{+8}_{-7}$&2$^{+3}_{-2}$&34$^{+7}_{-6}$&7.95$^{+1.69}_{-1.46}$&0.28$^{+0.40}_{-0.24}$&7.60$^{+1.66}_{-1.42}$&4.6$\pm$0.3&3.7$\pm$0.3&5.2$\pm$0.5&5.90$^{+1.33}_{-1.17}$&- - -&D\\
5&13$^{+6}_{-5}$&13$^{+5}_{-4}$&6&2.85$^{+1.36}_{-1.12}$&1.70$^{+0.65}_{-0.51}$&1.32&1.0$\pm$0.1&0.9$\pm$0.2&1.1$\pm$0.1&0.44$^{+0.21}_{-0.18}$&- - H&A\\
6&22$^{+7}_{-6}$&4$^{+4}_{-2}$&19$^{+6}_{-5}$&6.00$^{+1.82}_{-1.53}$&0.59$^{+0.55}_{-0.36}$&5.07$^{+1.71}_{-1.41}$&3.0$\pm$0.7&2.2$\pm$0.4&5.7$\pm$1.2&2.90$^{+1.10}_{-0.99}$&- - -&C\\
7&20$^{+6}_{-5}$&12$^{+5}_{-3}$&9$^{+4}_{-3}$&4.70$^{+1.35}_{-1.10}$&1.53$^{+0.60}_{-0.46}$&2.08$^{+1.04}_{-0.78}$&1.8$\pm$0.6&1.5$\pm$0.1&3.5$\pm$0.5&1.37$^{+0.57}_{-0.53}$&- - -&B\\
8&8$^{+5}_{-4}$&7$^{+4}_{-3}$&1$^{+4}_{-1}$&1.82$^{+1.17}_{-0.92}$&0.85$^{+0.51}_{-0.37}$&0.33$^{+0.93}_{-0.33}$&1.6$\pm$0.9&1.4$\pm$0.3&1.8$\pm$3.8&0.47$^{+0.40}_{-0.35}$&- - -&A\\
9&7&2$^{+3}_{-2}$&5&1.74&0.27$^{+0.48}_{-0.27}$&1.23&5.2$\pm$4.8&2.9$\pm$7.1&7.6$\pm$7.1&1.46&F - H&C\\
10&6$^{+4}_{-3}$&1$^{+2}_{-1}$&5$^{+4}_{-2}$&1.11$^{+0.74}_{-0.50}$&0.08$^{+0.26}_{-0.08}$&0.98$^{+0.72}_{-0.47}$&5.1$\pm$0.8&4.4$\pm$1.4&5.4$\pm$0.3&0.91$^{+0.62}_{-0.44}$&- - -&E\\
 \hline
\end{tabular}
\begin{tablenotes}
\item  \underline{Notes:}  Table \ref{tab:catalog2} is published in its entirety in the electronic edition of the Astrophysical Journal.  A portion is shown here for guidance regarding its form and content.  See Appendix \ref{app:secondtable} for detailed column descriptions of the electronic version.
\end{tablenotes}
\end{threeparttable}
\label{tab:catalog2}
\end{sidewaystable}

\begin{table*}[t]
\centering
\footnotesize
\caption{Catalog of Point and Extended Sources: Infrared Counterparts}
\begin{threeparttable}
\begin{tabular}{ccccccccccccc} \hline \hline
No. & VVV Source Name & R.A. & Dec. & $\Delta_{\mathrm{X-IR}}$ & $p_{\mathrm{noise}}$ & Reliability \\
& & (deg) & (deg) & (arcsec) & & \\
(1) & (2) & (3) & (4) & (5) & (6) & (7) \\
\hline
1&515727792649&248.117752&-47.631649&1.89&3.07e-03&0.3216\\
2&515726841264&248.172806&-47.677017&1.84&5.29e-07&0.8660\\
3&515726837733&248.185730&-47.693638&4.13&5.29e-07&0.3508\\
4&515727238897&248.203003&-47.505127&1.05&1.71e-04&0.7804\\
5&515726847521&248.212341&-47.693485&1.52&1.71e-04&0.9482\\
6&515727540494&248.220947&-47.700108&0.76&1.71e-04&0.9142\\
7&515726868309&248.246140&-47.638630&0.64&5.29e-07&0.9777\\
8&515726918176&248.247345&-47.468082&0.78&9.52e-06&0.4607\\
9&515726930863&248.262817&-47.429794&1.29&5.29e-07&0.0657\\
10&515727577185&248.276459&-47.544682&1.59&2.81e-04&0.8825\\
\hline
\end{tabular}
\begin{tablenotes}
\item  \underline{Notes:}  Table \ref{tab:catalog3} is published in its entirety in the electronic edition of the Astrophysical Journal.  A portion is shown here for guidance regarding its form and content.  See Appendix \ref{app:thirdtable} for detailed column descriptions of the electronic version.
\end{tablenotes}
\end{threeparttable}
\label{tab:catalog3}
\end{table*}

In order to refine the astrometry of our observations, we searched the VISTA Variables in the Via Lactea (VVV) Survey catalog \citep{minniti10} for infrared counterparts to the X-ray sources we detected\footnotemark\footnotetext{See http://www.eso.org/sci/observing/phase3/data\_releases/vvv\_dr1.html for the first data release used in this paper.}  For each \textit{Chandra} observation, we made a list of VVV sources within 12$^{\prime}$ of the observation aim point with less than a 0.137\% probability of being a noise fluctuation.  We then determined the reliability of each IR counterpart based on the positional uncertainties of the X-ray and IR sources, the distance between the X-ray and IR source, and the density of IR sources following the treatment of \citet{suther92} but without making any assumptions about the probability distribution function in magnitude of the true IR counterparts.\footnotemark\footnotetext{Our calculation takes into account the probability that the NIR counterpart of an X-ray source is undetected in the VVV survey. In \citet{suther92} this null probability is the quantity (1-\textit{Q}). Since we do not know $Q$ a priori, we guess its value and then refine our guess iteratively until the $Q$ value meets the criterion in Equation (7) of \citet{suther92}. In this way, we find $Q$ = 0.85.}  The 1$\sigma$ positional uncertainty of sources in the VVV catalog is typically $0\farcs07$.  
We determined the positional uncertainty of the \textit{Chandra} sources using the parameterization of the statistical error as a function of offset angle and net counts\footnotemark\footnotetext{These are the net counts reported by \texttt{wavdetect}, not those determined by aperture photometry.} in Equation 5 of \citet{hong05}.  These statistical errors were combined in quadrature with a systematic error of 0\farcs7 (95\% error\footnotemark\footnotetext{See http:/cxc.harvard.edu/cal/ASPECT/celmon.}) due to Chandra guide star alignment uncertainties. 
We used VVV matches with a reliability greater than 0.9 and \texttt{reproject\_aspect} to derive a linear and rotational astrometric correction for each \textit{Chandra} observation, reducing the systematic astrometric errors to $\leq 0\farcs54$.  We applied the same corrections to the source positions in our source lists.  We replaced the $0\farcs7$ systematic errors with the average residuals from the astrometric transformation derived from the VVV counterparts to the X-ray sources.  Table \ref{tab:reproject} presents the refined astrometry for each observation.

We detected a total of 1658 sources but since each of the 27 observations partially overlaps with at least three other survey observations, we checked for duplicate sources between the different observations in order to only have one entry per source in our final catalog (see Tables \ref{tab:catalog}, \ref{tab:catalog2}, and \ref{tab:catalog3} for a sample).  A source was considered a true duplicate if the distance between the two sources was smaller than the quadrature sum of the positional uncertainties of the two sources.  If the distance between two sources was larger than the quadrature sum of the positional uncertainties but smaller than the regular sum of the uncertainties, then the sources were flagged for manual inspection.  Some sources that met the true duplicate criterion but were unusual in some respect (e.g. one source was flagged as extended while the other was not) were also flagged by the algorithm for manual inspection.  Whether or not the sources flagged for manual inspection were determined to be duplicates, they were flagged with ``id" for ``inspected duplicate" in the catalog.\footnotemark\footnotetext{Flags for each source are provided in column 13 of Table \ref{tab:catalog}.  Descriptions of all flags are provided in Appendix \ref{app:firsttable}.}.  In total, 38 sets of sources were flagged for manual inspection and we determined 28 of them were true duplicates.  The catalog entries for duplicates were combined so that exposure times and net counts were summed, source positions were weight-averaged, and the ObsIDs, offsets from the aim point, source region radii, 90\% PSF sizes, and flags of the duplicate sources were all listed. If a source was determined to be variable on long timescales (see \S \ref{sec:variability}), its derived photometric properties were averaged, but if it was determined to be constant, they were weight-averaged.  After combining the entries of duplicate sources, our catalog contains 1415 sources. 

\subsection{Photometry}
\label{sec:photometry}

\begin{figure*}[t]
	\centering
	\subfigure[Flag ``m1" case.]{\includegraphics[width=2.2in]{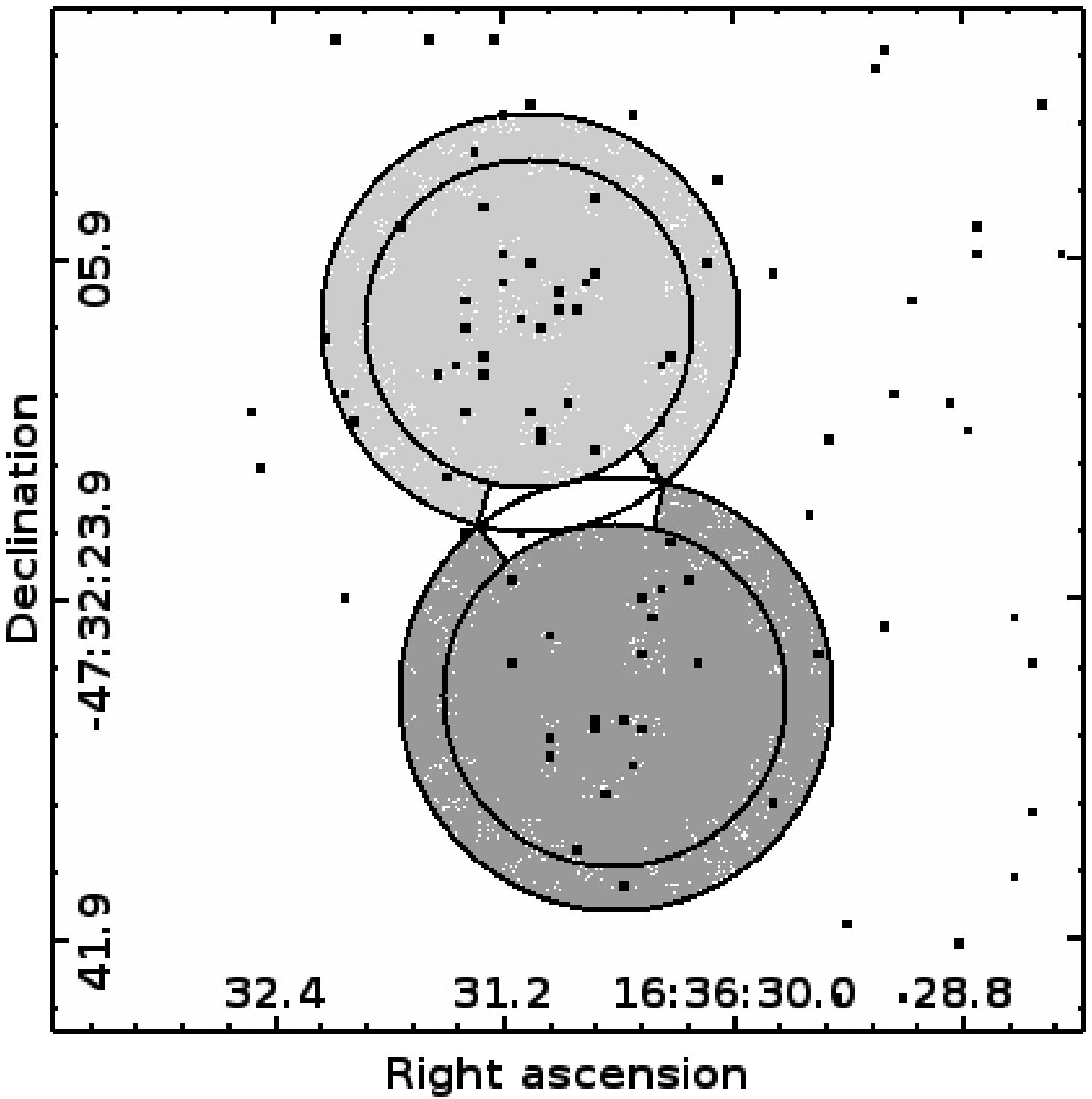}
	\label{fig:pie1}}
	\subfigure[Flag ``m2" case.]{\includegraphics[width=2.05in]{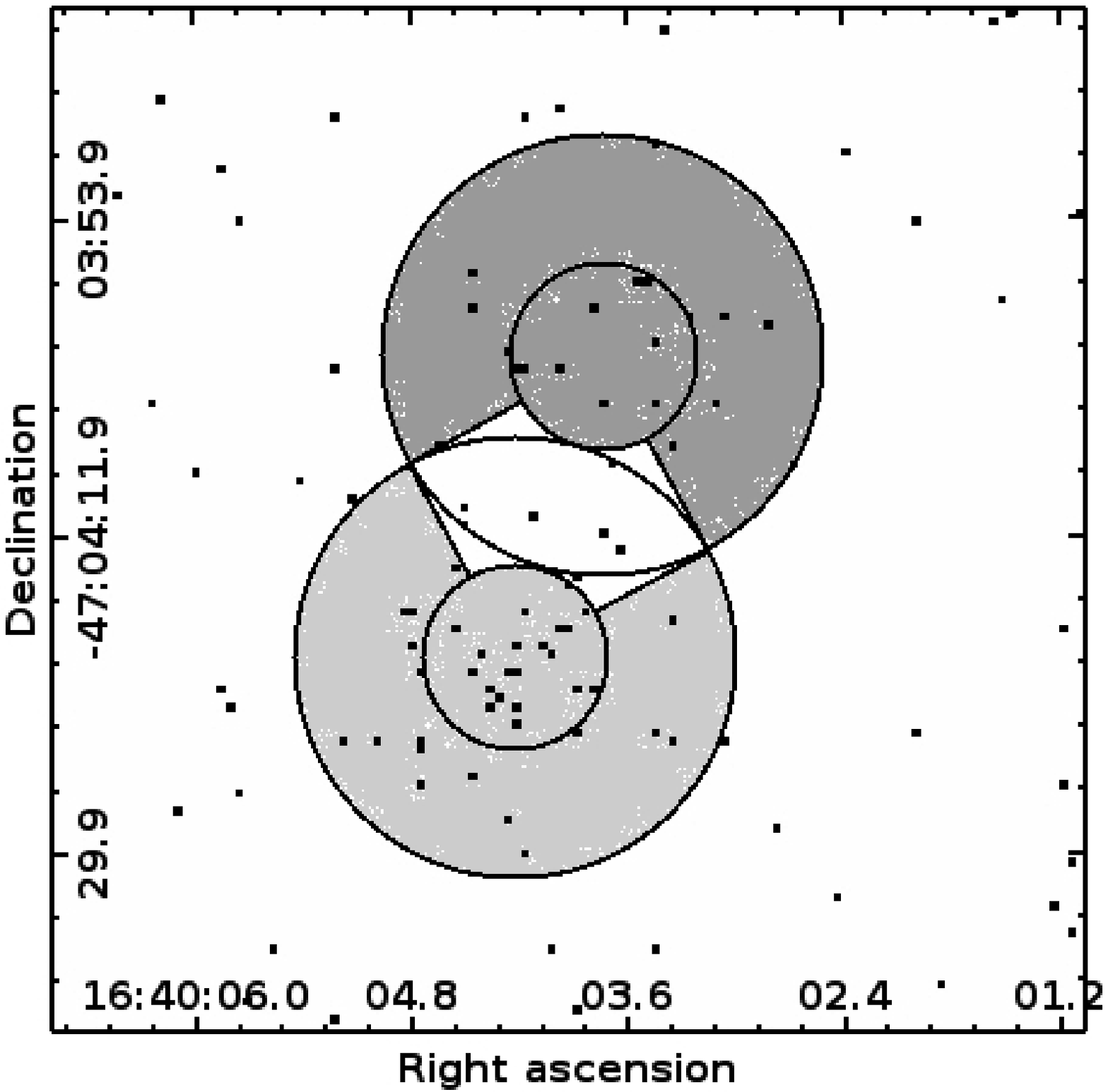}
	\label{fig:pie2}}
	\subfigure[Flag ``m3" case.]{\includegraphics[width=2.15in]{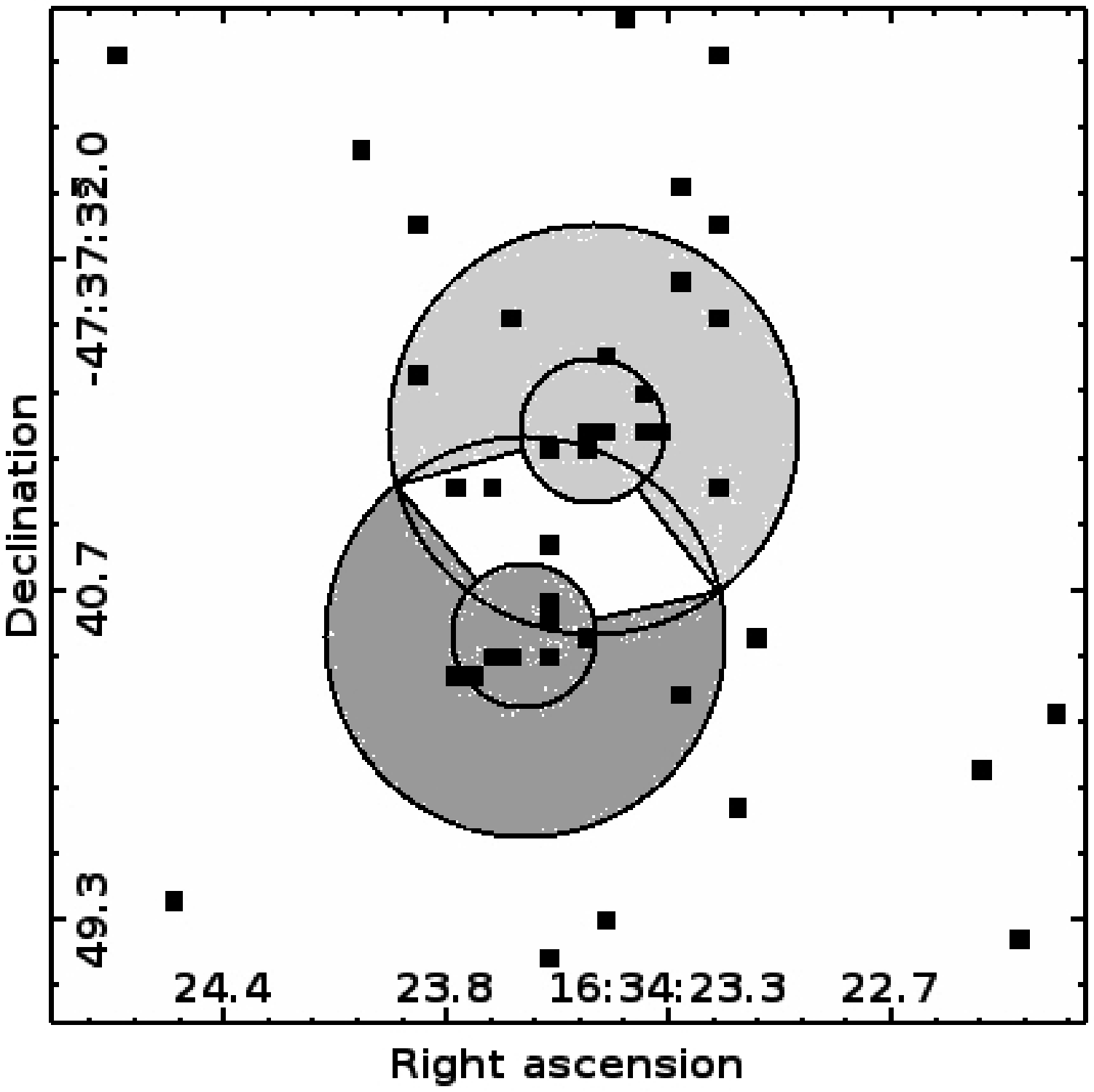}
	\label{fig:pie3}}
	\caption{Examples of modified overlapping regions.}
\label{fig:pie}
\end{figure*}

We used \texttt{dmextract} to compute photometric quantities for each of the X-ray sources in our catalog.  In most cases, we defined an aperture region for each source as a circle with radius ($r_{90\%}$) equal to the 90\% ECF PSF for 4.5 keV photons.  Given the relatively low source counts for the vast majority of sources, this radius is well-optimized to obtain the highest S/N ratio.  However, if the semi-major axis of the source region provided by \texttt{wavdetect} was more than twice as large as $r_{90\%}$ in all images in which the source was detected, then the semi-major axis was used as the radius of the aperture region and the source was flagged with ``e" for ``extended", to denote that it may be an extended source.  In some cases, these ``e" source regions surrounded another source region; these sources were additionally flagged with an ``s" for ``surrounding".  We nonetheless included these sources in our analysis of point-like sources, but they only constitute 3\% of sources detected at $\geq3\sigma$.  We also modified the aperture regions of overlapping sources; following the method of \cite{hong05}, we defined a source region as the sum of a circular core and a pie sector of an annular shell that excludes the common sector with the neighbor's source region.  The core size was determined empirically to maximize the source photons included and to minimize contamination from neighbors, as described in Table \ref{tab:pie}, which also lists the flags associated with each type of source region modification from the standard one.  As was implemented in \cite{hong05}, if the source region overlapped with more than one neighbor, the core size was determined by the nearest neighbor and the pie sector excluded all common sectors with the neighbors' aperture regions.  Figure \ref{fig:pie} shows some examples of these modified overlapping source regions.  We manually modified 12 aperture regions of sources with multiple nearby neighbors.  Finally, we manually created source and background regions for the extended sources in observations ObsID 12508, 12516, 12523, 12525, 12526, and 12528 that were not detected by \texttt{wavdetect} and flagged these sources with a ``c" for ``created". 

The background region for each source was defined as an annulus with an inner radius equal to $r_{96\%}$ and an outer radius equal to $2r_{96\%}$.  In the few instances when a source flagged as ``e" had a source radius larger than $r_{96\%}$, then the inner radius of the annulus was set equal to the source radius, and the outer radius was twice as large as this inner radius.  As we did when making the background maps, we generated images with punched-out $r_{96\%}$ source regions, which were again defined as circles with radii equal to $r_{96\%}$.  When extracting photometric information from the background regions, we used these punched-out images so that contamination from neighboring sources was avoided.  When calculating the background region area, we corrected the annular area for any segments that were excluded due to overlap with punched-out source regions.  

\begin{table*}
\vspace{-0.3in}
\centering
\footnotesize
\caption{Aperture Region Definitions}
\begin{threeparttable}
\begin{tabular}{cccccc} \hline \hline
Source Region & Condition & Core Radius & Refined Source & Background & Flag \\ 
 Overlap && $r_c$ & Region & Region &  \\
\hline
No....&$\Delta\geq r_{90\%}+r'_{90\%}$&$r_{90\%}$&$r\leq r_c = r_{90\%}$&$r_{96\%}< r < 2r_{96\%}$ and $r'' > r''_{96\%}$& ... \\
&&&&for all neighbors&\\
Yes...& $\Delta\geq 1.5r'_{90\%}$,&$\Delta-r'_{90\%}$&$r\leq r_c$ and pie sector&Same as above&m1\\
&$\Delta<r_{90\%}+r'_{90\%}$&&with $r_c<r\leq r_{90\%}$&&\\
Yes...&$\Delta < 1.5r'_{90\%},$&$\Delta-r'_{90\%}$&Same as above&Same as above&m2\\
&$\Delta \geq r_{68\%}+r'_{90\%}$&&&&\\
Yes...&$\Delta < r_{68\%}+r'_{90\%}$&$\Delta/3$&Same as above&Same as above&m3\\
\hline
\end{tabular}
\begin{tablenotes}
\item  \underline{Notes:} Parameter $\Delta$ is the distance between the source and its nearest neighbor, and $r'_{90\%}$ is the 90\% PSF radius of the nearest neighbor.  $r$ refers to the distance from the source and $r''$ refers to the distance from neighbors.  The PSF radii are calculated for 4.5 keV photons.  For sources flagged as potentially extended (``e"), these criteria remain the same, except that instead of using $r_{90\%}$ and $r_{96\%}$, the radii listed in the catalog are used.
\end{tablenotes}
\end{threeparttable}
\label{tab:pie}
\end{table*}

\begin{figure}[b]
\includegraphics[width=0.47\textwidth]{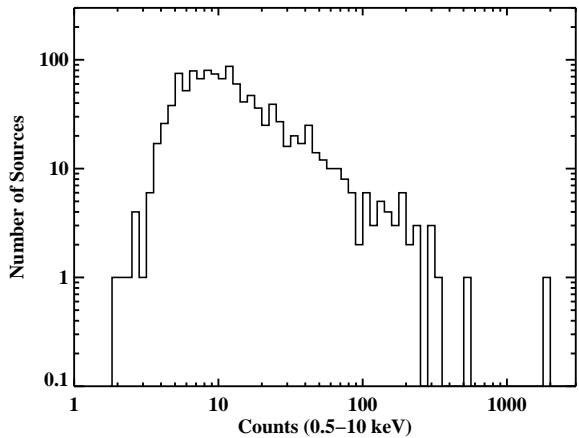}
\caption{Histogram of 0.5-10 keV net counts for $\geq3\sigma$ sources.  The brightest source with 14,720 counts is not shown. \label{fig:counthist}}
\end{figure}

\begin{figure*}[t]
\makebox[\textwidth]{ %
\centering
\includegraphics[width=1.1\linewidth]{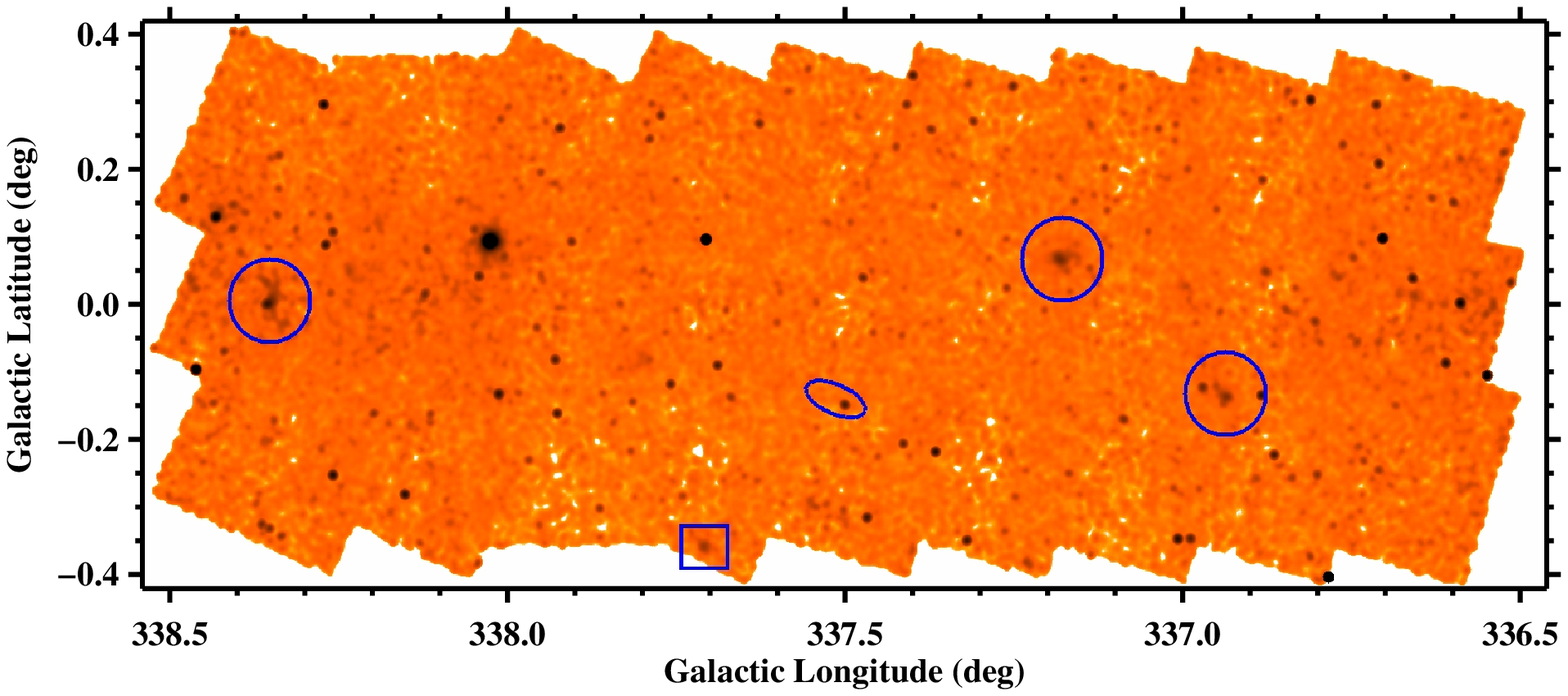}}
\caption{Mosaic image of the 0.5-10 keV band showing locations of extended sources identified by eye.  Circles indicate confirmed SNRs, while the square indicates a candidate SNR, and the ellipse indicates a PWN candidate. (A color version of this figure is available in the online journal.)}
\label{fig:extended}
\end{figure*}

Having defined source and background regions, we used \texttt{dmextract} to calculate the source core, source shell, and background region counts ($C_{\mathrm{core}}$, $C_{\mathrm{pie}}$, and $C_{\mathrm{bkg}}$) in the full, hard, and soft energy bands, their areas ($A_{\mathrm{core}}$, $A_{\mathrm{pie}}$, and $A_{\mathrm{bkg}}$), and their mean effective areas ($E_{\mathrm{core}}$, $E_{\mathrm{pie}}$, and $E_{\mathrm{bkg}}$) by including exposure maps in the call to \texttt{dmextract}.  We calculated the 1$\sigma$ Gaussian errors in the measured counts (both in the source and background regions) using the recommended approximations for upper and lower limits in \citet{gehr}.  For sources with apertures consisting of a circular core and a pie sector of an annular shell, the total source region counts, area, and effective area were calculated in the following manner, assuming azimuthal symmetry of the PSF:
\begin{eqnarray}
C_{\mathrm{src}} &=& C_{\mathrm{core}} + \frac{A_{\mathrm{ann}}}{A_{\mathrm{pie}}}C_{\mathrm{pie}} \\
A_{\mathrm{src}} &=& A_{\mathrm{core}} + A_{\mathrm{ann}} \\
E_{\mathrm{src}} &=& E_{\mathrm{core}} + \frac{A_{\mathrm{ann}}}{A_{\mathrm{pie}}}E_{\mathrm{pie}}
\end{eqnarray}
where $A_{\mathrm{ann}} = \pi(r_{\mathrm{outer}}^2-r_{\mathrm{inner}}^2)$ is the total area of the annular shell. For all other sources, the source region simply consists of the circular core region, and thus $C_{\mathrm{src}} = C_{\mathrm{core}}$, $A_{\mathrm{src}} = A_{\mathrm{core}}$, and $E_{\mathrm{src}} = E_{\mathrm{core}}$.  The total observed source region counts include contributions from the source and from the background.  The background counts within the source region were estimated and subtracted as shown below to estimate the true source counts: 
\begin{equation}
C_{\mathrm{net}} = C_{\mathrm{src}} - f C_{\mathrm{bkg}}, \hspace{0.3in} f = \frac{A_{\mathrm{src}} E_{\mathrm{src}}}{A_{\mathrm{bkg}} E_{\mathrm{bkg}}} .
\end{equation}
If the estimated background counts were equal to or greater than the source region counts, then we calculated the 90\% upper confidence limit to the net source counts based on the method described in \citet{kraft91}.  The photon flux for each energy band was calculated by dividing the net source counts by the mean source region effective area and the exposure time.  Since the mean effective area was determined from the exposure maps, these photon fluxes will not be accurate for all sources because in making the exposure maps we assumed a source spectral model with $\Gamma = 2.0$ and $N_{\mathrm{H}} = 5.0 \times 10^{22}$ cm$^{-2}$.  To determine the extent to which we may be under or overestimating the fluxes of sources with different spectral properties, we made exposure maps for one observation using different spectral models spanning the range of $\Gamma$ and $N_{\mathrm{H}}$ covered by our sources.  We find that the mean effective areas vary by $\lesssim$ 20\% in the full band and $\lesssim$ 5\% in the soft and hard bands, making our derived photon fluxes uncertain by the same percentages.  

We also computed the probability that the sources in our catalog could be noise fluctuations of the local background using the following formula derived in Appendix A of \citet{weiss07}:
\begin{multline}
P(\geq C_{\mathrm{src}}|C_{\mathrm{bkg}};C_{\mathrm{net}}=0) = \\ \sum_{c = C_{\mathrm{src}}}^{C_{\mathrm{bkg}}+C_{\mathrm{src}}}\frac{(C_{\mathrm{bkg}}+C_{\mathrm{src}})!}{c!(C_{\mathrm{bkg}}+C_{\mathrm{src}}-c)!}\left(\frac{f}{1+f}\right)^c\left( 1-\frac{f}{1+f}\right)^{C_{\mathrm{bkg}}+C_{\mathrm{src}}-c} .
\label{eq:probnoise}
\end{multline}
We determine the significance of a source based on this probability and the Gaussian cumulative distribution function. For sources detected in multiple observations, these probability values from individual observations were multiplied together, and the source significance was determined from this combined probability.  The photometric values used in our data analysis are included in our catalog, a sample of which can be seen in Table \ref{tab:catalog2}.  Although in our catalog we include all detected sources, in our analysis we only use sources detected at $\ge$ 3$\sigma$ in the full, soft, or hard energy band and refer to these sources as the full sample.  Figure \ref{fig:counthist} shows the histogram distribution of the total 0.5-10 keV counts detected for these sources; as can be seen, most of these sources have fewer than 100 counts, which is roughly the number of photons required to determine their spectral and variability properties accurately enough to determine the nature of the X-ray source.  Our catalog contains 1130 point-like sources detected at $\geq3\sigma$ in at least one of the three energy bands and 5 extended sources which we identified by eye and are shown in Figure \ref{fig:extended}.  Three of these extended sources are confirmed supernova remnants (G337.2+0.1, G337.8-0.1, and HESS J1640-465), another is a possible SNR based on its morphology (CXOU J163942.3-471257), and one has a jet-like morphology and is probably a pulsar wind nebula (CXOU J163802.6-471345).  New results about these extended sources are discussed in \citet{jakobsen13} and \citet{jakobsen14}.  

\begin{figure}[b]
\includegraphics[width=0.47\textwidth]{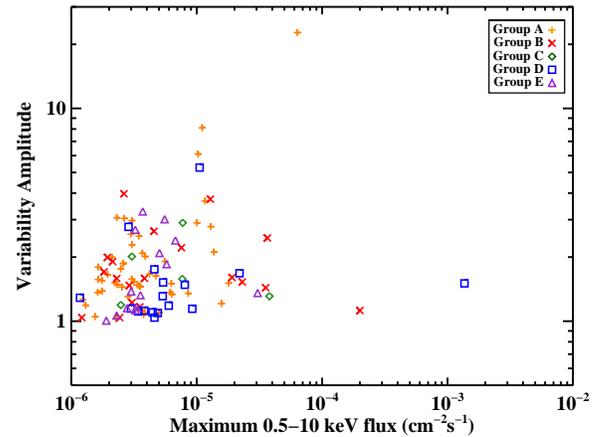}
\caption{Variability amplitudes in the 0.5-10 keV band (maximum photon flux divided by the minimum photon flux) versus maximum photon flux for sources that are detected in multiple observations and found to vary between observations at $\geq3\sigma$ confidence.  The median 1$\sigma$ fractional errors are +0.36/-0.29 for the maximum flux and +0.58/-0.23 for the variability amplitude.  Different symbols represent different quantile groups. (A color version of this figure is available in the online journal.)}
\label{fig:varamp}
\end{figure}

\begin{figure*}[t]
\makebox[\textwidth]{ %
	\hspace{-5pt}
		\subfigure[No. 750]{%
		\hspace{-20pt}\includegraphics[width=0.4\linewidth]{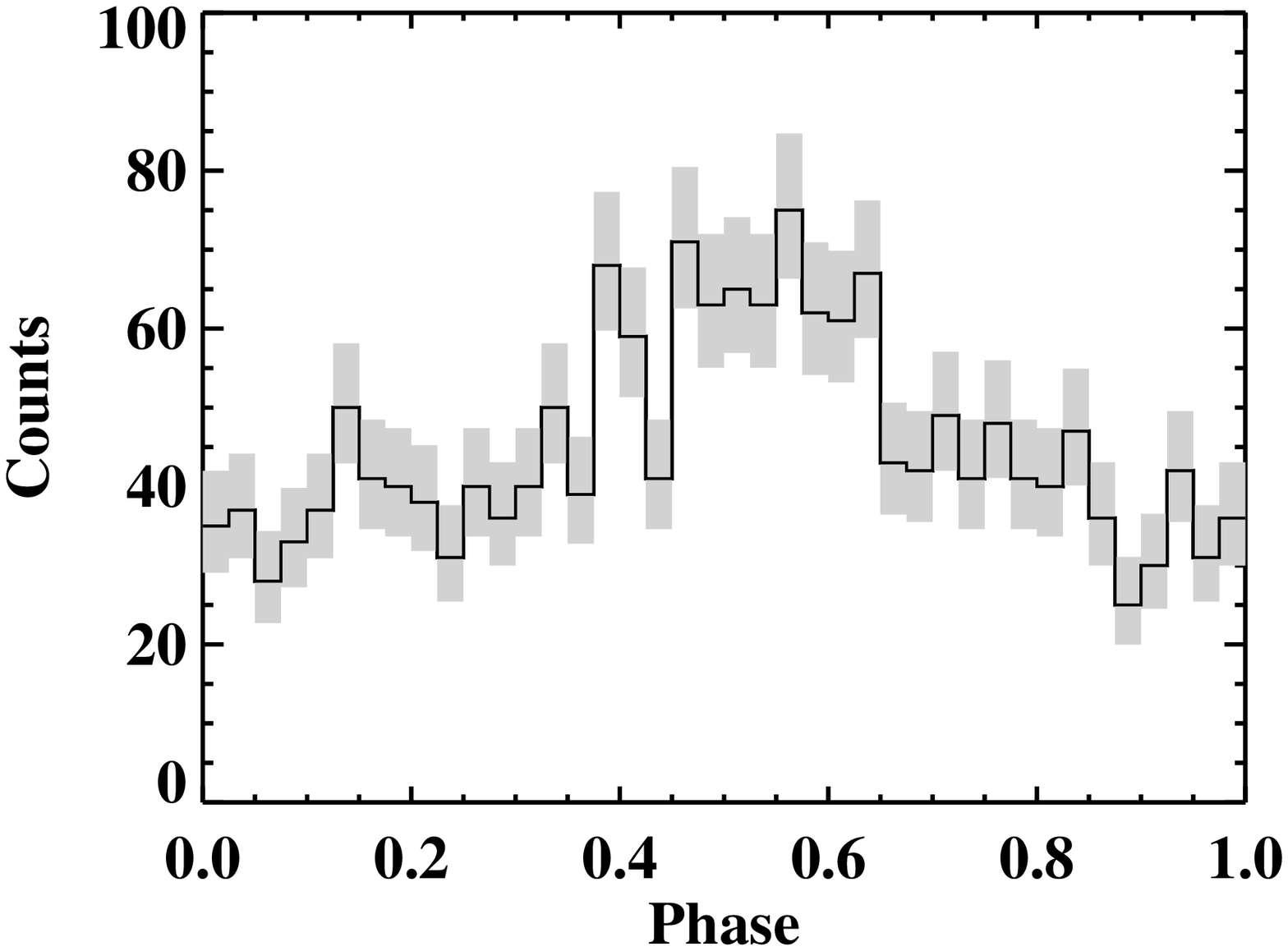}
		\label{fig:lightcurve750}}
		\subfigure[No. 961]{%
		\hspace{-20pt}\includegraphics[width=0.4\linewidth]{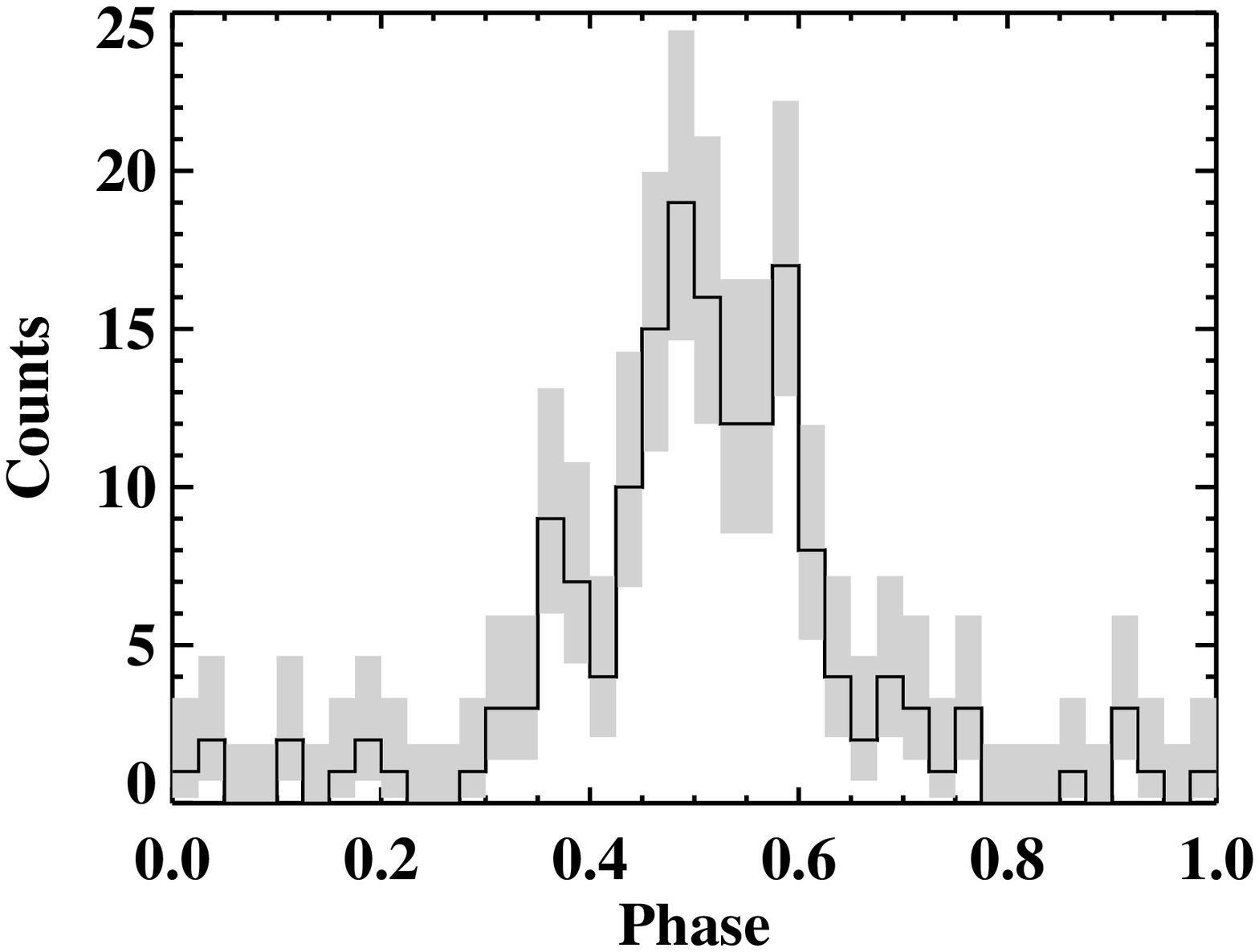}
		\label{fig:lightcurve961}}
		\subfigure[No. 999]{%
		\hspace{-20pt}\includegraphics[width=0.4\linewidth]{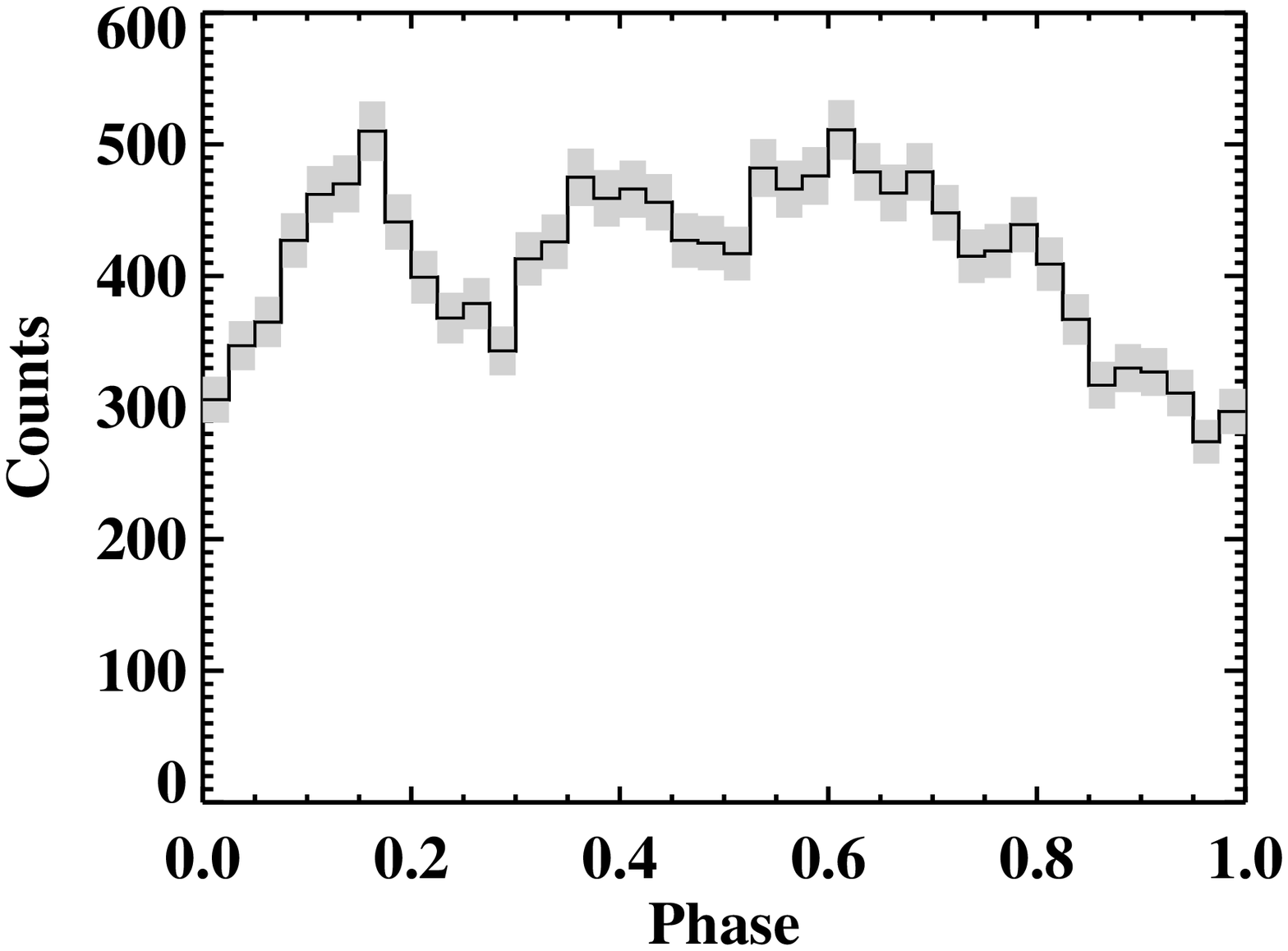}
		\label{fig:lightcurve999}}}
\caption{Phase-folded lightcurves of periodic sources labeled with source catalog number.  Periods are presented in Table \ref{tab:periodicity}.  Gray bars show 1$\sigma$ errors.}
\label{fig:lightcurve}
\end{figure*}

\subsection{Source Variability}
\label{sec:variability}

\begin{table}
\centering
\footnotesize
\caption{Sources Showing Periodic Variability}
\begin{threeparttable}
\begin{tabular}{cccccc} \hline \hline
No. & Source & $C_{net}$ & Period & ${Z_2^2}$ & Variability \\
 & (CXOU J) & FB & (s) & & Flags \\
(1) & (2) & (3) & (4) & (5) & (6) \\
\hline
750 & 163750.8-465545 & 1790$\pm$40 & 7150$\pm$50 & 95 & vs, vl \\
961 & 163855.1-470145 & 160$\pm$10 & 5660$\pm$20 & 213 & vp, vl \\
999 & 163905.6-464212 & 14720$\pm$120 & 906$\pm$1 & 368 & vs, vl \\
\hline
\end{tabular}
\begin{tablenotes}

\item (1) Catalog source number.

\item (2) \textit{Chandra} source name.

\item (3) Net source counts in the full 0.5-10 keV band.

\item (4) Most probable period determined by the $Z_2^2$-test.

\item (5) Result of $Z_2^2$-test.

\item (6)``vl" - variability on long timescales (hours-days), ``vp" - variability on short timescales (sec-hour) at $\geq 95$\% confidence, ``vs" - variability on short timescales at $\geq 99.73$\% confidence 
\end{tablenotes}
\end{threeparttable}
\label{tab:periodicity}
\end{table}

X-ray sources can be variable on timescales from milliseconds to years, so we tried to characterize the variability of the sources in our catalog to help classify them.  We determined whether a source was variable on short timescales (seconds to hour) by comparing the arrival times of events with a constant event rate using the K-S test.  Sources that have $\ge$99.73\% chance of not being constant are flagged with ``vs" (short variability), while those that have $\ge$95\% chance of not being constant are flagged with ``vp" (probable short variability).  The K-S test is more reliable for sources with more counts; we only consider the K-S test to be reliable for sources with at least 40 counts.  Of the 80 sources with more than 40 counts (in a single observation), 27 (16) show short-timescale variability with $\ge$95\% (99.73\%) confidence.

We also checked whether sources detected in multiple observations demonstrated long-term variability (hours-days) by determining whether the source photon flux in two observations differed by more than 3$\sigma$ in the full, soft, or hard energy band; these sources were flagged with ``vl" (long variability). In cases in which a source's flux is measured in one observation but only an upper limit can be obtained in another, we consider the source to be variable on long timescales if the measured flux and upper limit are inconsistent at $>3\sigma$ confidence. We found 219 sets of sources detected in two or more observations with a combined significance $\geq3\sigma$, 105 (48\%) of which show long-term variability.  Of the 758 sources with $\geq3\sigma$ confidence located in regions where multiple observations overlap, 373 are not detected in multiple observations.  This is not surprising because the fluxes of all but seven of these 373 sources are lower than the flux to which 100\% of our survey area is sensitive (see \S \ref{sec:sensitivity}).  We calculated the variability amplitude of each source displaying long-term variability, which we defined as the ratio of the maximum photon flux to the minimum photon flux in the 0.5-10 keV band.  The variability amplitude does not seem to correlate with the photon flux of these variable sources, as shown in Figure \ref{fig:varamp}.  

Finally, we searched for a coherent signal with period 6.8 s $< P <$ 10 ks in sources with more than 50 counts in at least one observation using the $Z_n^2$-test \citep{buccheri83}, which depends on the sum of the Fourier powers of the first $n$ harmonics.  Since it is not feasible to try an infinite number of $n$ values, \citet{buccheri83} suggested using $n=2$ as a general test; we decided to use both $n=2$ and $n=1$ since the latter is equivalent to the traditional Rayleigh test.  For each source, photon arrival times were corrected to the Solar System barycenter using the JPL DE405 ephemeris and our catalog coordinates, and for sources detected in multiple observations, photon arrival times from different observations were combined.  We found three sources with significant $Z_2^2$ values; these sources have significant $Z_1^2$ values as well.  Table \ref{tab:periodicity} provides the periods and $Z_2^2$ values of these sources; the uncertainties in the periods were calculated using the method described in \citet{Ransom02}.  In Figure \ref{fig:lightcurve} we present the phase-folded lightcurve of each periodic source.  Source 999 is a previously discovered HMXB; the period we measure is consistent with the period found by \citet{bodaghee06}.  The other two sources are most likely magnetic cataclysmic variables (CVs), as discussed in \S \ref{sec:groupB}.

\subsection{Infrared Counterparts}

We searched for infrared counterparts to our X-ray sources in the VVV catalog.  First, we created a list of non-duplicate VVV sources lying within our surveyed area; we considered an entry in the VVV catalog to be duplicate if the angular difference between the source positions was less than the $3\sigma$ positional uncertainty of the sources (approximately $0\farcs21$). Then, we determined the reliability of each counterpart as described in \S \ref{sec:detection}, and consider a match good if its reliability is $\geq$ 90\% and the noise probability of the VVV source is $\leq$ 0.31\%.  The latter constraint excludes low-significance detections and very bright sources, which saturate the array and have less accurate positions and magnitudes.  Table \ref{tab:catalog3} provides information about the closest VVV source to each \textit{Chandra} source and its reliability as a counterpart.  

\begin{figure}[b]
\hspace{-0.4in}
\vspace{-0.4in}
\includegraphics[width=0.55\textwidth]{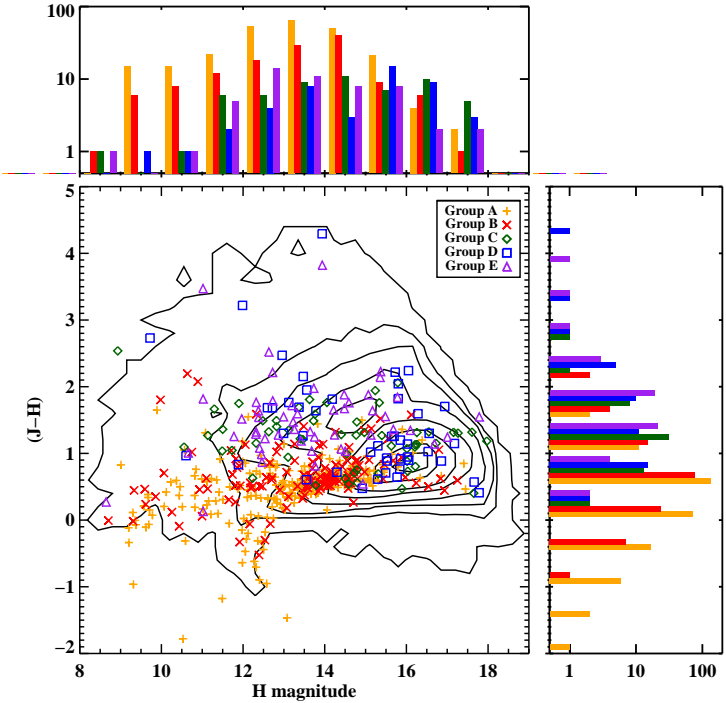}
\caption{Near-IR color-magnitude diagram showing the distribution of all VVV sources in our surveyed area (black contours) and the reliable counterparts to X-ray sources in our survey (symbols).  The black contours, from outside inwards, encircle 99\%, 95\%, 90\%, 80\%, 60\%, 40\%, 20\%, and 10\% of the VVV sources.  For the counterparts represented with symbols, the median 1$\sigma$ error is 0.005 magnitudes for the $H$ magnitude and 0.007 magnitudes for the $(J-H)$ color. (A color version of this figure is available in the online journal.)}
\label{fig:vistacolormag}
\end{figure}

We have found reliable counterparts for 52\% of X-ray sources detected at $\geq3\sigma$. X-ray sources without reliable counterparts may have IR counterparts below the sensitivity limits of the VVV survey, have large positional uncertainties, and/or be in a particularly crowded region in which multiple IR sources are equally likely counterparts.  Figure \ref{fig:vistacolormag} shows the $(J-H)$ vs. $H$ magnitude of the reliable counterparts (the $K_s$ magnitudes have not yet been made publicly available).  In this color-magnitude diagram, the distribution of reliable counterparts does not match that of all the VVV sources located within our field-of-view, indicating that the majority of these counterparts are not random associations.  

\subsection{Quantile Properties}
\label{sec:quantile}

Robust identifications of X-ray sources typically require multiwavelength information, but the X-ray data itself can provide clues to the nature of a source.  First, we tried to classify the sources using a hardness ratio and a soft and hard X-ray color.  However, we found that this method was ineffective for distinguishing between sources with intrinsically hard spectra and very absorbed sources.  This distinction is important since the HII regions near which we expect to find younger populations of X-ray sources, the primary target of this survey, are at a distance of $\sim$11 kpc and thus the X-ray emission from such sources would be attenuated by large amounts of intervening gas.  

\begin{figure}
\hspace{-0.4in}
\includegraphics[width=0.55\textwidth]{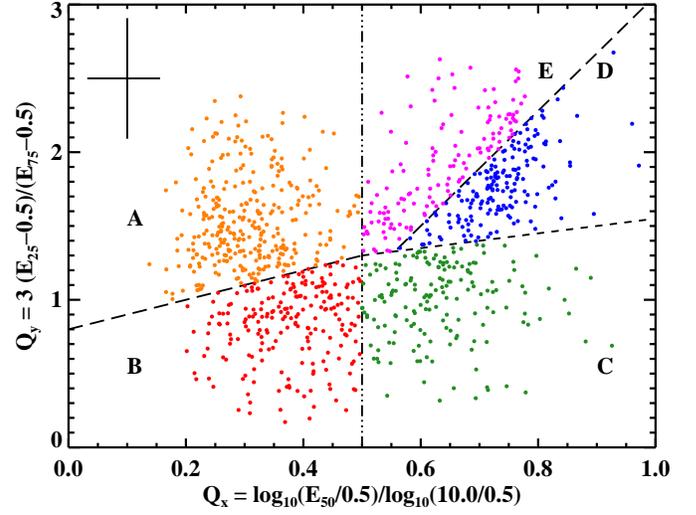}
\caption{Quantile diagram for sources detected at $\geq3\sigma$ in the 0.5-10 keV band.  Sources were split into five spectral groups defined by the black lines.  Short dashes at $N_{\mathrm{H}}\approx2\times10^{22}$ cm$^{-2}$ roughly divide nearby ($\lesssim6$ kpc) and distant ($\gtrsim6$ kpc) sources.  Long dashes roughly separate thermal and nonthermal sources.  The dash-dotted line at $Q_x=0.5$ subdivides sources detected in the soft energy band from those detected in the hard energy band.  Median 1$\sigma$ errors are shown in the upper left. (A color version of this figure is available in the online journal.)}  
\label{fig:quantile}
\end{figure}

\begin{figure*}[t]
\centering
\includegraphics[angle=90,width=0.8\textwidth]{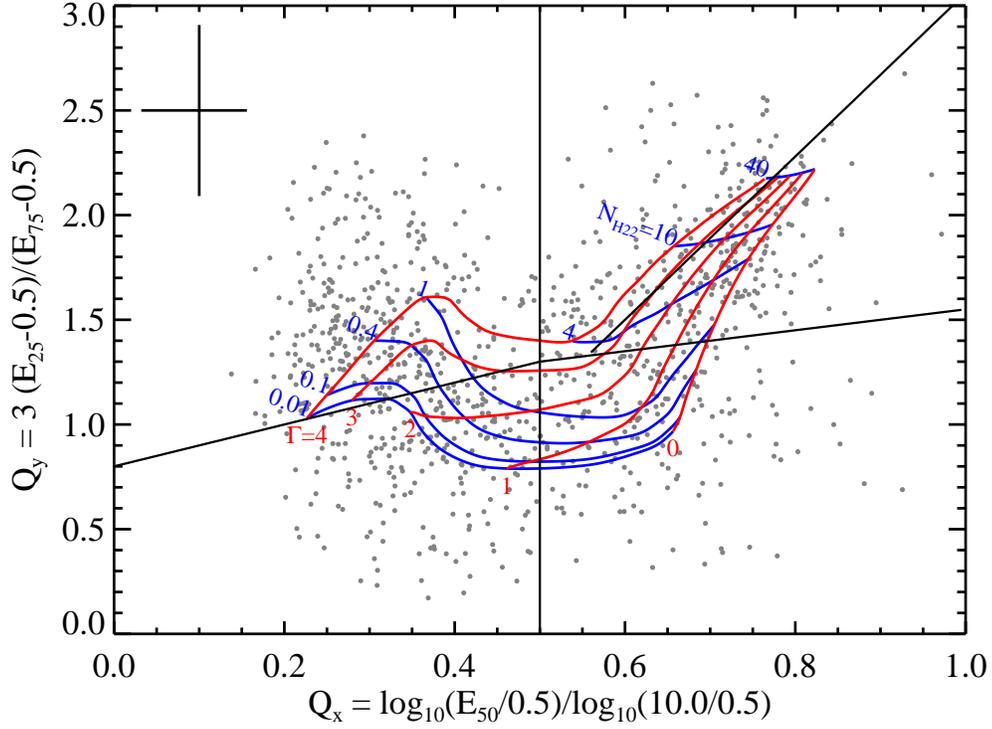}
\caption{Quantile diagram or sources detected at $\geq3\sigma$ in the 0.5-10 keV band.  A grid of power-law spectra attenuated by interstellar absorption is overlaid.  Red (primarily vertical) lines represent values of the photon index $\Gamma = 0, 1, 2, 3,$ and 4 from right to left.  Blue (primarily horizontal) lines represent values of the hydrogen column density $N_{\mathrm{H}} = 10^{20}, 10^{21}, 10^{21.6}, 10^{22}, 10^{22.6}, 10^{23},$ and $10^{23.6}$ cm$^{-2}$ from bottom to top.  Black solid lines separate the five spectral groups defined in \S \ref{sec:specanalysis}. (A color version of this figure is available in the online journal.)}
\label{fig:specgroup}
\end{figure*}

\begin{figure}[b]
\vspace{-0.2in}
\hspace{-0.4in}
\includegraphics[angle=90,width=0.55\textwidth]{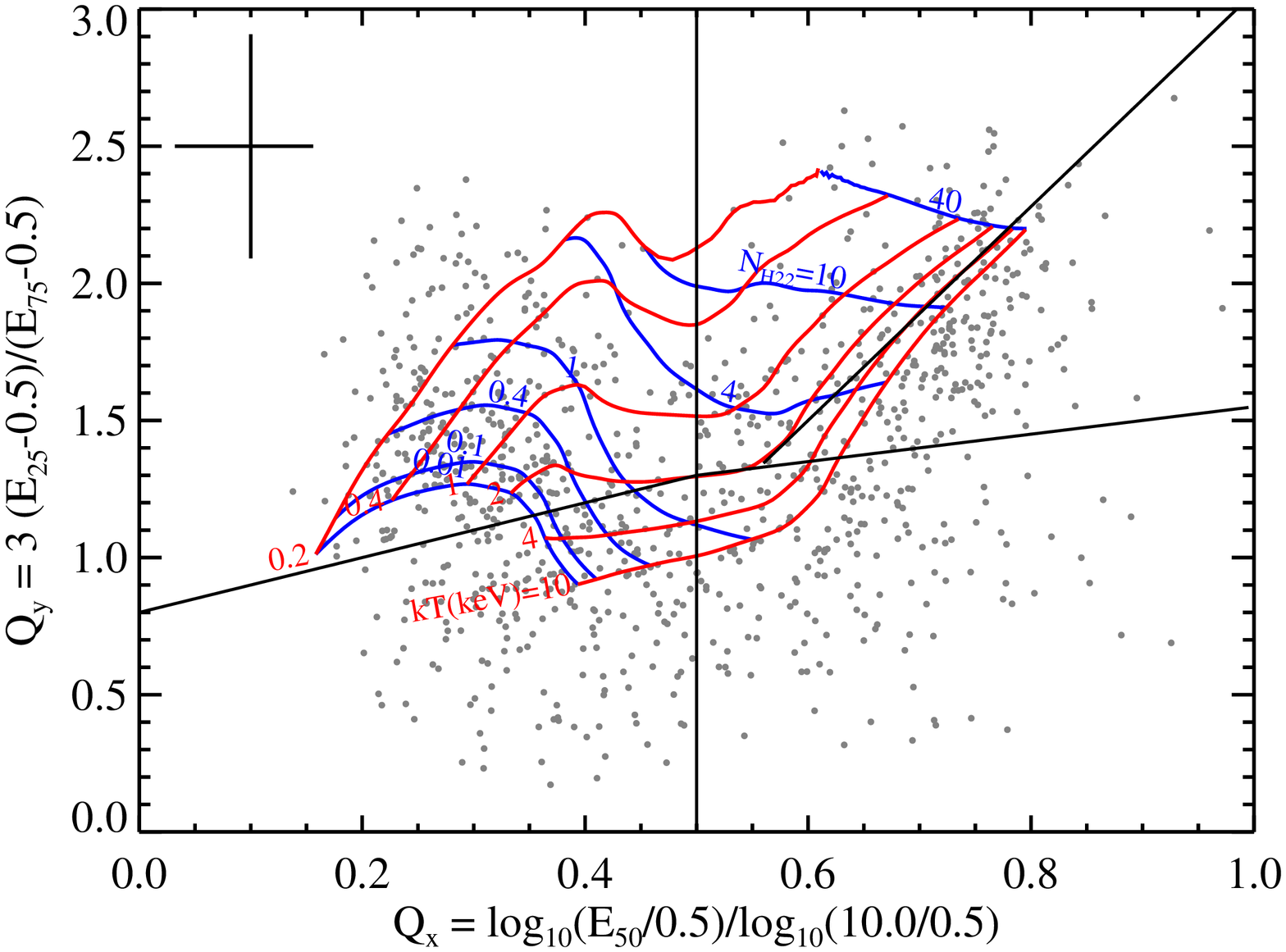}
\caption{Quantile diagram or sources detected at $\geq3\sigma$ in the 0.5-10 keV band.  A grid of thin thermal plasma spectra attenuated by interstellar absorption is overlaid.  Red (primarily vertical) lines represent values of the photon index $kT = 0.2, 0.4, 1, 2, 4,$ and 10 keV from left to right.  Blue (primarily horizontal) lines represent values of the hydrogen column density $N_{\mathrm{H}} = 10^{20}, 10^{21}, 10^{21.6}, 10^{22}, 10^{22.6}, 10^{23},$ and $10^{23.6}$ cm$^{-2}$ from bottom to top.  Black solid lines are the same as in Figure \ref{fig:specgroup}. (A color version of this figure is available in the online journal.)}
\label{fig:specgroup2}
\end{figure}

Therefore, we instead employed quantile analysis to classify the X-ray sources.  Quantile analysis, first introduced by \citet{hong04}, uses the median energy and other quantile energies of a source as proxies for its spectral hardness and spectral shape.  Its main advantage is that it does not require subdivision of the full energy range into different bands, making it free of the selection effects inherent in the hardness ratio and colors methods and yielding meaningful results even for low-count sources.  The fundamental quantities required in quantile analysis are $E_{x}$, the energy below which $x\%$ of the source counts reside.  We made event files for each source and background region and passed them to \texttt{quantile.pro}, an IDL program developed by J. Hong.\footnotemark\footnotetext{Available at http://hea-www.harvard.edu/ChaMPlane/quantile}.  The other input to this code is the ratio of source and background areas and effective areas:
\begin{equation}
\textrm{ratio} = \frac{(A_{\mathrm{core}}+A_{\mathrm{pie}})(E_{\mathrm{core}}+E_{\mathrm{pie}})}{A_{\mathrm{bkg}}E_{\mathrm{bkg}}}
\end{equation}
With this code, we computed $E_{25}$, $E_{50}$, and $E_{75}$, which are included in our catalog and can be seen in Table \ref{tab:catalog2}. These parameters were then combined into two quantities, $Q_x = \log(E_{50}/E_{min})/\log(E_{max}/E_{min})$ and $Q_y = 3 (E_{25}-E_{min})/(E_{75}-E_{min})$, where $E_{min}$ and $E_{max}$ are 0.5 and 10 keV, respectively.  $Q_x$ measures the hardness of the spectrum, while $Q_y$ indicates how broad or narrow the spectrum is.  Figure \ref{fig:quantile} shows diagrams of $Q_x$ and $Q_y$ for all sources in our catalog detected at $\ge$ 3$\sigma$ in the full energy band.  To faciliate interpretation of this diagram, we have overlaid grids for a power-law model and a thermal bremsstrahlung model, both attenuated by interstellar absorption (see Figures \ref{fig:specgroup} and \ref{fig:specgroup2}).

\section{Spectral Analysis}
\label{sec:specanalysis}

One of the goals of our survey is to understand the nature of the X-ray sources in our field.  Analyzing the X-ray spectrum of a source can provide important clues about its physical nature, but most of our sources have too few counts to permit meaningful spectral analysis.  Therefore, to identify the dominant populations of X-ray sources in our survey, we divided them into spectral groups based on their quantile properties (see \S \ref{sec:quantile}) and analyzed the stacked spectrum of each quantile group.  Our goals in making group divisions were to combine enough sources together to reduce the errors caused by poor statistics in spectral fitting, but also to maintain the maximum spectral diversity in our sample. 

\begin{figure}[t]
\includegraphics[width=0.47\textwidth]{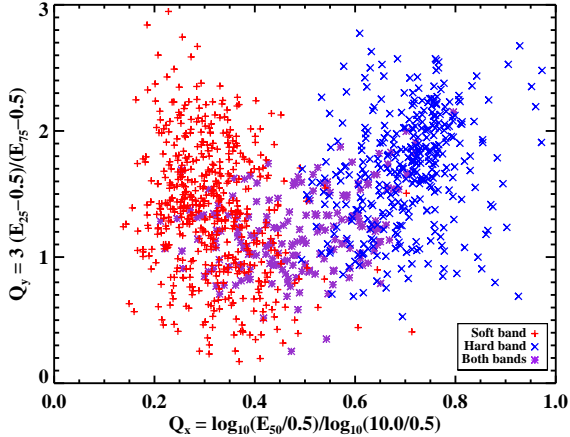}
\caption{Quantile diagram showing soft sources in red crosses (detected at $\geq3\sigma$ in SB but not HB), hard sources in blue X's (detected at $\geq3\sigma$ in HB but not SB), and bright sources in purple asterisks (detected at $\geq3\sigma$ in both SB and HB).}
\label{fig:quanthard}
\end{figure}

As can be seen in Figure \ref{fig:quantile}, the sources are loosely confined to a U-shaped region in the $Q_x$-$Q_y$ diagram, with overdensities towards the upper right and middle left.  However, apart from these slight overdensities, the sources do not split up into visibly discernible groups, so we decided to split up the sources into groups with physically-motivated dividing lines roughly following the $N_{\mathrm{H}} \approx 2 \times 10^{22}$ cm$^{-2}$ and $\Gamma \approx$ 3 grid lines, as shown in Figure \ref{fig:specgroup}.  The dividing $N_{\mathrm{H}}$ line corresponds to the average $N_{\mathrm{H}}$ out to a distance of $\sim$ 6 kpc in our survey region, as determined from the sum of $N_{\mathrm{HI}}$ estimated from the Leiden/Argentine/Bonn (LAB) Survey \citep{kalberla05} and $N_{\mathrm{H}_2}$ estimated from the MWA CO survey \citep{bronfman89};\footnotemark\footnotetext{To estimate $N_{\mathrm{HI}}$ ($N_{\mathrm{H}_2}$) to a distance of 6 kpc, we calculated the line-of-sight velocity of an object at this distance in circular motion around the Galaxy, integrated the brightness temperature measured in the LAB (MWA) survey from 0 km/s to this velocity, and multiplied the total brightness by $N_{\mathrm{HI}}$/$I_{\mathrm{H}}$= $1.8\times10^{18}$ cm$^{-2}$ K$^{-1}$ km$^{-1}$ s ($N_{\mathrm{H}_2}$/$I_{\mathrm{CO}}=2\times10^{20}$ cm$^{-2}$ K$^{-1}$ km$^{-1}$ s derived in \citet{dame01}).  The sum of $N_{\mathrm{HI}}$ and $N_{\mathrm{H}_2}$  will actually be a lower limit of $N_{\mathrm{H}}$ since this calculation assumes the emission lines are optically thin.} this line roughly divides sources in the foreground and in the Scutum-Crux and near Norma spiral arms from sources in the far Norma arm.  The dividing $\Gamma$ line instead splits thermal and nonthermal sources.  We further subdivided sources along $Q_x = 0.5$, because, as shown in Figure \ref{fig:quanthard}, this $Q_x$ value roughly separates sources detected in the soft band from sources detected in the hard band.  The equations of the dividing lines, in counterclockwise order from the top left of the diagram, are:
\begin{eqnarray}
Q_y &=& Q_x +0.8 \nonumber \\
Q_x &=& 0.5 \\
Q_y &=& 0.5Q_x + 1.05 \nonumber \\
Q_y &=& 3.9Q_x -0.84 \nonumber
\end{eqnarray} 
Varying the quantile divisions by $\sim$0.1 dex leads to no significant change in the best-fitting spectral parameters of the stacked spectum of each quantile group (\S \ref{sec:discussion}), and the maximum likelihood slopes of the number-flux distributions of the sources in each quantile group remain consistent at the 2$\sigma$ level or better (\S \ref{sec:logNlogS}).  Thus, our results are robust to $\sim$0.1 dex variations in the quantile group definitions.  

\begin{figure*}[t]
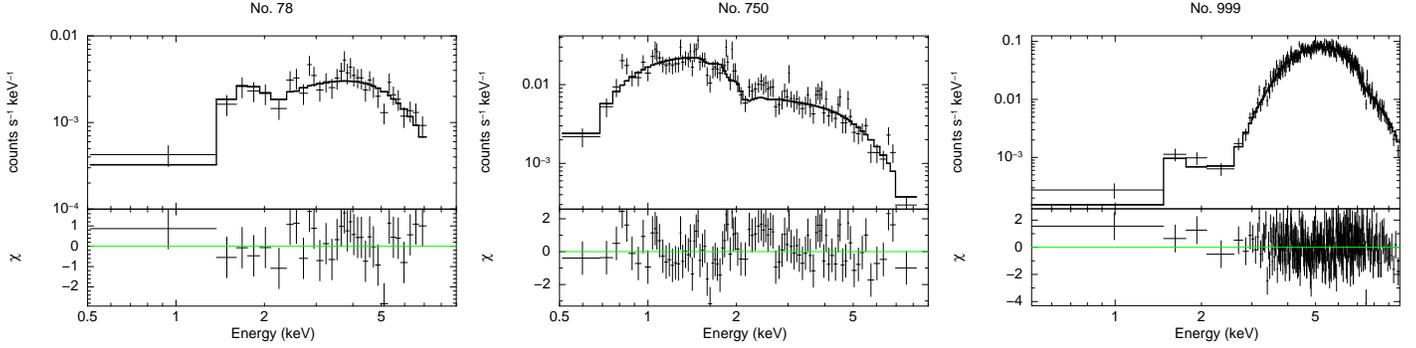

\makebox[\textwidth]{ %
	\centering
	\subfigure{
		\hspace{-10pt}\includegraphics[angle=270,width=0.35\textwidth]{Src78_new.ps}}
	\subfigure{
		\hspace{-5pt}\includegraphics[angle=270,width=0.35\textwidth]{Src750_new.ps}}
	\subfigure{
		\hspace{-5pt}\includegraphics[angle=270,width=0.35\textwidth]{Src999_new.ps}}}
\caption{X-ray spectra of sources with more than 500 net counts in the 0.5-10 keV band.  Title labels provide source catalog number.  Grouped data and the best-fit model convolved with the instrumental response are shown in the upper panels.  Lower panels show the data residuals. Table \ref{tab:indspectra} provides the best-fit parameters.}
\label{fig:indspectra}
\end{figure*}

For each source, we used \texttt{specextract} to extract source and background spectra and build associated ARFs and RMFs.  Then the spectra of sources within each quantile group were combined with \texttt{combine\_spectra}.  Sources with more than 500 net counts were excluded from the stacked spectra to prevent individual sources from excessively influencing the combined spectrum.  The spectra of these three individual sources are shown in Figure \ref{fig:indspectra} and their best-fit spectral parameters are provided in Table \ref{tab:indspectra}.  Sources 78 and 999 fall in quantile group D while source 750 falls in quantile group B, and they are best fit by absorbed power-laws.  

\begin{table*}
\vspace{-0.2in}
\centering
\footnotesize
\caption{Spectral Fit Results for Individual Bright Sources}
\begin{threeparttable}
\begin{tabular}{ccccccccc} \hline \hline
No. & Source & $C_{net}$ & $N_{\mathrm{H}}$ & $\Gamma$ &  $f_X$ FB &  $\chi^2$/dof & Galactic $N_{\mathrm{H}}$ & Unabsorbed $f_X$ FB \\
& CXOU J& FB & (10$^{22}$cm$^{-2}$)&&(10$^{-13}$erg cm$^{-2}$s$^{-1}$) & & (10$^{22}$cm$^{-2}$) & (10$^{-13}$erg cm$^{-2}$s$^{-1}$) \\
(1) & (2) & (3) & (4) & (5) & (6) & (7) & (8) & (9) \\
\hline
78&163355.1-473804&530$\pm$20&2.4$^{+1.1}_{-0.9}$&0.5$\pm$0.4&8.8$^{+0.8}_{-0.9}$&31/31&7.0&10.0$^{+0.9}_{-1.0}$\\
750&163750.8-465545&1790$\pm$40&0.12$^{+0.09}_{-0.08}$&1.2$\pm$0.1&9.9$\pm$0.4&124/95&9.4&10.3$\pm$0.5\\
999&163905.4-464212&14720$\pm$120&49$\pm$2&0.9$\pm$0.1&268$^{+5}_{-4}$&347/314&8.1&301$^{+6}_{-5}$\\
\hline
\end{tabular}
\begin{tablenotes}
\item  \underline{Notes:} Quoted errors are 1$\sigma$ unless specified in notes. 

\item (1) Catalog source number.

\item (2) \textit{Chandra} source name.

\item (3) Net source counts in the full 0.5-10 keV band.

\item (4) Hydrogen column density from spectral fit with 90\% uncertainties.

\item (5) Power-law photon index from spectral fit with 90\% uncertainties.

\item (6) 0.5-10 keV flux determined from spectral model.

\item (7) Chi-square of best-fit model over degrees of freedom.

\item (8) Estimated line-of-sight $N_{\mathrm{H}}$ through the Galaxy based on $N$(H) from the LAB survey and $N$(H$_2$) from the MWA CO survey (see \S \ref{sec:specanalysis}).

\item (9) 0.5-10 keV flux corrected for line-of-sight $N_{\mathrm{H}}$.
\end{tablenotes}
\end{threeparttable}
\label{tab:indspectra}
\end{table*}

\begin{figure*}
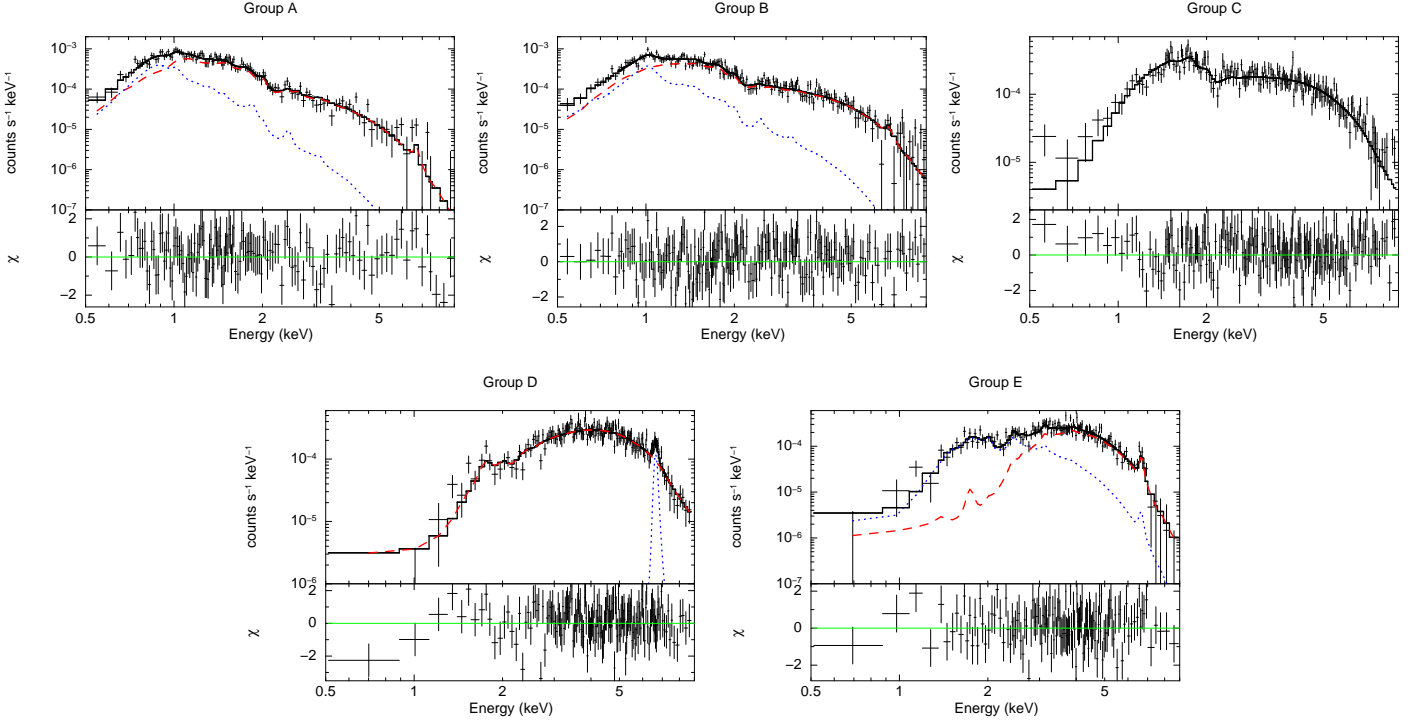

\makebox[\textwidth]{ %
	\centering
	\subfigure{
		\hspace{-10pt}\includegraphics[angle=270,width=0.35\textwidth]{GroupA.ps}}
	\subfigure{
		\hspace{-5pt}\includegraphics[angle=270,width=0.35\textwidth]{GroupB.ps}}
	\subfigure{
		\hspace{-5pt}\includegraphics[angle=270,width=0.35\textwidth]{GroupC.ps}}}

	\centering
	\subfigure{
		\includegraphics[angle=270,width=0.35\linewidth]{GroupD.ps}}
	\subfigure{
		\includegraphics[angle=270,width=0.35\linewidth]{GroupE.ps}}
\caption{Stacked spectra and the best-fit models for each quantile group.  Best-fit parameters can be found in Table \ref{tab:stacked}. (A color version of this figure is available in the online journal.)}
\label{fig:stacked}
\end{figure*}

Each of the stacked spectra was fit with an interstellar absorption model (\texttt{tbabs}, \citealt{wilms00}, with cross-section from \citealt{verner}) convolved with a power-law model (\texttt{pegpwrlw}) and an optically thin thermal plasma model (\texttt{vapec}, \citealt{smith01}, with abundances frozen to values from \citealt{gudel07} relative to \citealt{anders89}, scaled to \citealt{wilms00}).  If an Fe line was visible between 6 and 7 keV, we added a Gaussian component to the power-law model.  If neither the power-law nor thermal model produced a fit with reduced $\chi^2$ less than 1.2, then a second component was added to the model; both power-law and thermal second components were tried in all such cases and the best fit was determined by the minimum reduced $\chi^2$.  For each quantile group, we made and fit a stacked spectrum first only using sources detected at $\geq3\sigma$ confidence in the 0.5-10 keV band and then only using sources detected at $\geq$ 3$\sigma$ confidence in the 2-10 keV band.  The stacked spectra and their best fits are shown in Figure \ref{fig:stacked} and Table \ref{tab:stacked}.  

Following the example of \citet{ebisawa05}, we studied how the spectral parameters of sources with or without IR counterparts differ.  We split up the sources detected at $\geq3\sigma$ in the full band in each quantile group into two groups, based on whether or not they have a VVV counterpart with $\geq90$\% reliability.  A stacked spectrum for each of these subgroups was made and fit with the best-fitting model for its parent quantile group.  For the case of a power-law model (with or without a Gaussian component), the power-law index, column density, and the normalization of the power-law and Gaussian components were left free, while for two-temperature thermal models, only the normalizations of the two components were allowed to vary.  In addition, since the strength of the Fe emission line for group E sources with and without IR counterparts appeared different, we decided to measure the equivalent width of the Fe line in the group E stacked spectra by fitting the 5-9 keV band of these stacked spectra with a power-law plus Gaussian line model. The results of this spectral analysis are shown in Table \ref{tab:stackedir}.  Finally, we also studied how the spectral parameters of a given quantile group vary with source brightness. The sources within each quantile group were organized by photon flux and combined into subgroups containing 800-1000 total source counts. The analysis of the stacked spectra for these subgroups was done in the same way as for the subgroups based on the presence/absence of IR counterparts.  All the brightness trends that we found can be explained by changes in the relative fraction of sources with and without IR counterparts as a function of flux.  Thus, we only discuss the dependence of the spectral fitting results on the presence/absence of an IR counterpart.  

\begin{table*}
\centering
\footnotesize
\caption{Spectral Fit Results for Stacked Sources}
\begin{threeparttable}
\begin{tabular}{cccccccccc} \hline \hline
\multicolumn{10}{c}{Power-law fit parameters} \\
\hline
Quantile & Energy & \# Sources & $N_{\mathrm{H}}$ & $\Gamma$ & Line E & Eq. Width & $f_X$ 2-10 keV & $\chi_{\nu}^2$/dof & $\epsilon^{-1}$ \\
Group& Band &&(10$^{22}$cm$^{-2}$)& & (keV) & (eV) & (10$^{-14}$erg cm$^{-2}$s$^{-1}$) & & (10$^{-9}$erg/ph.) \\
(1) & (2) & (3) & (4) & (5) & (6) & (7) & (8) & (9) & (10) \\
\hline
\multirow{2}{*}{C} & full & 170 & 1.4$^{+0.3}_{-0.2}$ & 1.1$\pm$0.1 & -- & -- & 2.94$^{+0.08}_{-0.13}$ & 1.02/180 & 8.66\\
& hard & 136 & 1.8$\pm0.3$ & 1.2$\pm$0.1 & -- & -- & 3.2$^{+0.10}_{-0.2}$ & 0.99/158 & 8.60\\
\hline
\multirow{2}{*}{D} & full & 164 & 7$\pm1$ & 0.7$\pm0.2$ & 6.65$^{+0.04}_{-0.05}$ & 300$^{+60}_{-50}$ & 7.1$^{+0.1}_{-0.4}$ & 1.06/153 & 11.47\\
& hard & 168 & 6.6$^{+0.9}_{-0.8}$ & 0.6$\pm0.2$ & 6.60$\pm$0.04 & 300$^{+50}_{-40}$ & 6.83$^{+0.09}_{-0.31}$ & 1.16/186 & 11.13\\
\hline \hline
\multicolumn{10}{c}{Thermal fit parameters} \\
\hline
Quantile & Energy & \# Sources & $N_{\mathrm{H}}$ & $kT_1$ & $N_{\mathrm{H}}$ & $kT_2$ & $f_{X,low}/f_{X,high}$ & $\chi_{\nu}^2$/dof & $\epsilon^{-1}$\\
Group& Band &&(10$^{22}$cm$^{-2}$)& (keV) & (10$^{22}$cm$^{-2}$) & (keV) &  & & (10$^{-9}$erg/ph.) \\
(1) & (2) & (3) & (4) & (11) & (4) & (11) & (12) & (9) & (10) \\
\hline
\multirow{2}{*}{A} & full & 312 & $<$0.43 & 2.4$^{+1.5}_{-0.4}$ & $<$0.58 & 0.75$^{+0.05}_{-0.12}$ & 0.04$^{+0.05}_{-0.01}$ & 0.95/124 & 2.60\\
& hard & 41 & 0.7$\pm0.4$ & 2.4$^{+0.6}_{-0.4}$ & 0.3$\pm0.2$ & 0.39$\pm$0.06 &  0.0092$\pm$0.0006 & 1.01/106 & 5.78 \\
\hline
\multirow{2}{*}{B} & full & 210 & 0.3$^{+0.4}_{-0.1}$ & 7$\pm3$ & $<$0.14 & 1.2$^{+0.1}_{-0.2}$ & 0.04$\pm0.02$ & 1.08/186 & 3.83\\
& hard & 70 & 5$^{+9}_{-3}$ & 4$^{+3}_{-2}$ & 0.29$\pm$0.08 & 2.1$^{+0.9}_{-0.3}$ & 0.74$\pm$0.05 & 1.09/134 & 7.33\\
\hline
\multirow{2}{*}{E} &  full & 131 & 25$^{+7}_{-5}$ & 1.8$^{+0.5}_{-0.3}$ & 5.4$\pm0.9$ & 1.4$^{+0.8}_{-0.4}$ & 0.25$^{+0.10}_{-0.07}$ & 1.01/125 & 16.5 \\
& hard & 128 & 24$^{+7}_{-6}$ & 1.8$^{+0.5}_{-0.3}$ & 6$\pm1$ & 1.2$^{+0.9}_{-0.4}$ & 0.20$\pm$0.01 & 0.99/122 & 9.51 \\
\hline
\end{tabular}
\begin{tablenotes}
\item  \underline{Notes:} All quoted errors are 90\% unless otherwise indicated.

\item (1) Quantile groups defined in \S \ref{sec:quantile}.

\item (2) Stacked spectrum contains all sources that are detected at $\geq3\sigma$ in given energy band with fewer than 500 counts.

\item (3) Number of sources in stacked spectrum.

\item (4) Hydrogen column density associated with model component in next column.

\item (5) Power-law photon index.

\item (6) Energy of Gaussian line component.

\item (7) Equivalent width of Gaussian line component (1$\sigma$ errors).

\item (8) 2-10 keV absorbed flux (1$\sigma$ errors).  

\item (9) Chi-square of best-fit model over degrees of freedom.

\item (10) Photon flux to unabsorbed energy flux conversion factor.

\item (11) Temperature of thin thermal plasma component.  

\item (12) Flux ratio in 2-10 keV band of low $kT$ component over high $kT$ component (1$\sigma$ errors).
\end{tablenotes}
\end{threeparttable}
\label{tab:stacked}
\end{table*}

\begin{table*}[t]
\centering
\footnotesize
\caption{Spectral Comparison of Sources with and without IR Counterparts}
\begin{threeparttable}
\begin{tabular}{ccccccccc} \hline \hline
\multicolumn{9}{c}{Power-law stacked spectra} \\
\hline
Quantile & With/Without & \# Sources & $N_{\mathrm{H}}$ & $\Gamma$ & Line Eq. & \multicolumn{2}{c}{$f_X$ 2-10 keV} & $\chi_{\nu}^2$/dof \\
Group& IR Counterparts &&(10$^{22}$cm$^{-2}$)&  & Width (eV) & \multicolumn{2}{c}{(10$^{-14}$ erg cm$^{-2}$ s$^{-1}$)} & \\
(1) & (2) & (3) & (4) & (5) & (6) & \multicolumn{2}{c}{(7)} & (8)\\
\hline
\multirow{2}{*}{C} & IR & 62 & 1.6$\pm0.3$ & 1.3$\pm0.2$ & -- & \multicolumn{2}{c}{2.8$\pm0.2$} & 0.93/72\\
& No IR & 108 & 1.4$\pm0.4$ & 0.9$\pm0.2$ & -- & \multicolumn{2}{c}{3.0$\pm$0.2} & 1.08/120\\
\hline
\multirow{2}{*}{D} & IR & 49 & 6$^{+2}_{-1}$ & 0.8$^{+0.4}_{-0.3}$ & 330$^{+150}_{-100}$ & \multicolumn{2}{c}{5.1$^{+0.1}_{-0.7}$} & 0.89/55\\
& No IR & 115 & 7$\pm1$ & 0.6$\pm0.3$ & 270$^{+80}_{-50}$ & \multicolumn{2}{c}{8.6$^{+0.2}_{-0.6}$} & 1.12/115\\
\hline \hline
\multicolumn{9}{c}{Thermal stacked spectra} \\
\hline
Quantile & With/Without & \# Sources & $f_X$ 2-10 keV & $f_{X,low}/f_{X,high}$ & $\chi_{\nu}^2$/dof &  $\Gamma$ & Line Eq. & $\chi_{\nu}^2$/dof \\
Group& IR Counterparts & & (10$^{-14}$ erg cm$^{-2}$ s$^{-1}$) & & & & Width (eV) & \\
(1) & (2) & (3) & (7) & (9) & (8) & (10) & (11) & (12)\\
\hline
\multirow{2}{*}{A} & IR & 246 & 0.49$\pm$0.02 & 0.034$\pm$0.004 & 0.99/97 & -- & -- & --\\
& No IR & 66 & 0.26$\pm$0.02 & 0.036$\pm$0.006 & 1.71/49 & -- & -- & --\\
\hline
\multirow{2}{*}{B} & IR & 135 & 1.37$\pm$0.03 & 0.034$\pm$0.003 & 1.22/157 & -- & -- & --\\
& No IR & 75 & 0.55$\pm$0.04 & 0.07$\pm$0.01 & 1.33/57 & -- & -- & --\\
\hline
\multirow{2}{*}{E} & IR & 64 & 2.50$\pm$0.10 & 0.29$\pm$0.02 & 1.08/73 & 5$\pm2$ & 1700$^{+1000}_{-100}$ & 0.20/11\\
& No IR & 67 & 2.14$\pm$0.09 & 0.21$\pm$0.02 & 1.06/64 & 2$^{+2}_{-1}$ & $<$316 & 1.25/11\\
\hline
\end{tabular}
\begin{tablenotes}
\item \underline{Notes:} All quoted uncertainties are 90\% unless stated otherwise.  

\item (1) Quantile groups defined in \S \ref{sec:quantile}.

\item (2) Stacked spectrum of sources with or without IR counterparts.

\item (3) Number of sources in stacked spectrum.

\item (4) Hydrogen column density associated with model component in next column.

\item (5) Power-law photon index.

\item (6) Equivalent width of Gaussian line component at energy 6.65 keV, line center determined from stacked spectrum of all FB, group D sources (1$\sigma$ errors).

\item (7) 2-10 keV absorbed flux (1$\sigma$ errors).  

\item (8) Reduced chi-square of best-fit model over degrees of freedom.

\item (9) Flux ratio in 2-10 keV band of low $kT$ component over high $kT$ component (1$\sigma$ errors).

\item (10) Power-law photon index of 5-9 keV band when fitting with power-law plus Gaussian line model.

\item (11) Equivalent width of Gaussian line component at 6.65 keV, line center determined from stacked spectrum of all FB group E sources (1$\sigma$ errors).

\item (12) Reduced chi-square of best-fit model for 5-9 keV band over degrees of freedom.
\end{tablenotes}
\end{threeparttable}
\label{tab:stackedir}
\end{table*}

\section{Discussion of X-ray Populations}
\label{sec:discussion}

Although it is difficult to determine the nature of individual sources in our survey, it is possible to make reasonable inferences about the classes of X-ray sources that dominate each quantile group defined in \S \ref{sec:quantile}.  The stacked spectra, variability, and IR counterparts of the sources in each group provide important clues to the nature of the sources.  In order to facilitate our understanding of the reliable counterparts, we estimated the $J$ and $H$ magnitudes of main sequence and supergiant high-mass (O,B) and low-mass (G,K,M) stars at distances of 100 pc, 1 kpc, 4 kpc, 10 kpc, 12 kpc with corresponding $N_{\mathrm{H}}$ values\footnotemark\footnotetext{For each distance listed, the corresponding $N_{\mathrm{H}}$ value was calculated as described in \S \ref{sec:specanalysis}.} of 0.0, 0.3, 1.0, 3.0, and 5.0 $\times 10^{22}$ cm$^{-2}$. These $J$ and $H$ magnitude estimates were based on the absolute $V$ magnitudes from \citet{wegner07}, the intrinsic colors from \citet{ducati01}, the $N_{\mathrm{H}}-A(V)$ relationship derived by \citet{guver09}, and the $A(\lambda)/A(V)$ relations from \citet{cardelli89}.  Unless explicitly stated otherwise, we refer readers to \citet{muno04} and references therein for an overview of the spectral and timing properties of Galactic X-ray sources.  

\subsection{Group A}
\label{sec:groupA}
The group A stacked spectrum is best fit by a two-temperature thermal plasma model.  The column density associated with each temperature component is low ($\lesssim10^{21}$ cm$^{-2}$), suggesting most sources in this group are foreground sources, located at a distance $\lesssim$1 kpc.  Both components have low temperatures ($kT \lesssim2$ keV), and the sources in this group have 0.5-10 keV luminosities\footnotemark\footnotetext{In the remainder of this section, $L_X$ refers to 0.5-10 keV luminosity.} between $L_X = 10^{27}$-$10^{31}$ ergs s$^{-1}$, assuming they are located at a distance between 100 pc and 1 kpc.  The very soft component ($kT = 0.75$ keV) most likely originates from the magnetic coronae of low-mass stars, which are the faintest sources of X-ray emission. X-ray active low-mass stars typically have $kT <$ 1 keV and $L_X<10^{29}$ ergs s$^{-1}$.  The higher $kT$ component is more consistent with X-ray emission from coronally active binaries (ABs) such as RS CVn systems.  These binaries usually have $kT\approx0.1-2$ keV and $L_X=10^{29}-10^{32}$ ergs s$^{-1}$.  Other types of sources which could contribute to the group A X-ray emission are symbiotic binaries and massive stars.  Symbiotic binaries, which consist of a mass-losing cool supergiant and white dwarf companion, are often considered a subtype of CVs and tend to have $kT = 0.3-1.3$ keV and $L_X=10^{30}-10^{33}$ ergs s$^{-1}$ \citep{murset97}.  X-rays can be produced in the shocks in the winds of high-mass stars, either in isolation or in a colliding-wind binary, and typically have $kT = 0.1-6$ keV; in fact, three group A sources (\#622, 469, and 298) are coincident with previously identified O and B stars, HD149452, HD149358, and ALS 3666, respectively.  Based on the optical band spectral and photometric information available on these sources, we estimate they are located at a distance of approximately 1-4 kpc, farther than the majority of sources in this group appear to be based on the column density of the group A stacked spectrum.

Group A contains an enhanced fraction of variable sources compared to other quantile groups. Ten of 26 (38\%) group A sources with $\geq$ 40 counts show variability on $\sim$hour timescales (see \S \ref{sec:variability}) at $\geq$ 95\% confidence.  Of the 83 group A sources detected in multiple observations, 50 (60\%) are found to be variable between observations, a higher percentage than is found for sources in the full sample detected in multiple observations.  The significant variability seen in this group of sources is consistent with the flaring behavior of low-mass active stars and interacting binaries.  As can be seen in Figure \ref{fig:varamp}, the fluxes of some group A sources vary by large factors ($>5$) in between observations; hour-long flares with amplitudes as large as a factor of 10 have been observed in RS CVns and could be the origin of these large variability amplitudes. 

In the VVV catalog, we found reliable infrared counterparts to 67\% of the group A sources. The only significant difference between the X-ray properties of group A sources with and without IR counterparts is that those without counterparts tend to have lower X-ray fluxes; thus sources without IR counterparts may be intrinsically dimmer in both the X-ray and NIR bands or they may located at larger distances.  As can be seen in Figure \ref{fig:vistacolormag}, these counterparts have blue $(J-H)$ colors, in agreement with our inference that group A sources are mostly foreground sources.  In this color-magnitude diagram, there is a tight cluster of counterparts with $H$ magnitudes between 13 and 15, and a looser cluster of counterparts with $H$ magnitudes between 8 and 13.  The former is consistent with the colors and magnitudes of low-mass main sequence stars at a distance of $\approx$ 1 kpc, while the latter is likely a mixture of low-mass main sequence stars at $\approx$ 100 pc and cool giants at $\approx$ 1 kpc; the small number of counterparts with negative $(J-H)$ are more likely to be high mass stars at $\approx$ 1 kpc.  The fact that the majority of counterparts are most likely to be low-mass stars, either on the main sequence or in a giant/supergiant phase, is consistent with the suggestion that the dominant X-ray populations in group A are X-ray active low-mass stars and interacting binaries.  

\subsection{Group B}
\label{sec:groupB}

Similar to group A, the group B stacked spectrum is also best-fit by a two-temperature thermal model with low hydrogen column densities.  However, the temperature of the hotter component is significantly higher ($kT\approx 7$ keV) for the B sources than for the A sources.  The origin of the low-temperature component may be low-mass X-ray active stars, active binaries, and symbiotic binaries, the dominant sources in group A, while the high-temperature component is more typical of CVs.  The majority of CVs are close binary systems with a white dwarf accreting matter from a low-mass main-sequence stellar companion, although some have been found to have giant donors \citep{kuulkers06}.  CVs are typically subdivided into three main categories based on the magnetic field strength of the WD they host.  The majority of CVs have weak magnetic fields ($B\lesssim10^4$ G) which do not significantly affect the accretion flow from the Roche lobe-filling donor and are called nonmagnetic CVs.  About 20\% of CVs are polars, exhibiting strong magnetic fields ($B\gsim10^{6.5}$ G) which prevent the formation of an accretion disk; about 5\% are intermediate polars (IPs), having intermediate strength magnetic fields which channel material from the inner accretion disk onto the magnetic poles.  Nonmagnetic CVs and polars have $\Gamma\approx1-2$ or $kT\approx1-25$ keV and $L_X = 10^{29}-10^{32}$ erg s$^{-1}$.  IPs tend to be more luminous ($L_X = 10^{31}-10^{33}$) and display harder emission ($\Gamma < 1$) than other CVs.  Assuming that the group B sources lie at a distance between 100 pc and 1 kpc based on their low $N_{\mathrm{H}}$, they span the luminosity range $10^{27.4}-10^{31.5}$ ergs s$^{-1}$, consistent with the luminosities of low-mass stars at the faint end and with the luminosities of active binaries and CVs at the bright end.  

Further evidence that a CV population exists in group B is provided by one of the brightest sources in our survey, source 750, which is coincident with ASCA source AX J1637.8-4656 \citep{sugizaki01}
This source has $\Gamma$ = 1.15 (see Table \ref{tab:indspectra}), and, assuming a distance of 1 kpc from its low $N_{\mathrm{H}}$, $L_X\approx8\times10^{31}$ ergs s$^{-1}$.  In addition to having a luminosity and photon index consistent with that of a CV, this source is variable on both short and long timescales.  As can be seen in Table \ref{tab:periodicity}, this source was determined to be periodic by the $Z_n^2$ test \citep{buccheri83}, with a best period of roughly 7100 s.  Periodic X-ray emission has been observed from magnetic CVs and is believed to be associated with the spin period of the white dwarf, which can range from $10^2$ to $10^4$ seconds \citep{scaringi10}.  Finally, the near-IR spectrum obtained of the IR counterpart of this source exhibits strong emission lines suggestive of an accretion disk \citep{rahoui14}.  Thus source 750 is likely a CV with an intermediate strength magnetic field that is not fully disrupting the accretion disk.

As mentioned in \S \ref{sec:variability}, source 961, another group B source, is also likely to be a magnetic CV.  Its $\approx$5700 second period is consistent with the spin and orbital periods of CVs.  As can be seen in Figure \ref{fig:lightcurve961}, its pulse profile shows large brightness variations from approximately zero to twenty counts; such variations could either result from pulsations due to emission from accretion spots at the WD magnetic poles as the WD rotates or from eclipsing of the WD by its companion.  Its location in the quantile diagram indicates it has $\Gamma \approx1.5$, which is more typical of polars and nonmagnetic CVs than IPs.  Although it is likely a nearby source based on its low $N_{\mathrm{H}}$ (estimated to be $\sim10^{21}$ cm$^{-2}$ from its quantile parameters), it has no stellar IR counterpart in the VVV survey; we do find an infrared counterpart to this source in the VVV survey, but it is morphologically classified as a galaxy, and therefore it cannot be the true counterpart since the X-ray spectrum of the \textit{Chandra} source shows very little absorption.  Since the source is likely located $\lesssim$ 1 kpc, the stellar IR counterpart must be very faint to avoid detection, thus making it unlikely that this source has a significant accretion disk which indicates that, of all CV types, this source is most likely a polar.  Thus, there is evidence that both sources 750 and 961 are magnetic CVs, and their location in the quantile diagram supports the hypothesis that group B may contain a significant CV population. 

Group B has a comparable percentage of variable sources to group A.  Eight of the 17 (47\%) group B sources with $\geq$40 counts are variable on short timescales, and 19 of the 52 (37\%) group B sources detected in multiple observations are variable on long timescales.  The fact that this latter percentage is lower than that of group A sources may be partly explained by the significant number of CVs, probably present in group B but not group A, which undergo flaring episodes less frequently than flare stars and thus tend to be fairly constant on day-week timescales.  

We found reliable IR counterparts for 61\% of group B sources, which largely overlap in the color-magnitude diagram with the group A counterparts, indicating that they have similar stellar types and are located at similar distances.  This similarity is not surprising since the X-ray properties of the group B sources indicate that they are dominated by the same X-ray populations as group A plus a population of CVs, which tend to have near-IR properties similar to low-mass main sequence or evolved stars \citep{hoard02}.  As can be seen in Table \ref{tab:stackedir}, group B sources without IR counterparts have a lower average flux and a more significant contribution from the low $kT$ component compared to group B sources with IR counterparts.  These trends suggest that X-ray active low-mass stars make up a relatively larger fraction of group B sources without counterparts than of B sources with counterparts.  

\subsection{Group C}
\label{sec:groupC}

The group C stacked spectrum is best-fit by an absorbed power-law with $\Gamma\approx1.1$ and $N_{\mathrm{H}}=1.4\times10^{22}$ cm$^{-2}$, which suggests that these sources are located at a distance of 3-5 kpc, in the Scutum-Crux and near Norma spiral arms.  The luminosities spanned by group C sources are $L_X = 10^{31}-10^{32.7}$ ergs s$^{-1}$, assuming a distance of 4 kpc.  Possible classes of X-ray sources present in this group are magnetic and nonmagnetic CVs, hard-spectrum symbiotic binaries, low-mass X-ray binaries (LMXBs), and HMXBs.  IPs and HMXBs tend to have $\Gamma<1$ while nonmagnetic CVs, polars, symbiotic binaries, and LMXBs tend to have $\Gamma>1$.  CVs are the most numerous accreting sources, so they are most likely the dominant population.  
 
Group C sources show the lowest levels of variability of any group.  Only 2 of 12 (17\%) sources with $\geq 40$ counts exhibit short-timescale variability and only 5 of 26 (19\%) sources detected in multiple observations are variable on long timescales.  These results are consistent with magnetic CVs and LMXBs dominating group C, since although they show periodic variations and occasional outburts, they generally have stable emission.

Reliable IR counterparts were found for 35\% of group C sources.  They have redder $(J-H)$ colors than the counterparts of group A and B sources, confirming that they are more distant than the group A or B counterparts.  Their $H$ magnitudes are consistent with high-mass stars and evolved low-mass stars located at $\sim$4 kpc. Considering the extinction resulting from the $N_{\mathrm{H}}$ measured for these sources and their likely distances, most main-sequence, low-mass counterparts of group C sources would be undetectable in the VVV survey.  The stacked spectrum of group C sources with IR counterparts has a softer power-law index ($\Gamma\approx1.3$) than the sources lacking IR counterparts ($\Gamma\approx0.9$).  Thus, the majority of group C sources with counterparts may be symbiotic binaries and CVs with subgiant and giant companions, while those without counterparts may primarily be CVs with main-sequence companions, especially IPs given the lower photon index of these sources.  The presence of some type II AGN among the group C sources lacking IR counterparts could also help explain their lower photon index, but, as discussed in \S \ref{sec:AGN}, very few AGN are likely to be found in group C.

\subsection{Group D}
\label{sec:groupD}

The group D stacked spectrum has a very hard photon index ($\Gamma\approx0.7$), a prominent Fe line, and a high $N_{\mathrm{H}}$ indicating that these sources typically lie on the far side of the Galaxy, near, in, or beyond the far Norma arm.  The Fe emission is well-modeled by a Gaussian centered at 6.65 keV with an equivalent width of approximately 300 eV; this emission likely results from the blending of lines at 6.4 keV and 6.7 keV, arising from low-ionization Fe and He-like Fe respectively.  The presence of this strong, non-redshifted Fe line suggests that many of the sources in this group must be Galactic; otherwise, if this group were dominated by AGN, their spread in redshift would result in a smearing out of the Fe line.

Two classes of X-ray sources that are frequently observed having spectra with $\Gamma < 1$ are IPs and HMXBs.  Fe line emission has been observed from both of these types of sources.  Although the luminosity range spanned by group D ($L_X =10^{32}-10^{33.7}$ ergs s$^{-1}$ assuming $d = 10$ kpc) extends to higher luminosities than are typically observed for IPs, roughly 80\% of group D sources have $L_X \lesssim10^{33}$ ergs s$^{-1}$ for $d = 10$ kpc, a reasonable luminosity range for IPs.  Thus, IPs could be the dominant population among faint group D sources.  In contrast, HMXBs can have X-ray luminosities as high as $L_X \sim 10^{34}$ ergs s$^{-1}$ during quiescence and $L_X \sim 10^{38}$ ergs s$^{-1}$ during outburst.  In fact, one of the group D sources, source 999, is a previously discovered HMXB \citep{sugizaki01, bird04, bodaghee06} and has $\Gamma\approx0.9$ and $L_X\approx10^{35.3}$ ergs s$^{-1}$.  

Five of 14 (36\%) group D sources exhibit short-timescale variability, and although 17 of 35 (49\%) group D sources detected in multiple observations display long-timescale variability, they have the lowest mean and median variability amplitude of any group.  These modest levels of variability are consistent with, although not necessarily proof of, group D being dominated by a population of IPs, which tend to have fairly stable emission.  

Only 29\% of group D sources have reliable IR counterparts.  The ($J-H$) color and $H$ magnitudes of these counterparts are consistent with being high-mass and evolved low-mass stars at distances between 8-12 kpc.  The low fraction of detected counterparts is more easily explained by a dominant population of IPs rather than HMXBs, since a large fraction of the low-mass companions of the white dwarfs in IPs would have $J$ and $H$ magnitudes greater than the VVV sensitivity limit when located at distances of 8-12 kpc.  However, while most massive stellar counterparts of HMXBs in group D should be above the VVV sensitivity limit, they can occasionally be so obscured by circumbinary gas and dust that they are much fainter than otherwise expected \citep{bodaghee12a}; in fact, our counterpart-matching algorithm does not find the faint massive counterpart of HMXB IGR J16393-4643 \citep{bodaghee12b} in the VVV catalog.  

The X-ray spectral differences between group D sources with and without IR counterparts are not statistically significant.  Nonetheless, if the lower photon index and lower Fe equivalent width of the sources without IR counterparts is a real trend, it could be explained if the sources without counterparts are primarily IPs and type II AGN (see \S \ref{sec:AGN}), while those with counterparts include some hard-spectrum symbiotic binaries \citep{luna13}, whose spectra are a bit softer and whose red giant companions should be detectable by the VVV survey.  

\subsection{Group E}

The group E stacked spectrum is best-fit by a two-temperature thermal model, making it significantly different from the group D power-law spectrum, even though some of the same classes of X-ray sources must be present in both groups D and E since they are not sharply separated in the quantile diagram.  Both temperature components have $kT$ = 1-2 keV but very different hydrogen column densities, the lower of which is similar to that of group D, and the higher of which is $N_{\mathrm{H}} \approx 2.4\times10^{23}$ cm$^{-2}$.  This $N_{\mathrm{H}}$ value is roughly 3 times higher than the maximum $N_{\mathrm{H}}$ value measured through the Galaxy along a line of sight within our surveyed area, indicating that some E sources are obscured by large amounts of local absoption and/or may be imbedded in the molecular clouds of the far Norma arm.  Assuming the same typical distance of 10 kpc as we did for group D, these sources span the luminosity range $L_X = 10^{32}-10^{33.7}$ ergs s$^{-1}$.  A significant fraction of these sources may be associated with the shocks produced in the winds of high-mass stars; these sources typically have $kT = 0.1-6$ keV, $L_X \sim 10^{33}-10^{35}$ ergs s$^{-1}$, and their emission can be significantly absorbed by circumstellar material.  The photometric and spectral properties of group E are also consistent with symbiotic binaries.  In addition, a small number of magnetars could be present in this group.  In fact, one of the group E sources is a previously discovered magnetar, SGR 1627-41, and data from this survey was used in an in-depth study of this magnetar by \citet{an12}.    

Only two of the nine (22\%) group E sources with $\geq$40 counts exhibits variability on short timescales.  However, 14 of the 24 (58\%) group E sources detected in multiple observations are found to vary between observations.  If group E is indeed dominated by high-mass stellar populations, these trends in X-ray variability are similar to those found for high-mass stars in the Galactic center survey \citep{mauerhan10}.

Although group E sources are roughly located at the same distance as group D sources based on the $N_{\mathrm{H}}$ measured in their stacked spectra, a higher fraction of group E sources 48\%) have reliable IR counterparts.  At distances of $\approx$10 kpc, only high-mass stars and cool giants have $J$ and $H$ magnitudes above the sensitivity limit of the VVV survey.  The higher percentage of IR counterparts is consistent with group E having a larger fraction of high-mass stellar X-ray sources and symbiotic binaries than group D, since the high-mass and giant counterparts of these sources are brighter than the primarily main-sequence low-mass counterparts of IPs.  A significant fraction of group E sources without IR counterparts are also likely to be type I AGN, as discussed in the following section.

\subsection{AGN Contribution}
\label{sec:AGN}

In addition to the Galactic classes of sources described in the previous sections, we expect a significant population of AGN to be present in our catalog.  Using the AGN count-distribution from the COSMOS survey \citep{cappelluti09} and taking into account the sensitivity variations across the survey area (see \S \ref{sec:sensitivity}) and the incompleteness of our detection method (see \S \ref{sec:recovery}), we estimate that roughly 150 AGN could be present in our catalog.  X-ray emission from AGN is attenuated by the integrated column density through the whole galaxy and typically has $\Gamma \approx 1.7$ \citep{molina09} for type I AGN or $\Gamma \lesssim 1$ for type II AGN with a reflection component.  Thus, AGN are most likely to be found in groups D and E based on the regions of quantile space they occupy.  Due to the spread in $N_{\mathrm{H}}$ values across our field of view and the large median error bars on $Q_x$ and $Q_y$, some group C sources may also be AGN.

We expect to find most of the AGN in our sample among the sources without NIR counterparts.  Only 2\% of AGN in the Chandra COSMOS Survey \citep{civano12} with $f_X > 1\times10^{-15}$ erg cm$^{-2}$ s$^{-1}$ have $H$ magnitudes $\leq$ 18 --the sensitivity limit of the VVV survey--and we would expect an even smaller percentage of the AGN in NARCS to be detected in the VVV survey due to the higher extinction in the Galactic plane compared to the COSMOS field; the integrated $N_{\mathrm{H}}$ through the galaxy in the NARCS region varies between 3-8$\times10^{22}$cm$^{-2}$, which corresponds to extinction values of $A(J) \approx 4-10$ mag and $A(H) \approx 3-7$ mag.

Based on the spectral properties of group E sources without IR counterparts, it seems likely that many of them are AGN.  Their stacked spectrum shows a more prominent high-column density component compared to the spectrum of group E sources with counterparts; this enhancement could be due to a large number of AGN, which suffer from extinction by their local environment and host galaxy as well as the Milky Way ISM.  In addition, fitting the 5-9 keV stacked spectra of group E sources with and without counterparts with a power law plus Fe line model shows that the sources lacking IR counterparts have a harder spectrum with $\Gamma \approx 2$ and a lower Fe line equivalent width.  This photon index is typical of type I AGN and the lower Fe equivalent width is expected for a group of extragalactic sources whose Fe line would be smeared out due to their redshift distribution.  Thus, a large number of type I AGN among the group E sources lacking IR counterparts can explain the difference between the stacked spectra of group E sources with and without IR counterparts. 

There is weaker evidence for the presence of AGN among the group C and group D sources lacking IR counterparts.  The stacked spectrum of group D sources without counterparts is harder ($\Gamma\approx0.6$) and has a lower Fe line equivalent width that the spectrum of sources with counterparts; however, while these trends are consistent with the presence of type II AGN with a reflection component among the sources lacking counterparts, these differences are not statistically significant.  Thus, these trends may be real and indicative of an AGN population, or they may be statistical fluctuations, in which case the group D sources lacking counterparts are probably just faint and/or distant versions of the group D sources with counterparts.  The stacked spectrum of group C sources without counterparts is harder ($\Gamma\approx0.9$) than that of the sources with counterparts; this difference could be driven by a population of type II AGN among the sources lacking counterparts.  However, it is unlikely that many AGN would be found in group C, since the average column density of sources in this group is $N_{\mathrm{H}}\approx10^{22}$ cm$^{-2}$, which is low for an AGN whose light would be shining through the entire Galaxy.  Thus, it appears more likely that the differences between the group C sources with and without counterparts are due to different populations of Galactic sources (as discussed in \S \ref{sec:groupC}), although a small AGN contribution cannot be ruled out.  

In order for our survey to contain the expected number of AGN, the majority of group E sources lacking IR counterparts and 30\%-50\% of group C and D sources without counterparts must be AGN.  Given that there is good evidence for the former and that the latter cannot be ruled out, it is possible that about 150 AGN are present in our catalog, as expected from other surveys.

\section{Computing the Number-Flux Distribution}
\label{sec:methodology}

Having determined the X-ray populations which likely dominate each of the quantile groups, we sought to compare the populations in our survey to predictions based on surveys of other regions of the Galaxy.  A useful tool in comparing the populations of different surveys is the number count distribution.  In addition, for a particular population of sources located at similar distances, the number-flux distribution is closely related to the luminosity function of the sources and thus can also shed light on the physics which determines the brightness of these sources.  

At faint fluxes, this calculation is complicated by the nonuniform sensitivity across the \textit{Chandra} image, the incompleteness of the source-detection algorithm, and the Eddington bias, which is caused by fluctuations in the source and/or background making a faint source appear brighter.  To help correct for these effects, we used a method similar to that developed by \citet{georgakakis} and adapted by \citet{lehmer12}, which uses a Bayesian approach with maximum-likelihood optimizations.  We decided to compute the number count distribution in the 2-10 keV band, because (1) most foreground, thermal sources are not detected in this band, allowing us to concentrate on the populations in the spiral arms, (2) photons in this energy band are less likely to be absorbed by dust along the line-of-sight, resulting in more robust conversion factors between the net counts of a source and its unabsorbed energy flux, and (3) it will enable comparisons to published number count distributions from previous surveys of Galactic X-ray populations, which are primarily in the 2-10 or 2-8 keV band.  In the remainder of this section, whenever we refer to catalog sources, we only mean sources detected at $\geq 3\sigma$ in the 2-10 keV band.

\subsection{Sensitivity Curves}
\label{sec:sensitivity}

Near the flux limit of a survey, a source of a given flux can only be detected over a limited fraction of the total solid angle covered by the survey due to the inhomogeneous background and nonuniform PSF across the \textit{Chandra} image.  In order to account for this varying sensitivity in our number count distribution, we calculate the effective solid angle as a function of source flux, which we refer to as the sensitivity curve.  This sensitivity curve depends on the significance threshold we choose to select our sources.  In computing the number count distribution of sources in our survey, we select point-like sources that have been detected by \texttt{wavdetect} in any energy band as described in \S \ref{sec:detection} and have $P(\geq C_{\mathrm{src}})$ $\le$ 0.00137 = $P_{\mathrm{thresh}}$ in the 2-10 keV band as determined by Equation \ref{eq:probnoise}, which is the probability required for a 3$\sigma$ detection.  This selection procedure will not include all real sources with $P(\geq C_{\mathrm{src}})$ $\le P_{\mathrm{thresh}}$ because of \texttt{wavdetect}'s complex source detection criteria (see \citealt{freeman}).  Correcting for this detection incompleteness is discussed in \S \ref{sec:recovery}.  

To compute the sensitivity curve, we follow the method described in \citet{georgakakis}, which should allow us to extrapolate the number-flux distribution to fluxes roughly an order-of-magnitude fainter than the formal survey flux limit--the flux to which $\gsim$90\% of the image is sensitive.  First, we determine the minimum number of counts, $C_{\mathrm{lim}}$, required for a detection, such that $P(\geq C_{\mathrm{lim}})$ = $P_{\mathrm{thresh}}$, at each location in the image.  We use the background maps (see \S \ref{sec:detection}) to determine the mean expected background counts, $\langle C_{\mathrm{bkg}}\rangle$, within circular regions with radii equal to the local 90\% ECF radius.  The cumulative probability that the observed counts will exceed $C_{\mathrm{lim}}$ within a particular region is
\begin{equation}
P(\geq C_{\mathrm{lim}}) = \gamma(C_{\mathrm{lim}}, \langle C_{\mathrm{bkg}}\rangle)
\label{eq:probnoise2}
\end{equation}
where $\gamma(a,x)$ is the lower incomplete gamma function, defined as
\begin{equation}
\gamma(a,x) = \frac{1}{\Gamma(a)}\int_0^x e^{-t}t^{a-1}dt
\end{equation}
Equation \ref{eq:probnoise2} is a simplication of Equation \ref{eq:probnoise} for situations in which the mean background within an aperture region is well determined.  Setting $P(\geq C_{\mathrm{lim}}) = P_{\mathrm{thresh}} = 0.00137$, we invert Equation \ref{eq:probnoise2} numerically to find $C_{\mathrm{lim}}$ for a region with mean expected background $\langle C_{\mathrm{bkg}}\rangle$.  For each observation, we perform this procedure for different regions, which combined cover the full image area, thus obtaining a 2D image of $C_{\mathrm{lim}}$ known as a sensitivity map. 

Then we can compute the probability of detecting a source of a given flux $f_X$ and spectral shape within each region of the sensitivity map.  The total observed counts in the region are the sum of the source and background contributions, which can be expressed as:
\begin{equation}
C_{\mathrm{src}} = C_{\mathrm{net}} + \langle C_{\mathrm{bkg}}\rangle = f_X t_{\mathrm{exp}} E_{\mathrm{src}} \eta \epsilon + \langle C_{\mathrm{bkg}}\rangle
\label{eq:count2flux}
\end{equation}
where $t_{\mathrm{exp}}, E_{\mathrm{src}}, \eta,$ and $\epsilon$ are the exposure time, mean effective area, ECF, and unabsorbed energy flux to observed photon flux conversion factor, respectively.  $\epsilon$ includes a correction factor for extinction along the line-of-sight due to the amount of $N_{\mathrm{H}}$ determined by the spectral fits of each quantile group, but not exceeding 5.5$\times 10^{22}$ cm$^{-2}$, the average $N_{\mathrm{H}}$ integrated through the entire Galaxy in our surveyed area; we assume that larger values of $N_{\mathrm{H}}$ are likely due to both interstellar and intrinsic absorption, and we do not wish to correct for absorption that may be intrinstic or very local to the source.  The energy flux to photon flux conversion factor depends on the source spectrum, and we used the same $\epsilon$ for all sources in the same quantile group; the conversion factors are listed in Table \ref{tab:stacked}.  For a region with particular values of $\langle C_{\mathrm{bkg}}\rangle$ and $C_{\mathrm{lim}}$, the probability of detecting a source of flux $f_X$ is given by
\begin{equation}
P_{f_X}(\geq C_{\mathrm{lim}}) = \gamma(C_{\mathrm{lim}},C_{\mathrm{src}}) .
\end{equation}

\begin{figure}[t]
\includegraphics[width=0.48\textwidth]{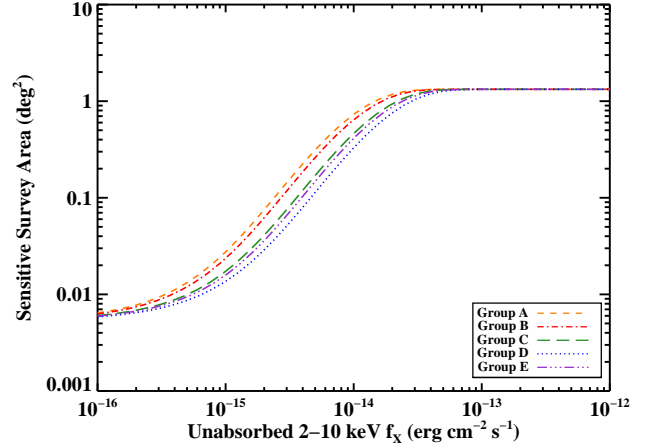}
\caption{Angular area of our survey that is sensitive to a given source flux. (A color version of this figure is available in the online journal.)}
\label{fig:sensitivitycurve}
\end{figure}

We calculate the sensitivity curve for each observation by summing the $P_{f_X}(\geq C_{\mathrm{lim}})$ distributions of individual regions, each weighted by the solid angle (in degrees) of each region.  Then we added together the sensitivity curves of all the individual observations, and divided the combined sensitivity curve by a factor of 1.54 so that the maximum value of the sensitivity curve was equal to the total survey area (1.3268 deg$^2$); this division was necessary because the observations partially overlap.  This combined sensitivity curve, $A(f_X,\epsilon)$ (shown in Figure \ref{fig:sensitivitycurve}), is an approximation which overweights the overlapping regions, which generally have worse sensitivity since they are at large off-axis angles.  We estimate errors in the sensitivity curve to be ~0.1 dex, which are satisfactory for our purposes.  We choose to only compute number counts to a flux limit of $1\times10^{-15}$ erg cm$^{-2}$ s$^{-1}$, to which 1-3\% of the total solid angle is sensitive; below this flux, too few sources are detected by \texttt{wavdetect} to make reliable predictions.  Note that the sensitivity curve as a function of source counts is the same for all spectral groups, but since different spectral groups have different count-to-flux conversions, they have different sensitivity curves as a function of flux.  

\subsection{Recovery Fraction Correction}
\label{sec:recovery}

\begin{figure}[b]
\centering
\includegraphics[width=0.48\textwidth]{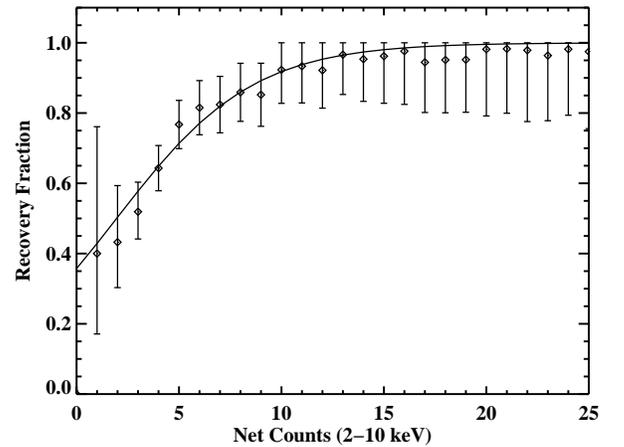}
\caption{Fraction of simulated sources that are detected at $\geq3\sigma$ in the 2-10 keV band and are also detected by our detection procedure as a function of net 2-10 keV counts.  The solid line shows our best-fit model for this recovery fraction, which we use in \S \ref{sec:maxlikelihood} to correct for incompleteness in our number-count computation.}
\label{fig:recoveryfunction}
\end{figure}

The sensitivity curve will only successfully correct for the incompleteness of our catalog if all sources are detected above a specific selection probability.  However, due to the complex criteria of \texttt{wavdetect}, some faint sources that would meet the probability criterion are not detected.  Employing an approach similar to that of \citet{lehmer12} to compute and correct for detection incompleteness, we generated 200 mock images in the full, soft, and hard energy bands of ObsID 12519, which we took to be representative of all observations since they have very similar exposure and background maps.  To make each set of mock observations, we added 60 sources to the full, soft, and hard unsmoothed background images (see \S \ref{sec:detection}), since this was the average number of sources detected in an individual observation.  Each source was assigned a random position on the sky and a random number of total counts between 3 and 50, taken from a power-law count distribution with an index of $-$1.6 (a compromise between approximating the count distribution we measure and having a statistically significant sample of sources within each count bin).  The total counts were then randomly divided between the soft and hard band.  We approximated the PSF at the location of each source as an azimuthally-symmetric Rayleigh distribution in the radial direction, the normalization parameters of which were determined from the size of the PSF for 4.5 keV photons (for the full and high energy band mock images) or 1.5 keV (for the low energy band mock images) and a range of ECFs.  The counts for each source were then distributed according to this approximation of the local PSF.  

We determined the photometric properties of the 60 sources in each set of mock observations as described in \S \ref{sec:photometry}.  We also produced a sourcelist including photometric properties for each set of mock observations using our standard pipeline beginning with the second round of \texttt{wavdetect} using the background and exposure maps for ObsID 12519.  Then we calculated, as a function of the input source counts in a particular energy band, what fraction of input sources that satisfy our 3$\sigma$ threshold in that energy band were detected by \texttt{wavdetect}.  This recovery fraction is shown in Figure \ref{fig:recoveryfunction} for the 2-10 keV band, and it is well-fit by the analytic form $F_{\mathrm{rec}}(C) = 1/(1+\mathrm{exp}[-\delta\{C-\xi\}])$, where $C$ are the input source counts in a given band, and $\delta$ and $\xi$ are fitting constants that vary with energy band.  For the 2-10 keV band, $\delta$ = 0.30$\pm$0.07 and $\xi$ = 1.96$\pm$0.79.

\subsection{Flux Probability Distributions}

\begin{figure}[b]
\includegraphics[width=0.48\textwidth]{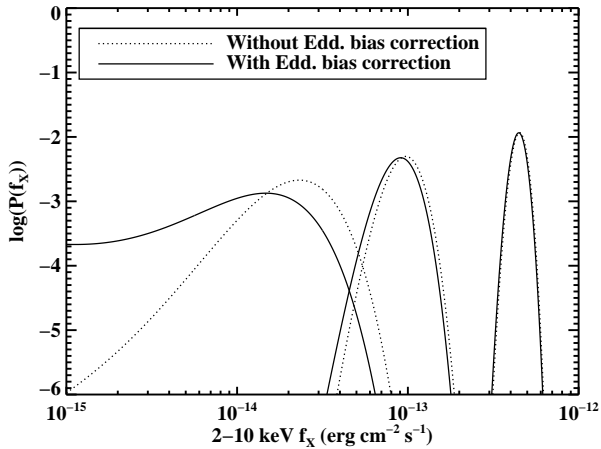}
\caption{Example flux probability density distributions, before and after implementing the Eddington bias correction based on the best-fit power-law index $\beta$ reported in Table \ref{tab:dNdS}.  From left to right, these sources are detected at confidence levels of $3.2\sigma$, $8.5\sigma$, and $29.8\sigma$, respectively, in the 2-10 keV band.}
\label{fig:fluxdist}
\end{figure}

In addition to correcting for incompleteness in our catalog, we take into account the fact that the observed counts in a given source aperture can be attributed to a source with a range of possible fluxes, rather than assigning a single flux value to each source.  For $C_{\mathrm{src}}$ total counts in the source aperture and $C_{\mathrm{bkg}}$ counts in the background aperture, the probability distribution of source counts, $C_{\mathrm{net}}$ within the source aperture is given by (derived from \citet{weiss07} with $\psi_T \rightarrow 1$ and $\psi_R \rightarrow 0$)
\begin{equation}
P(C_{\mathrm{net}}|C_{\mathrm{src}},C_{\mathrm{bkg}}) = \frac{1}{Z}\sum_{i=0}^{C_{\mathrm{src}}} \frac{(C_{\mathrm{src}}+C_{\mathrm{bkg}}-i)!}{C_{\mathrm{bkg}}!(C_{\mathrm{src}}-i)!} \omega_B^{C_{\mathrm{bkg}}}\omega_S^{C_{\mathrm{src}}-i}\frac{C_{\mathrm{net}}^ie^{-C_{\mathrm{net}}}}{i!}
\end{equation}
where the partition function, $Z$, is
\begin{equation}
Z = \sum_{i=0}^{C_{\mathrm{src}}}\frac{(C_{\mathrm{bkg}}+i)!}{C_{\mathrm{bkg}}!i!}\omega_B^{C_{\mathrm{bkg}}}\omega_S^i
\end{equation}
and $\omega_B$ or $\omega_S$ is the probability that a background event occurs in the background or source aperture, respectively:
\begin{equation}
\omega_B = \frac{A_{\mathrm{bkg}}E_{\mathrm{bkg}}}{A_{\mathrm{bkg}}E_{\mathrm{bkg}}+A_{\mathrm{src}}E_{\mathrm{src}}}, \hspace{0.3in}\omega_S = \frac{A_{\mathrm{src}}E_{\mathrm{src}}}{A_{\mathrm{bkg}}E_{\mathrm{bkg}}+A_{\mathrm{src}}E_{\mathrm{src}}} .
\end{equation} 
$P(C_{\mathrm{net}})$ is normalized and converted into $P(f_X)$, using the relationship included in Equation \ref{eq:count2flux}.  Finally, assuming that the differential counts of sources within each quantile group obey a power law of the form $dN/df_X \propto f_X^\beta$, we corrected for the Eddington bias by multiplying each source flux distribution, $P(f_X)$, by $f_X^\beta$.  Examples of the resulting flux distributions are shown in Figure \ref{fig:fluxdist}.

\subsection{Cumulative Number-Flux Computation}
\label{sec:maxlikelihood}

The number count distribution is equal to the sum of the flux probability distributions of individual sources, divided by the sensitivity curve calculated in \S \ref{sec:sensitivity} and the recovery fraction function determined in \S \ref{sec:recovery}:
\begin{equation}
N(>f_X) = \int_{f_X}^\infty \left[\sum_{i=1}^{N_{\mathrm{src}}}\frac{P_i(f_X)}{A(f_X,\epsilon_i)F_{\mathrm{rec},i}}\right]df_X
\end{equation}
However, the number count distribution depends on the power-law index $\beta$ through $P(f_X)$.  We estimate $\beta$ using a maximum likelihood (ML) method with power-law differential number-flux Bayesian priors.  The probability of source $i$ being present in our catalog is
\begin{equation}
p_i = \frac{\int P_i(f_X) df_X}{\int dN/df_X|_{\epsilon_i} A(f_X,\epsilon_i) F_{\mathrm{rec},i} df_X}
\end{equation}  
Therefore, the total likelihood of obtaining the sources in our catalog is $\prod_i p_i$.  We find the best-fit power-law index for each quantile group by maximizing the total likelihood for each quantile group model separately.  The normalization, $K$, of each group model was found by computing the differential number counts in 20 flux bins,
\begin{equation}
\frac{dN}{df_X} = \left(\int_{f_{X,\mathrm{min}}}^{f_{X,\mathrm{max}}} \left[\sum_{i=1}^{N_{\mathrm{src}}}\frac{P_i(f_X)}{A(f_X,\epsilon_i)F_{\mathrm{rec},i}}\right]df_X\right)/\left(f_{X,\mathrm{max}}-f_{X,\mathrm{min}}\right)
\end{equation}
calculating 20 corresponding normalizations, and then weight-averaging these normalization values.  We calculate the statistical errors of the number counts using the bootstrap method; we resample our list of catalog sources, determine new best-fit $\beta$ and $K$ parameters, and recompute the number count distribution.  

\section{The Number Count  (log\textit{N}-log\textit{S}) Distribution}
\label{sec:logNlogS}

\begin{figure}[t]
\hspace{-0.2in}
\includegraphics[width=0.5\textwidth]{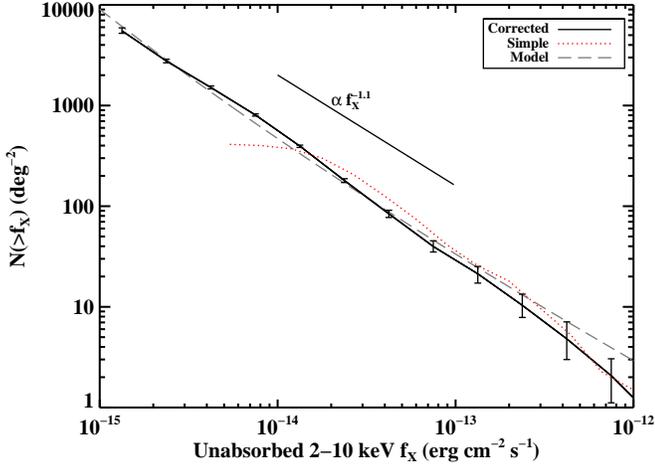}
\caption{Cumulative number-flux distribution for all sources detected at $\geq3\sigma$ in the 2-10 keV band.  The red dotted line is calculated by assigning a single flux value to each detected source and not correcting for any systematic errors.  The bars represent 1$\sigma$ uncertainties.  The gray dashed line represents the sum of the maximum-likelihood models for each quantile group computed as described in \S \ref{sec:methodology}.  The black solid line is the result of using a flux probability distribution for each source and correcting for the Eddington bias, the variations in sensitivity across the surveyed area, and the incompletness of our detection procedure.  A line with a power-law index equal to $-$1.1 is shown as a visual aid. (A color version of this figure is available in the online journal.)}
\label{fig:logNcomparison}
\end{figure}

\begin{figure}
	\vspace{+0.1in}
	\hspace{-0.2in}
	\subfigure{
		\includegraphics[width=0.5\textwidth]{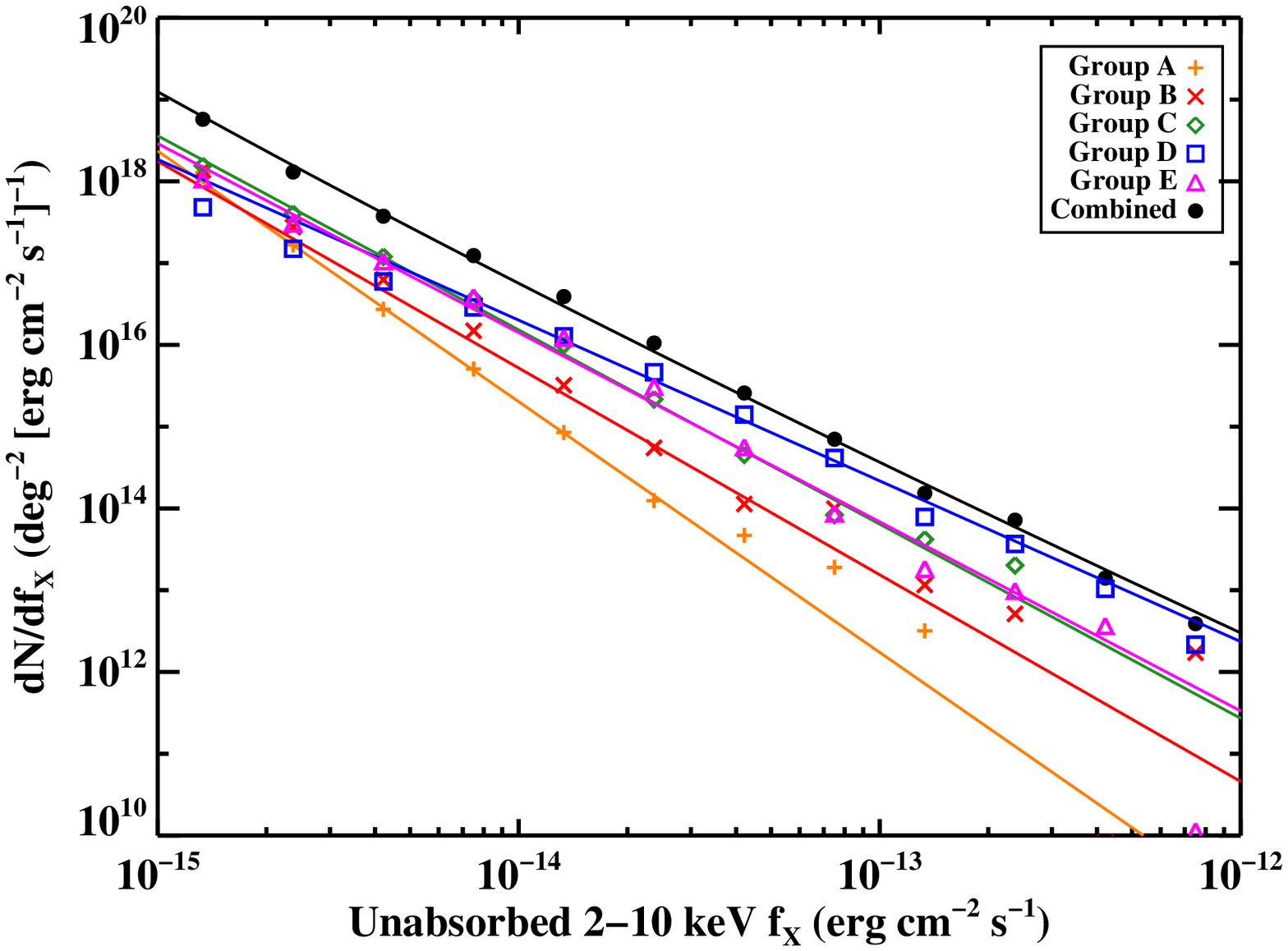}}
	
	\hspace{-0.2in}
	\subfigure{
		\includegraphics[width=0.5\textwidth]{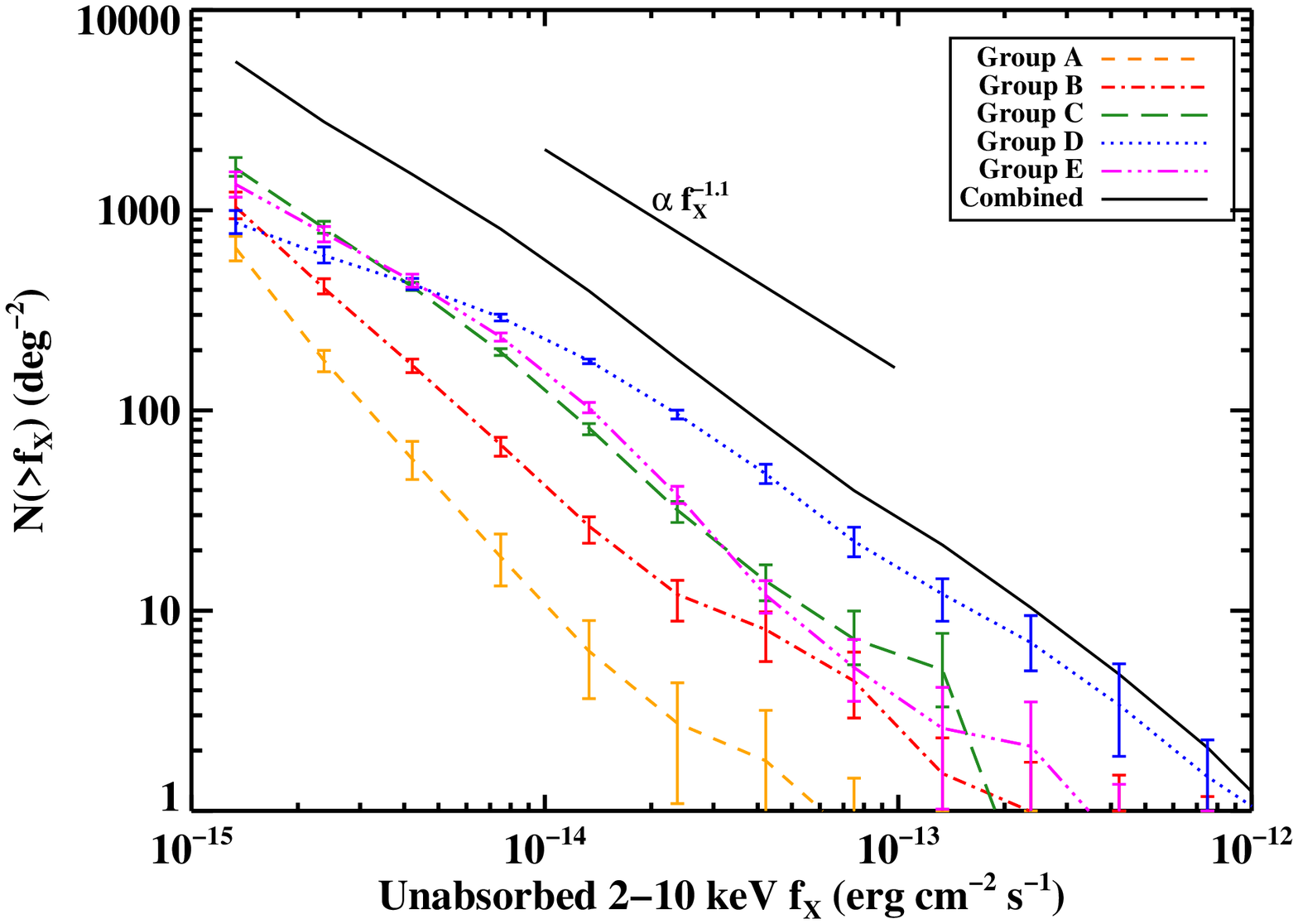}}
\caption{Upper panel shows the differential number counts versus 2-10 keV flux calculated as described in \S \ref{sec:methodology}.  The solid lines represent the maximum-likelihood simple power-law models while the points represent the corrected data with 1$\sigma$ errors.  Lower panel shows the cumulative number counts versus 2-10 keV flux.  Lines represent the corrected data with 1$\sigma$ errors.  Both $dN/dS$ and log$N$-log$S$ are shown for the five quantile groups and for all groups combined. (A color version of this figure is available in the online journal.)}
	\vspace{+0.1in}
\label{fig:logNlogS}
\end{figure}

Figure \ref{fig:logNcomparison} compares the number-flux distribution calculated using the methodology described in \S \ref{sec:methodology} with the ``simple" distribution constructed using a single flux value for each detected source and without corrections for the Eddington bias, sensitivity curve, or recovery fraction.   As can be seen in the figure, we can compute the  number-flux distribution down to a flux limit roughly an order-of-magnitude below the nominal flux limit of the survey, the point at which the ``simple" distribution turns over ($f_X\approx2\times10^{-14}$ erg cm$^{-2}$ s$^{-1}$)  The power-law parameters describing the differential count distribution for each quantile group in the 2-10 keV band that are found to have maximum likelihood are provided in Table \ref{tab:dNdS}, and the differential and cumulative count distributions are shown in Figure \ref{fig:logNlogS}.  The combined cumulative distribution for all groups has a power-law index of roughly $-$1.1.  This index is similar to those found for the cumulative distributions of sources in other surveys, which vary from $-$1.0 to $-$1.5 \citep{muno09,hong09,ebisawa05}.  The group D sources dominate in the 2-10 keV band down to a flux limit of $\approx5\times10^{-14}$ erg cm$^{-2}$ s$^{-1}$, below which group C and E sources dominate.  We modified the quantile groups divisions by $\sim$0.1 dex and re-calculated the number-flux distributions; the power-law indices of the differential count distributions of the modified quantile groups vary by $<2\sigma$, and the log$N$-log$S$ distributions also remain consistent at the $2\sigma$ level or better.  

Groups B, C, and E have differential count distributions with power-law indices that are consistent with each other at $\lesssim1\sigma$ confidence, with $\beta\approx-$2.4.  The similarity between the group B and C slopes was expected since the majority of hard sources in these groups are likely to be a mixture of different types of CVs.  However, it is somewhat surprising that the group E slope is so similar to those of groups B and C since we expect group E to contain a significant fraction of high-mass stars and AGN in addition to white dwarf binary systems.  Perhaps group E does not contain quite as many high-mass stellar X-ray sources as we expect based on the spectral properties of this group, or perhaps the flux distribution of X-ray sources associated with high-mass stars is similar to that CVs.

The group A power-law index is significantly steeper than $\beta\approx-$2.4, which may be because this group is made up of very different X-ray populations, such as low-mass X-ray active stars, coronally active binaries, and high-mass stars.  However, this very steep power-law index may be a result of poor statistics, since only a small number of group A sources detected in the 2-10 keV band.  In addition, since only about 5 group A sources have fluxes higher than fluxes at which the sensitivity curve and recovery fraction corrections become important, its differential count distribution parameters will be more severely impacted than any other group by any systematic imperfections in these corrections.  However, even if the maximum likelihood results for this group are not reliable, the results for all other groups are independent and since group A only contributes $\lesssim$10\% of sources at all fluxes, it also has little impact on the combined group distribution.  

\begin{table}[t]
\centering
\vspace{+0.1in}
\footnotesize
\caption{Maximum Likelihood Parameters for $dN/df_X$ Distributions in 2-10 keV Band}
\begin{threeparttable}
\begin{tabular}{ccc} \hline \hline
Quantile & $\beta$ & $K_{norm}$\\
Group & & (10$^{14}$ deg$^{-2}$ (erg cm$^{-2}$ s$^{-1}$)$^{-1}$) \\
(1) & (2) & (3) \\
\hline
A & -3.1$\pm0.3$ & 20$^{+4}_{-8}$\\
B & -2.5$\pm0.1$ & 52$^{+8}_{-11}$\\
C & -2.37$^{+0.05}_{-0.08}$ & 150$^{+10}_{-20}$\\
D & -1.97$\pm0.05$ & 200$^{+10}_{-30}$\\
E & -2.31$\pm0.05$ & 140$^{+20}_{-50}$\\
\hline
\end{tabular}
\begin{tablenotes}
\item \underline{Notes:} All quoted errors are 1$\sigma$ statistical.

\item (1) Quantile groups defined in \S \ref{sec:quantile}. 

\item (2) Power-law index of $dN/df_X$ distribution.

\item (3) Normalization of $dN/df_X$ distribution.
\end{tablenotes}
\end{threeparttable}
\vspace{+0.1in}
\label{tab:dNdS}
\end{table}

The other power-law index that significantly differs from those of groups B, C, and E is that of group D, which is significantly flatter.  It is not too surprising that the group D slope is different since this group appears to be dominated by a single class of CVs, intermediate polars, rather than a mixture of magnetic and nonmagnetic CVs.  As can be seen in Figure \ref{fig:logNlogS}, only the group D differential-count distribution deviates at $>3\sigma$ confidence from a simple power-law model at fluxes $<10^{-13}$ erg cm$^{-2}$ s$^{-1}$; while other group distributions deviate significantly from a simple power-law model above this flux, there are simply not enough bright sources in each group to constitute a statistically significant sample. The group D distribution deviates from the simple power-law model at fluxes $\lesssim10^{-14}$ erg cm$^{-2}$ s$^{-1}$ by as much as 8$\sigma$ at the faintest fluxes. The turnover in the group D distribution remains significant even when we modify the quantile divisions by $\sim$0.1 dex. Although our sensitivity curve, recovery fraction, and Eddington bias corrections may still not perfectly correct for all these systematic errors, it is unlikely that this group D deviation is simply due to a systematic error since it is the only group displaying this turnover at faint fluxes.  

Thus, the turnover at faint fluxes in the group D number-count distribution is likely indicative of a real break in the power-law distribution.  Such a break could result if the sources in group D have a break in their luminosity function, a minimum luminosity, or a high enough luminosity to be seen through the entire galaxy.  
As discussed in \S \ref{sec:AGN}, 30-50\% of group D sources without IR counterparts may be AGN, which have a number-count distribution that is shallower at fluxes $<10^{-14}$ erg cm$^{-2}$ s$^{-1}$ \citep{cappelluti09}, and therefore, the break in the AGN distribution could at least partly explain the break in the group D distribution. However, if the turnover in the group D distribution is primarily due to AGN, it is surprising that a significant turnover is not also seen in the group E distribution, since there is stronger evidence for AGN being present in group E than group D. Since group D appears to be dominated by IPs, this break in the number-count distribution could be an indication of a break in the luminosity function of IPs, although we emphasize that this is a possibility but it cannot be confirmed with this data alone.  A break in the IP luminosity function could be due to the propeller effect, a centrifugal barrier to accretion at low mass accretion rates that results when the magnetosphere of a compact object has a higher angular velocity than the accretion flow at the Alfven radius \citep{illa75}.  The propeller effect has been invoked to explain the turnover of the HMXB luminosity function at the faint end \citep{shty05}, the variability of supergiant fast X-ray transients \citep[SFXTs,][]{bozzo08}, and the state transitions in some LMXBs \citep{zhang98}.  Although this effect has primarily been used to explain the behavior of low-luminosity accreting neutron stars, similar physical mechanisms may be important in accreting white dwarf systems, even in CVs with weak magnetic fields \citep{matthews06}.  We only present this interpretation of the break in the group D number-count distribution as a speculative hypothesis; a theoretical study of the propeller effect in IPs is beyond the scope of this paper, and multiwavelength follow-up of group D sources that will help to confirm whether they indeed are primarily IPs is ongoing.

\subsection{Comparison to Expectations Based on Previous Surveys}

Having calculated the number-count distribution of the NARCS X-ray sources, we want to compare it to the expected distribution based on other Galactic surveys since any significant discrepancies would indicate that the X-ray populations in this region might be unusual in some way.  Thus, we estimated the expected contributions of ABs, CVs, LMXBs, HMXBs, and AGN to the observed number-count distribution.

In the hard X-ray band, CVs are the most numerous Galactic X-ray sources and they are thought to be the main contributors to the observed Galactic Ridge X-Ray Emission, the large-scale background emission of the Galaxy.  CVs are low-luminosity sources ($L_X \lesssim 10^{33}$ erg s$^{-1}$) and trace the old stellar population of the Milky Way.  Another significant population of low-luminosity sources that follow the stellar mass distribution are ABs.  Thus, to calculate the expected flux distribution of ABs and CVs in NARCS, both their luminosity functions and a model of the Galactic stellar mass distribution are required.  \citet{sazonov06} measured the combined luminosity function of ABs and CVs in the local vicinity of the Sun over the luminosity range $L_X = 10^{27}-10^{34}$ erg s$^{-1}$.  Since the local AB/CV cumulative emissivity per unit stellar mass was found to be consistent with that measured elsewhere in the Galaxy \citep{revnivtsev06,revnivtsev07,krivonos07,revnivtsev08}, we used the AB/CV luminosity function per unit stellar mass from \citet{sazonov06} to estimate the AB/CV flux distribution in NARCS.  We utilize a stellar mass model similar to that used by \citet{sazonov06}, which is an exponential disk with a central hole:
\begin{equation}
\rho \propto \mathrm{exp}\left[-\left(\frac{R_{\mathrm{m}}}{R}\right)^3-\frac{R}{R_{\mathrm{scale}}}-\frac{z}{z_{\mathrm{scale}}} \right]
\end{equation}
 where $R$ is the radial distance from the Galactic center, $z$ is the height above the plane, $R_{\mathrm{m}}$ is the radius of the hole in the Galactic disk, $R_{\mathrm{scale}}$ is the disk scale length, and $z_{\mathrm{scale}}$ is the scale height of CVs.  We assume $R_{\mathrm{m}}$ = 3 kpc \citep{binney97,freuden98}, but adopt a range of values for parameters that are not well constrained: 2.5-3.5 kpc for $R_{\mathrm{scale}}$ \citep{binney97,freuden98,hammer99} and 80-220 pc for $z_{\mathrm{scale}}$ \citep{revnivtsev08}.  For our stellar mass model we futher adopt a disk-to-bulge mass ratio of 2:1 and a range of values for the Galactic bulge mass of $1.3\pm0.5\times10^{10}$M$_{\Sun}$ \citep{dwek95}.  Using this model, the projected stellar mass contained in NARCS is roughly 1.4$\times10^{8} M_{\Sun}$.  

\begin{figure}[b]
\hspace{-0.2in}
\includegraphics[angle=90,width=0.5\textwidth]{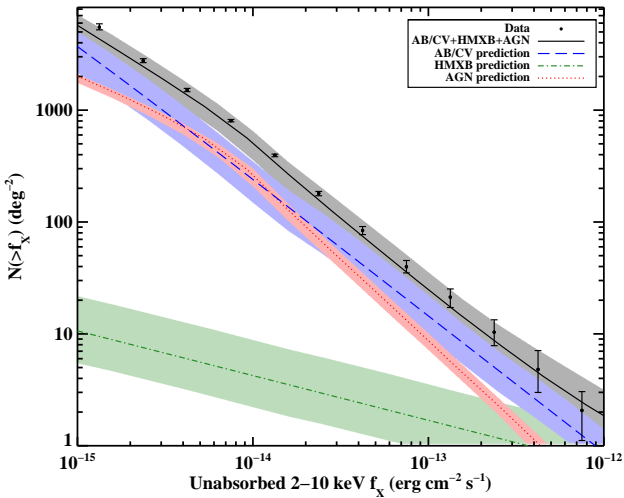}
\caption{The observed number-flux distribution compared to the combined estimates of the expected AB/CV, HMXB, and AGN flux distributions based on the luminosity functions of these populations from other surveys.  Estimated uncertainties for the predictions are shown as shaded regions. The AB/CV line shown is the average distribution of the ones we calculated by varying the parameters of the Galatic stellar mass model. (A color version of this figure is available in the online journal.)}
\label{fig:NSexpected}
\end{figure}

To compute the number of ABs/CVs expected above a given flux, $f_{\mathrm{lim}}$, in our survey, we first consider a small volume element $dV$ within the surveyed volume, and integrate the luminosity function per stellar mass from $L_{\mathrm{lim}} = f_{\mathrm{lim}} \times 4\pi d^2$, where $d$ is the distance to $dV$, to $L_{\mathrm{max}}$.  Multiplying by the stellar mass contained in $dV$ as determined from our stellar mass model distribution then gives the number of ABs/CVs with $f_X > f_{\mathrm{lim}}$ in $dV$.  Integrating the number per volume over all distances $d$ and all lines of sight through our survey, and then simply repeating this procedure for the range $10^{-15} < f_{\mathrm{lim}} < 10^{-11}$, we obtain the expected log$N$-log$S$ for ABs/CVs.  We repeat this calculation many times, choosing values randomly for each of the uncertain stellar mass model parameters.  

LMXBs are more luminous and far rarer than CVs, but they too scale with the old stellar mass.  Thus, to estimate the number of LXMBs expected in NARCS, we take the LMXB luminosity function from \citet{sazonov06} and apply the same method as for ABs/CVs, except that we use a scale height of $z_{scale}$ = 410 pc \citep{grimm02}.  We find that $\lesssim1$ LMXB is expected in our survey region.

Unlike CVs and LMXBs, the number of HMXBs is not correlated with the old stellar mass, and thus our estimate of the HMXBs in NARCS is independent of the Galactic stellar mass model.  Our predictions for the HMXB contribution to the number-count distribution is based on the luminosity function from \citet{lutovinov13}, which is derived from the number of Galactic HMXBs observed by \textit{International Gamma-Ray Astrophysics Laboratory} (\textit{INTEGRAL}, \citet{winkler03}).  Since the \textit{INTEGRAL} luminosity function was calculated in the 17-60 keV band, we converted it to the 2-10 keV band by assuming a typical spectral model for accreting pulsars that includes a power-law with a high-energy cutoff \citep{white83}:
\begin{equation}
f(E) \propto E^{-\Gamma} \times \begin{cases} 1, \hspace{1.2in} (E\leq E_{\mathrm{cut}}) \\ 
\mathrm{e}^{-(E-E_{\mathrm{cut}})/E_{\mathrm{fold}}}, \hspace{0.3in} (E>E_{\mathrm{cut}}) \end{cases} \\
\end{equation}
Using $\Gamma$ = 1, $E_{\mathrm{cut}}$ = 20 keV, $E_{\mathrm{fold}}$ = 10 keV, and intrinsic $N_{\mathrm{H}}$ = 5$\times10^{22}$ cm$^{-2}$ \citep{filippova05}, the conversion factor $f_{2-10\mathrm{keV}}/f_{17-60\mathrm{keV}} \simeq 0.5$.  Since out HMXB prediction is based on a survey of the HMXB population throughout the whole Galaxy, it does not take into account that, as discussed in \S \ref{sec:intro}, the Norma region appears to have an enhanced number of HMXBs due to its star formation activity.  Therefore, we may be underestimating the number of HMXBs in this region, but probably by no more than a factor of two based on the comparison of predictions and observations presented in \citet{lutovinov13} for the brightest HMXBs in the whole Norma arm.  

\begin{table}[b]
\centering
\footnotesize
\caption{Normalizations and Indices of log($N$)-log($S$) Distributions}
\begin{threeparttable}
\begin{tabular}{ccc} \hline \hline
Population & Normalization & Power-law Index \\
(1) & (2) & (3) \\
\hline
Observed & 630 & $\begin{cases} -1.14, f_X <10^{-14} \mathrm{\hspace{0.05in}erg \hspace{0.05in}cm}^{-2} \mathrm{\hspace{0.05in}s}^{-1} \\ -1.24, f_X >10^{-14} \mathrm{\hspace{0.05in}erg \hspace{0.05in}cm}^{-2} \mathrm{\hspace{0.05in}s}^{-1}\end{cases}$\\
AB/CV predicted & 250$\pm$100 & -1.21\\
HMXB predicted & 4$^{+4}_{-2}$ & -0.40 \\
AGN predicted & 260$^{+40}_{-50}$ & $\begin{cases} -0.90, f_X <10^{-14} \mathrm{\hspace{0.05in}erg \hspace{0.05in}cm}^{-2} \mathrm{\hspace{0.05in}s}^{-1} \\-1.46, f_X >10^{-14} \mathrm{\hspace{0.05in}erg \hspace{0.05in}cm}^{-2} \mathrm{\hspace{0.05in}s}^{-1} \end{cases}$\\
\hline
\end{tabular}
\begin{tablenotes}

\item \underline{Notes:}

\item (1) Population of sources, observed or predicted. 

\item (2) Number of sources with $f_X > 10^{-14}$ erg cm$^{-2}$ s$^{-1}$. 

\item (3) Power-law index of distribution.
\end{tablenotes}
\end{threeparttable}
\label{tab:NSdistdata}
\end{table}

One population of sources whose contribution is difficult to predict are isolated high-mass stars and colliding wind binaries (CWBs) because they are not evenly distributed throughout the Galaxy and their X-ray luminosity functions have yet to be determined.  However, we do expect high-mass stellar X-ray sources to be present in our survey due to the presence of massive-star forming complexes and HII regions along this line-of-sight, and we have already identified a small number of these sources.  As discussed in \S \ref{sec:groupA}, three group A sources with known massive counterparts are likely isolated high-mass stars.  In addition, \citet{rahoui14} find that five of 20 NARCS counterparts for which they obtained infrared spectra are high-mass stars; their X-ray properties favor the interpretation that three of these sources (\#239, 1168, and 1326) are quiescent HMXBs and two (\#1278 and 1279) are CWBs.  Ongoing spectroscopic follow-up of IR counterparts and improved constraints on the X-ray properties of NARCS sources with \textit{NuSTAR} observations will help to determine their total number and their flux distribution.  Given the rarity of high-mass stars even in regions of recent star formation like the Norma arm, we do not expect these sources to constitute a large fraction of the NARCS population; however, identifying even a small sample ($\sim$10) of such sources would be a significant contribution to the number of known high-mass X-ray sources (e.g. \citealt{mauerhan10, gagne11}) and our understanding of their X-ray properties.

Figure \ref{fig:NSexpected} shows our estimates for the expected flux distributions of ABs/CVs and HMXBs in our survey region (there are too few expected LMXBs to be shown), while Table \ref{tab:NSdistdata} provides their power-law indices and normalizations.  The observed AGN distribution from the COSMOS survey \citep{cappelluti09} is also included, attenuated by the average column density from the outer Norma arm to the outer edge of the Galaxy ($N_{\mathrm{H}} \approx 3\times10^{22}$ cm$^{-2}$).  Uncertainties in the predicted AGN distribution are determined by considering $N_{\mathrm{H}}$ values from zero to $8\times10^{22}$ cm$^{-2}$, the maximum measured along any line-of-sight in our surveyed area.

As can be seen, the sum of the predicted distributions of ABs/CVs, AGN, and HMXBs matches the calculated flux distribution of hard sources in NARCS.  The fact that the slope of the expected distribution matches the observed one so well, including some of the kinks in the slope, suggests that ABs, CVs, and AGN are the dominant populations in our survey and that their relative numbers are similar to the proportions predicted from other observations.  

\section{Summary and Future Work}
\label{sec:conclusion}

We have presented a catalog of 1129 $\geq3\sigma$ point-like sources and five extended sources detected in a \textit{Chandra} survey of a 2$^{\circ} \times 8^{\circ}$ region in the direction of the Norma spiral arm.  These sources span the luminosity range $L_X=10^{27}-10^{35}$ erg s$^{-1}$.  The systematic positional errors were reduced by matching X-ray sources to infrared VVV counterparts, so that the median positional error for sources in our catalog is $1\farcs26$ (95\% statistical plus systematic uncertainty).  The median number of counts for sources in our survey is 11, making most of them too faint to enable accurate determination of their spectral properties.  Therefore, to help classify the sources, we split them into five spectral groups based on their quantile properties.  The stacked spectra, photometric variability, and IR counterparts of the sources within each spectral group allowed us to identify the classes of X-ray sources that populate the foreground, the Scutum-Crux and near Norma arm, and the far Norma arm.  Foreground sources, which make up roughly 50\% of catalogued sources, are a heterogeneous group, probably containing X-ray active low-mass stars, interacting binaries, symbiotic binaries, and CVs.  The X-ray populations of the Scutum-Crux and near Norma arms are most likely dominated by a mixture of magnetic and nonmagnetic CVs.  The far Norma arm hard X-ray population is likely dominated by IPs, while the softer X-ray population probably includes high-mass stars (both isolated and in colliding wind binaries) and symbiotic binaries.

We also calculated the number-count distribution for sources in our survey down to a flux limit of $10^{-15}$ erg cm$^{-2}$ s$^{-1}$, correcting for the Eddington bias, the variations in sensitivity across the surveyed area, and the incompleteness of our detection procedure.  The observed distribution matches predictions based on AB, CV, AGN, and HMXB luminosity functions very well, lending further support to our conclusions that CVs are the dominant population in NARCS. Furthermore, the fact that the observed number-count distribution shows the same changes in slope as the predicted distribution, suggests that roughly a third of the NARCS sources detected in the hard energy band probably are AGN as predicted; we see some evidence for the presence of AGN in the stacked spectrum of group C, D, and E sources lacking NIR counterparts. However, it is unclear whether AGN can fully account for the flattening at faint fluxes that is seen in the number-count distribution of group D sources; additional X-ray observations would help to disentangle the relative fractions of AGN and IPs in this group and help determine which population is responsible for the break in the log\textit{N}-log\textit{S} distribution.

The analysis we have presented is primarily statistical in nature, but multiwavelength data could permit the classification and deepened understanding of individual sources. Our ongoing follow-up campaigns are focused on but not limited to group D X-ray sources, since any HMXBs in this survey are most likely to belong to this group.   Near-IR spectroscopic follow-up of counterparts to these X-ray sources is ongoing and will help us to determine which of these X-ray sources have high-mass versus low-mass counterparts.  A \textit{Nuclear Spectroscopic Telescope Array} (\textit{NuSTAR}, \citealt{harrison13}) survey of this region will help to constrain the hard X-ray emission from these sources.  The combination of this multiwavelength data should enable us to distinguish HMXBs, which have high-mass stellar counterparts and significant hard X-ray emission, from X-ray emission from shocks in the winds of high-mass stars, which have softer spectra, and IPs, which should have low-mass stellar or accretion disk signatures in the near-IR and significant hard X-ray emission.  This multiwavelength data set will help us to constrain the faint end of the HMXB luminosity function and be useful in advancing our understanding of other Galactic X-ray sources.

\acknowledgments
We thank the referee and the editor for their suggestions, which helped us improve the scientific quality of these results.  We thank G. K. Keating and C. Heiles for helpful discussions about various components of the statistical analysis carried out for this work.  We are also grateful to B. Lehmer for clarifications regarding the maximum likelihood number-flux computation, to J. Hong for providing his quantile analysis tools, and to L. Blitz and M. Ajello for valuable conversations.  The scientific results reported in this article are based on observations made by the \textit{Chandra X-ray Observatory}.  Our analysis of IR counterparts was based on data products from observations made with ESO Telescopes at the La Silla or Paranal Observatories under ESO programme ID 179.B-2002.  This research has made use of software provided by the Chandra X-ray Center (CXC) in the application packages CIAO and Sherpa.  This work was supported in part by NASA through \textit{Chandra} Award Number G01-12068A issued by the \textit{Chandra X-Ray Observatory Center}, which is operated by the Smithsonian Astrophysical Observatory for and on behalf of NASA under contract NAS8-03060.  In addition, FMF received support from the National Science Foundation Graduate Research Fellowship and the Berkeley Fellowship.  FEB acknowledges support from Basal-CATA PFB-06/2007, CONICYT-Chile FONDECYT 1141218 and "EMBIGGEN" Anillo ACT1101, and Project IC120009 "Millennium Institute of Astrophysics (MAS)" of Iniciativa Cient\'{\i}fica Milenio del Ministerio de Econom\'{\i}a, Fomento y Turismo.  

\bibliographystyle{jwapjbib}
\bibliography{refs}

\appendix
\section{Column Descriptions of Catalog Tables}

Below are detailed descriptions of the information provided in the catalog, published in its entirety in the electronic edition.  In these tables, when a value is presented along with its errors, the first column listed in the column range contains the value.  In the case of symmetric errors, the second column contains the error.  In the case of asymmetric errors, the second column contains the upper error and the third column contains the lower error. 

\subsection{Detection and Localization Table}
\label{app:firsttable}
\noindent (1) NARCS catalog source number.

\noindent (2) \textit{Chandra} source name.

\noindent (3) Observation(s) in which \texttt{wavdetect} detects the source.  The format of ObsID numbers is 125XX, where the last two digits are those provided in the catalog.  See \S \ref{sec:detection} for details about \texttt{wavdetect} usage.

\noindent(4-5) Right ascension and declination (J2000.0) of the source.  If the source is detected in multiple observations, the position reported is the weighted average of its positions in different observations.

\noindent(6) Positional uncertainty of the source.  For a source detected in a given observation, this uncertainty is equal to the quadrature sum of the 95\% statistical uncertainty based on Equation 5 of \citet{hong05} and the average systematic uncertainty of positions in that observation after astrometric refinement (see Column 5 in Table \ref{tab:reproject}).  For sources detected in multiple observations, the uncertainties associated with the source position in different obsevations were combined to provide the uncertainty of the weighted average of the source positions.

\noindent(7) Offset angular separation of the source from the center of the observation aim point.  For sources detected in multiple observations, a semicolon-separated list of the offset angle of the source from each observation aim point is provided; the order of offsets matches the order of ObsIDs reported in Column 3.  

\noindent(8-10) Significance of source in the full 0.5-10 keV band, the soft 0.5-2 keV band, and the hard 2-10 keV band.  It is calculated by finding the probability that the source is a noise fluctuation using Equation \ref{eq:probnoise} and using the Gaussian cumulative distribution function to determine the corresponding source significance.  If the source is detected in multiple observations, the reported significance is the sum in quadrature of the source significance in individual observations.

\noindent(11) Radius of the aperture source region.  For most sources, the aperture source region is defined as a circle with radius equal to the 90\% ECF for 4.5 keV photons (see Column 12).  For potentially extended sources, flagged with ``e" (see Column 13), the radius is instead equal to the semi-major axis of the aperture region defined by \texttt{wavdetect}.  In cases where two or more sources have overlapping circular regions, the regions are redefined as a circular core plus an annular pie sector following the guidelines in Table \ref{tab:pie}; in such cases, the radius provided in the catalog represents the outer radius of the pie sector.  For sources detected in multiple observations, a semicolon-separated list of the aperture region radius used in different observations is provided; the order of radii matches the order of ObsIDs reported in Column 3.  

\noindent(12) PSF radius for 90\% ECF for 4.5 keV photons at the detector location of the source.  The PSF radius varies with detector position, generally increasing with increasing offset angle from the observation aim point.  For sources detected in multiple observations, a semicolon-separated list of the PSF radius at the source detector position in different observations is provided; the order of PSF radii matches the order of ObsIDs reported in Column 3.  
  
\noindent(13) An alphabetical list of the possible flags:

\noindent``b" - ``blended": Blended source that is unblended in another observation.

\noindent``c" - ``created": Source noticed by eye but not detected by \texttt{wavdetect}.  The source aperture region was created manually based on the visible position and extent of the source.  The positional uncertainties calculated for such sources underestimate the true uncertainties, since the source is found by eye and not by \texttt{wavdetect}.  

\noindent``e" - ``extended": Possibly extended source.  The semi-major axis of the smallest aperture region defined by \texttt{wavdetect} for such sources is larger than twice the PSF radius reported in Column (12).  These sources are typically detected in images that have been binned by 4$\times$4 or 8$\times$8 pixels.   

\noindent ``id" - ``inspected duplicate": Possible duplicate source flagged for manual inspection.  A ``duplicate" source refers to a single source detected in multiple overlapping observations; sources were considered to be duplicates of one another if the distance between them was smaller than the quadrature sum of their positional uncertainties.  Sources were flagged for manual inspection if: a) they were separated by a distance greater than the quadrature sum of their positional uncertainties but smaller than the simple sum of their positional uncertainties, or b) they were separated by a distance smaller than the quadrature sum of their positional uncertainties but differed in a substantial way (e.g. one is flagged as possibly extended while another is not, one is found to have two duplicates by the distance criterion but these two duplicates of the first source are not found to be duplicates of one another by the distance criterion).  Generally, if sources flagged with ``id" showed consistent photon fluxes and quantile parameters, they were determined to be true duplicates.  

\noindent ``m1", ``m2", or ``m3" - ``modified": In cases where the circular source aperture region overlaps with the aperture region of another source, the source region is modified to reduce overlapping.  See Table \ref{tab:pie} for details.
  
\noindent ``nb" - ``near bright": Source near a very bright source which may be a spurious detection. 

\noindent``nd" - ``not detected": Source is located where at least two observations overlap but it is only detected in one observation.   

\noindent``s" - ``surrounding": A possibly extended source that completely surrounds one or more point sources.  The aperture regions of the surrounded sources are excluded from the aperture region of the source flagged with ``s".

\noindent``vl" - ``variable long": Source determined to be variable on long (hours-days) timescales.  The photon flux in at least one energy band (full, soft, or hard) varies by $\geq 3\sigma$ between different observations.

\noindent``vp" - ``variable probable": Source is probably variable on short (second-hour) timescales.  The K-S test finds the source lightcurve within a single observation to be inconsistent with a constant lightcurve at $\geq 95$\% confidence.

\noindent``vs" - ``variable short": Source is variable on short (second-hour) timescales.  The K-S test finds the source lightcurve within a single observation to be inconsistent with a constant lightcurve at $\geq 3\sigma$ confidence. 

For sources detected in multiple observations, a semicolon-separated list of the flags relevant for the source region in each observation is provided; the order of flags matches the order of ObsIDs reported in Column 3.  

\subsection{Photometry Table}
\label{app:secondtable}

\noindent(1) NARCS catalog source number.

\noindent(2-4) Net source counts in the full 0.5-10 keV band and corresponding 1$\sigma$ errors, calculated as described in \S \ref{sec:photometry}.  For cases in which the estimated background counts in a source aperture region were determined to be greater than or equal to the total number of counts in the source region, then the catalog presents the 90\% upper confidence limit to the net source counts based on the method described in \citet{kraft91}; in such cases, the error columns are left blank.  For sources detected in multiple observations, net counts from different observations were added together and errors combined in quadrature.  

\noindent(5-7) Net source counts in soft 0.5-2 keV band.  Same details as discussed for Columns 2-4 apply.

\noindent(8-10) Net source counts in hard 2-10 keV band.  Same details as discussed for Columns 2-4 apply.

\noindent(11-13) Photon flux in the full 0.5-10 keV band and corresponding 1$\sigma$ errors.  The photon flux was calculated by dividing the net counts by the exposure time and the mean effective area within the source region.  For sources with zero or negative net counts, the catalog provides the 90\% upper limit on the photon flux and leaves the error columns blank.  For sources detected in multiple observations, the average photon fluxes are reported; if a source was found to be variable between observations (flagged as ``vl") then its photon fluxes from individual observations were simply averaged, but otherwise its photon fluxes were weight-averaged.  

\noindent(14-16) Photon flux in the soft 0.5-10 keV band.  Same details as discussed for Columns 11-13 apply.

\noindent(17-19) Photon flux in the hard 2-10 keV band.  Same details as discussed for Columns 11-13 apply.

\noindent(20-21) The median energy of the source and corresponding 1$\sigma$ error.  It is determined from the total counts (not background corrected) in the source region.  For sources detected in multiple observations, the simple average of the energies from individual observations is reported if a source is found to be variable between observations or the weighted-average is reported otherwise.  

\noindent(22-23) The energy below which 25\% of the total source counts reside and corresponding 1$\sigma$ error.  Same details as discussed for Columns 20-21 apply.

\noindent(24-25) The energy below which 75\% of the total source counts reside and corresponding 1$\sigma$ error.  Same details as discussed for Columns 20-21 apply.

\noindent(26-28) The energy flux in the full 0.5-10 keV band and corresponding 1$\sigma$ errors.  This estimate of the energy flux is calculated by multiplying the full band photon flux and the median energy of the source provided in the catalog.  In cases where only an upper limit to the photon flux is available, the 90\% upper limit to the energy flux is reported and the error columns are left blank.    

\noindent(29) Photometric flags.  If the photometric values provided for a source are 90\% upper limits in the full, soft, or hard energy bands, this column displays an F, S, or H, respectively.  

\noindent(30) The spectral group defined using quantile diagrams to which the source belongs.  See \S \ref{sec:specanalysis} for details about quantile analysis and the spectral groups defined in this work.

\subsection{Table of Infrared Counterparts}
\label{app:thirdtable}

\noindent(1) NARCS catalog source number.

\noindent(2) Name of VVV source that is closest to the \textit{Chandra} source position, within $3\sigma$ of the position provided in Table \ref{tab:catalog}.  

\noindent(3) Right ascension (J2000.0) of the VVV source.

\noindent(4) Declination (J2000.0) of the VVV source.

\noindent(5) Angular separation between the \textit{Chandra} and VVV source.  

\noindent(6) Probability that the VVV source is a noise fluctuation, provided in the VVV catalog.

\noindent(7) Reliability of the VVV counterpart calculated according to the method of \citet{suther92}.  The reliability depends on the distance between the X-ray and IR sources, the positional uncertainties of the X-ray and IR sources, and the spatial density of IR sources.  The reliability is expressed as a fraction between zero and one; VVV sources with a higher reliability are more likely to be true IR counterparts to the \textit{Chandra} sources.



\end{document}